\documentclass[5p,times]{elsarticle}
\let\today\relax
\makeatletter
\def\ps@pprintTitle{%
	\let\@oddhead\@empty
	\let\@evenhead\@empty
	\def\@oddfoot{\footnotesize\itshape
		{} \hfill\today}%
	\let\@evenfoot\@oddfoot
}
\makeatother
\usepackage{multirow}
\usepackage{amssymb}
\usepackage{amsmath}
\usepackage{graphicx} 
\graphicspath{{images/}} 

\usepackage{xcolor}
\usepackage{todonotes}
\usepackage{xcolor}
\usepackage{balance}
\usepackage{subfigure}
\usepackage{comment}

\usepackage{url}

\usepackage{color, soul}
\usepackage{microtype}
\soulregister\cite7
\soulregister\ref7
\soulregister\pageref7

\begin{document}

\begin{frontmatter}
\title{SeCTIS: A Framework to Secure CTI Sharing}
\author[inst1]{Dincy R. Arikkat}
\ead{dincyrarikkat@cusat.ac.in}
\author[inst2]{Mert Cihangiroglu}
\ead{mert.cihangiroglu01@universitadipavia.it}
\author[inst3]{Mauro Conti}
\ead{mauro.conti@unipd.it}
\author[inst1]{Rafidha Rehiman K. A.}
\ead{rafidharehimanka@cusat.ac.in}
\author[inst4]{Serena Nicolazzo}
\ead{serena.nicolazzo@unimi.it}
\author[inst2]{Antonino Nocera}
\ead{antonino.nocera@unipv.it}
\author[inst3,inst1]{Vinod P.}
\ead{vinod.p@cusat.ac.in,vinod.puthuvath@unipd.it}
\affiliation[inst1]{organization={Department of Computer Applications},
                 addressline={Cochin University of Science and Technology},
                 country={India}}
\affiliation[inst2]{organization={Department of Electrical, Computer and Biomedical Engineering},
                 addressline={University of Pavia},
                 country={Italy}}
\affiliation[inst3]{organization={Department of Mathematics},
                 addressline={University of Padua},
                 country={Italy}}
\affiliation[inst4]{organization={Department of Computer Science},
                 addressline={University of Milan},
                 country={Italy}}
\begin{abstract}
The rise of IT-dependent operations in modern organizations has heightened their vulnerability to cyberattacks. As a growing number of organizations include smart, interconnected devices in their systems to automate their processes, the attack surface becomes much bigger, and the complexity and frequency of attacks pose a significant threat. Consequently, organizations have been compelled to seek innovative approaches to mitigate the menaces inherent in their infrastructure. In response, considerable research efforts have been directed towards creating effective solutions for sharing Cyber Threat Intelligence (CTI). Current information-sharing methods lack privacy safeguards, leaving organizations vulnerable to leaks of both proprietary and confidential data. To tackle this problem, we designed a novel framework called SeCTIS (Secure Cyber Threat Intelligence Sharing), integrating Swarm Learning and Blockchain technologies to enable businesses to collaborate, preserving the privacy of their CTI data. Moreover, our approach provides a way to assess the data and model quality, and the trustworthiness of all the participants leveraging some {\em validators} through Zero Knowledge Proofs. An extensive experimental campaign demonstrates our framework's correctness and performance, and the detailed attack model discusses its robustness against attacks in the context of data and model quality.
\end{abstract}


\end{frontmatter}
\section{Introduction}

With the advent of Industry 5.0, organizations tend to include smart technology in their systems with the objective of delegating repetitive and time-consuming activities to support devices, such as cobots or smart objects, and digital twins. As a consequence, the attack surface and the potential harm to the safety of cyber-physical systems is expanding exponentially~\cite{li2021comprehensive}. In the event of a cyber incident, having access to timely and relevant threat intelligence can greatly aid in reaction and mitigation efforts. It can provide valuable context about the attacker's Tactics, Techniques, and Procedures (TTPs), helping organizations prevent and remediate the incident more effectively. 

In this context, Cyber Threat Intelligence (CTI, hereafter) has emerged as a powerful tool to gain knowledge and insights about cyber threats and adversaries. CTI refers to systematically collecting, analyzing, and interpreting data related to vulnerabilities, threat reports, and attack trends observed across various sectors. 
CTI enables industries to understand the evolving threat landscape better, anticipate potential cyberattacks proactively, and carry out possible defenses 
\cite{liao2016acing}.

In the pervasive environment, connected smart objects and sensors produce an enormous amount of CTI in multiple forms and types. On the other hand, organizations cannot rely solely on their internally generated CTI to protect themselves; they need to benefit from the knowledge coming from external sources such as network traffic, hacker forums, APT reports, technical blogs, etc. 
CTI Sharing plays a crucial role in enabling organizations to disseminate both raw and processed information. This helps organizations access external threat intelligence and enhance their collective ability to defend against cyber threats \cite{johnson2016guide}. 

However, in practice, CTI Sharing is challenging due to a variety of factors~\cite{nweke2020legal}. First, participants may not want to disclose their identity to avoid damage to the organization's reputation. Unfortunately, this implies that if the source of certain data is unknown, the credibility of shared information is harmed, and trusting relationships among the participating entities cannot be easily established. Another factor linked to entities' trustworthiness that may discourage collaboration is the problem of incomplete or false information. This can contaminate or mislead the algorithms or the results of analysis. Furthermore, not all organizations are inclined to invest additional resources both for interoperability and to ensure that the shared CTI can be automated and easily reusable by the participating entities. Moreover, legal questions should be considered if the information to be shared contains materials protected under data protection and privacy law, antitrust law, or intellectual property law. For instance, in Germany, IP addresses are considered personal information; therefore, any disclosure of CTI including them must comply with German privacy laws~\cite{dunnett2022challenges}. Instead, in the UK, they can be freely shared. These legal and regulatory obligations can pose a significant barrier to international business cooperation. Although several solutions exist to provide organizations with an environment for sharing and consuming CTI~\cite{wagner2016misp,jolles2022building}, they are not designed to assess the privacy and trust of the participants~\cite{dunnett2022challenges}.

We propose a novel framework Secure CTI Sharing (SeCTIS) to contribute to this setting. SeCTIS is an architecture that allows organizations to share CTI data in a privacy-preserving way. Our framework collaboratively trains Machine Learning~(ML) models on CTI data and assesses both the quality of data and models and the trustworthiness of all participants. For this, SeCTIS integrates several technologies, such as Swarm Learning~(SL), Blockchain Smart Contracts, and Zero Knowledge Proof mechanisms.

\par SL has been recently introduced~\cite{han2022demystifying} as a novel paradigm for collaborative and privacy-preserving Machine Learning. Similar to Federated Learning~(FL), SL participants jointly train a global model and update their local model contribution. 
Instead of relying on a single aggregator, SL leverages a Blockchain to coordinate the model aggregation and securely onboard members. The aggregation of local model updates and the subsequent alignment of the local nodes are handled in a decentralized manner. In our framework, the different organizations represented by Swarm Edge Nodes compose a Swarm Network. They aim to collaboratively train a Global Model using their private CTI data and build the Local Models independently without revealing them to other participants. Only model parameters are shared via the Swarm Network. In this way, data security and confidentiality are preserved.

\par SeCTIS also includes some additional steps to assess the quality of the employed CTI data and model and compute the participants' trustworthiness. For this, in each iteration, a set of {\em validators} nodes is randomly selected among the participants to verify the performance of all the local model updates before their aggregation. To protect against malicious actions from the validators, a Zero-Knowledge Proof~(ZKP, hereafter) mechanism is also employed.
Finally, an elected {\em Swarm Aggregator} computes the reputation score of all local models using validator nodes and aggregates the parameter updates from the top-k local models. Subsequently, the aggregated Global Model is uploaded into IPFS and sent back to SL nodes to start the next iteration and continue until convergence. Through this mechanism, the quality of the model and the CTI data is assessed. SeCTIS also evaluates the reputation of participating organizations during the SL model training. Organizations' reputation is estimated by considering the quality of their contribution (local model updates) during each SL iteration.


In summary, the main contributions of this paper, intended to solve the major challenges in CTI Sharing, are as follows:
\begin{itemize}
    \item \textbf{Privacy-preserving CTI data sharing}: Our framework adopts an SL Network to generate a CTI Model in a distributed and collaborative way, guaranteeing organizations' data confidentiality.
    \item \textbf{Trust among participants and quality of CTI data}: SeCTIS provides a process based on {\em validator} nodes to assess  CTI data and model quality using reputation scores. In addition, through the ZKP mechanism, {\em validator} activities can be verified.
    \item \textbf{Interoperability and automation}: SL provides a middleware that can manage heterogeneous data formats and establish a unique methodology to be employed because only model parameters are exchanged to train the deep learning model. 
    \item \textbf{Legal liabilities}: Keeping CTI data confidential may reduce legal risks concerning information disclosure, making organizations more willing to participate in SL-sharing schemes.
\end{itemize}

Our paper is organized as follows. Section~\ref{sec:related} describes the main works related to our approach. Section \ref{sec:background} delves into the details about CTI, Blockchain, Federated Learning, Swarm Learning, and Zero-Knowledge Proof concepts that are essential to the understanding of our solution. Section~\ref{sec:approach} describes the main components of our framework and the steps we performed to secure CTI Sharing. In Section~\ref{sec:attackModel}, we present our attack model, demonstrating our approach as robust to possible attacks against data and model quality. Section~\ref{sec:experiment} deals with the experimental
campaign used to assess the performance of our solution, including the setup, results, and possible limitations of our system. An analysis of the security of validators' operations through ZKP is reported in Section~\ref{security-analysis-of-zkp-operations}. Section~\ref{sec:robustnessofzk}, instead, discusses some example attack scenarios and analyze the behavior of our solution. The performance impact of the inclusion of ZKP in our solution is evaluated in Section~\ref{sec:ZKPTime}. Finally, Section~\ref{sec:conclusion} concludes the work and presents possible future directions. 


\section{Related Work}
\label{sec:related}
In this section, we describe the related systems currently adopted to share CTI data and their characteristics.

Although lots of organizations still rely on informal means~(i.e., phone calls or emails) for sharing CTI-related information, recently, there has been a growing interest in dedicated platforms to facilitate the automated or semi-automated sharing of CTI data inside connected communities~\cite{wagner2019cyber}. 
Threat Intelligence Platforms~(TIPs, hereafter) are specialized software that helps industries collect and analyze real-time threat information from various sources to support defensive tactics. While numerous TIPs are available in the market, most are offered under commercial licenses. For instance, VirusTotal\footnote{\url{https://www.virustotal.com}}
is one of the leading CTI service
able to analyze suspicious files, domains, IPs, and URLs to detect malware and other breaches, produce threat reports, and automatically share them with the security community.
Another similar open-source solution, known as MISP (Malware Information Sharing Platform)~\cite{wagner2016misp}, aims to gather, store, and distribute cybersecurity IoCs and CTI reports, both within the security community and beyond. MISP offers a range of features, such as an indicator database, automated correlation, sharing capabilities, a user-friendly interface, and compatibility with various data formats and standards. 

The authors of~\cite{sauerwein2021threat} assess nine TIPs such as ThreatStream\footnote{\url{https://api.threatstream.com/}}, ThreatQ\footnote{\url{https://www.threatq.com/}}, ThreatConnect\footnote{\url{https://threatconnect.com/}}, Open Threat Exchange~(OTX)\footnote{\url{https://otx.alienvault.com/}}, MISP, IBM X-Force Exchange\footnote{\url{https://exchange.xforce.ibmcloud.com/}}, Falcon X Intelligence Crowdstrike\footnote{\url{https://go.crowdstrike.com/}}, Collective Intelligence Framework (CIF)\footnote{\url{https://csirtgadgets.com/collective-intelligence-framework}}, and Collaborative Research into Threats (CRITs)\footnote{\url{https://crits.github.io/}} by examining how they align with the CTI life cycle. Their investigative case studies uncovered that the current focus of these platforms, similar to the ones described previously, is primarily on the pre-processing and dissemination stages.

However, Joll{\`e}s et al.~\cite{jolles2022building} analyzed ThreatFox\footnote{\url{https://threatfox.abuse.ch}}, a free platform for IoCs Sharing similar to VirusTotal and MISP. Their findings revealed that building collaborative cybersecurity on an established network of trust is a crucial dynamic for this kind of platform.

A similar platform called ETIP (Enriched Threat Intelligence Platform) \cite{gonzalez2021etip} focuses on the collection and processing of structured data sourced from external sources, encompassing OSINT feeds, along with information originating from an organization's network infrastructure. ETIP includes the following components: {\em(i)} an input module responsible for the collection and standardization of IoCs from OSINT feeds and the monitoring infrastructure; {\em(ii)} an operational module, which produces enriched IoCs and evaluates threat data using a threat score; and {\em(iii)} an output module for the presentation of the outcomes and their sharing with external entities to strengthen cybersecurity defenses. This platform eliminates duplicate IoCs, creates composed IoCs, and assigns a threat score to each IoC to help Security Operations Center~(SOC) analysts prioritize security incident investigations.

Haque et al.~\cite{haque2021toward} underline the significance of adopting an automated strategy for sharing CTI while emphasizing the effectiveness of Relationship-Based Access Control for facilitating this sharing. Their approach aims to identify, generate, and disseminate structured CTI and implement these concepts through a prototype Automated Cyber Defense System in a cloud-based environment.
In response to the challenges related to trust in the source and integrity of threat intelligence data, Preuveneers et al. \cite{preuveneers2020distributed} improved the security framework TATIS~\cite{preuveneers2020tatis}. Both TATIS \cite{preuveneers2020tatis} and the framework in \cite{haque2021toward} guarantee that only authorized individuals can access sensitive data when it is being transferred between various threat intelligence systems. Moreover, in the proposal of \cite{preuveneers2020tatis}, encryption is applied using the Ciphertext-Policy Attribute-Based Encryption (CP-ABE) cryptographic scheme to protect the shared data.

Sharing CTI data can greatly improve IT security, but it faces several challenges, such as expenses, risks, and legal requirements. To overcome these issues, Riesco et al.~\cite{riesco2020cybersecurity} proposed an approach to encourage the sharing of CTI among various stakeholders by leveraging blockchain technology and smart contracts. Their research suggests creating a marketplace on the Ethereum blockchain where participants can exchange CTI tokens as digital assets, thereby incentivizing sharing while addressing potential storage limitations and transaction costs. 
Menges et al.~\cite{menges2021dealer} presented DEALER, a system promoting secure CTI sharing by providing incentives and addressing compliance concerns. Like our approach, DEALER relies on Blockchain technology and an InterPlanetary File System (IPFS) distributed hash table. It employs unbiased quality metrics for reputation assessment and protects buyers and sellers through dispute resolution and cryptocurrency rewards. However, the authors recommended limiting the platform's use to sharing noncritical data, as it may not be suitable for sharing highly sensitive and critical CTI. 
However, these studies~\cite{riesco2020cybersecurity,menges2021dealer} do not employ Federated Learning. The BFLS~\cite{jiang2023bfls} approach ensures the security of CTI data sharing by combining FL for training threat detection models and blockchain for decentralized aggregation. Specifically, the consensus protocol of the blockchain is enhanced to filter and select high-quality CTIs for participation in FL. Smart contracts are utilized to automate the aggregation and updation of models, ensuring efficient and secure CTI sharing. Sarhan et al.~\cite{sarhan2022hbfl} proposed a Hierarchical Blockchain-based Federated Learning~(HBFL) framework for collaborative IoT intrusion detection. The framework ensures secure and privacy-preserved collaboration by leveraging a permissioned blockchain and smart contracts. Moulahi et al~\cite{moulahi2023privacy} used blockchain technology and FL to safeguard data integrity and aggregation in detecting cyber-threats within Vehicular Ad Hoc Networks~(VANET) and Intelligent Transportation Systems~(ITS). Their approach involves uploading vehicle-generated models onto a blockchain-based smart contract for aggregation before returning them to the vehicles.

\par Table~\ref{tab:related} provides a summary analysis of the contributions of the existing related works compared to ours.
In this table, the core features of SeCTIS are shown, demonstrating that, to the best of our knowledge, a complete framework providing secure CTI Sharing~(in terms of data privacy, model quality, and participants' reputation) still does not exist in the present literature.

\begin{table*}
\caption{CTI Sharing Systems features}
\scriptsize
\centering
\begin{tabular}{|l|lllll|}
    \hline
    \textbf{System} & \textbf{Threat Detection} & \textbf{CTI Sharing} & \textbf{Data Privacy} & \textbf{Model Quality} & \textbf{Participant Reputation}\\
    \hline 
    VirusTotal & \checkmark & \checkmark & - & - &  - \\
    MISP \cite{wagner2016misp}& \checkmark & \checkmark & - & - &  - \\
    ThreatFox & \checkmark & \checkmark & - & - &  - \\
    ThreatStream & \checkmark & \checkmark & - & - &  - \\ ThreatQ & \checkmark & \checkmark & - & - &  - \\
    ThreatConnect & \checkmark & \checkmark & - & - &  - \\
    OTX & \checkmark & \checkmark & - & - &  - \\ 
    IBM X-Force Exchange & \checkmark & \checkmark & - & - &  - \\
    Falcon X Intelligence Crowdstrike & \checkmark & \checkmark & - & - &  - \\
    CIF & \checkmark & \checkmark & - & - &  - \\ 
    Collaborative CRITs & \checkmark & \checkmark & - & - &  - \\
    Haque et al. \cite{haque2021toward} & \checkmark & \checkmark & - & - & \checkmark\\
    ETIP \cite{gonzalez2021etip}& \checkmark & \checkmark & - & - & -\\
    DEALER\cite{menges2021dealer} & \checkmark & \checkmark & - &\checkmark  & -\\
    TATIS \cite{preuveneers2020tatis}& \checkmark & \checkmark & \checkmark  & - & \checkmark \\
    Riesco et al.~\cite{riesco2020cybersecurity} & \checkmark & \checkmark & - & - & -  \\ 
    BFLS~\cite{jiang2023bfls}& \checkmark & \checkmark & \checkmark  & \checkmark &  -\\
    HBFL~\cite{sarhan2022hbfl}& \checkmark & \checkmark & \checkmark  & - &  -\\
    Moulahi et al~\cite{moulahi2023privacy} & \checkmark & \checkmark & \checkmark  & - &  -\\\hline
    \textbf{SeCTIS} & \checkmark & \checkmark & \checkmark  & \checkmark & \checkmark \\
    \hline
\end{tabular}
\label{tab:related}
\end{table*}

\section{Background}
\label{sec:background}
In this section, we delve into some useful basic concepts to understand the framework described in our paper. In particular, we define the CTI scenario and the phases that compose its lifecycle. Then, we illustrate the fundamental notions of Blockchain. Moreover, we describe the main concepts related to FL and Swarm Learning, their workflow, and such approaches' principal differences and challenges. Finally, we provide details about the Zero-Knowledge Proof and Zero-Knowledge Machine Learning approaches.
Table \ref{tab:SystemSymbols} summarizes the acronyms used in the paper.

\begin{table}
\scriptsize
\centering
\caption{List of the acronyms used in the paper} \label{tab:SystemSymbols}
\begin{tabular}{|l|l|}
\hline
    \textbf{Acronyms} & \textbf{Description}\\
    \hline \hline
    CTI & Cyber Threat Intelligence \\\hline
    EVM & Ethereum Virtual Machine \\\hline
    FL & Federated Learning \\\hline  
    GM & Global Model \\\hline
    IoT &Internet of Things \\\hline
    IPFS & InterPlanetary File System \\\hline
    LM & Local Model \\\hline
    ML & Machine Learning \\\hline
    SC & Smart Contract \\\hline
    SL & Swarm Learning \\\hline
    SMC & Secure Multiparty Computation \\\hline
    TIP & Threat Intelligence Platforms \\\hline
    TTP & Tactic, Technique, and Procedure \\\hline
    ZKP & Zero-Knowledge Proof \\\hline
\end{tabular}
\end{table}

\subsection{Cyber Threat Intelligence}

With the term Cyber Threat Intelligence~(CTI)
, we refer to a set of data regarding security threats, threat actors, exploits, malware, vulnerabilities, and indicators of compromises that can help organizations, governments, and individuals in decision-making for proactive cybersecurity defense~\cite{shackleford2015s,sun2023cyber}.
CTI data is usually shared in a textual and unstructured form in several online data sources, such as blogs, forums, Online Social Networks (OSNs, for short), or Dark Net Marketplaces. Hence, an iterative process consisting of several phases should be followed to provide valuable insights and transform raw data into actionable intelligence. The main CTI phases
can be grouped into six stages \cite{arazzi2023nlp}, namely:

\begin{itemize}
    \item \textbf{Planning and Direction}: phase identifies the main stakeholders and defines the organization's objectives, priorities, and requirements.
    \item \textbf{Data Collection}: phase in which the data sources are identified and CTI data, including IoCs, malware samples, and network traffic logs, are collected through automated tools and manual research.
    \item \textbf{Data Processing}: a phase that involves the transformation and cleaning of raw data into a structured format.
    \item \textbf{Analysis}: phase in which patterns, trends, and potential threats are identified.
    \item \textbf{Dissemination}: phase that consists of sharing data with relevant stakeholders.
    \item \textbf{Feedback}: phase in which the effectiveness of the actions taken and the overall intelligence process are considered to refine and improve future iterations of the CTI lifecycle.
\end{itemize}

\subsection{Blockchain}

\begin{figure}[ht]
	\centerline{
        \includegraphics[scale=0.65]{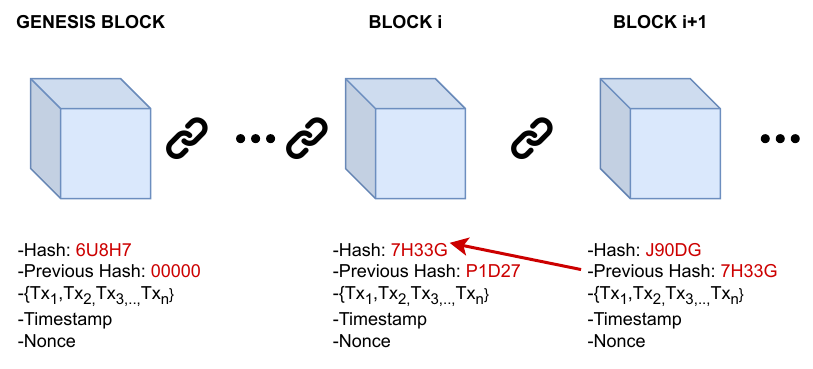}
    }
    \caption{Example of a Blockchain} \label{fig:Blockchain}
\end{figure}

Blockchain technology refers to a decentralized solution based on a distributed ledger mechanism. It records immutable transactions across multiple parties to provide tamper resistance and security without relying on any centralized trusted third party. This technology has been originally conceptualized as the underlying framework for Bitcoin, the first cryptocurrency, but its possible applications have developed far beyond this initial scenario~\cite{nakamoto2008bitcoin,arazzi2024novel,arazzi2023fully}.
Transactions, as visible in Figure~\ref{fig:Blockchain}, are grouped into blocks and are the fundamental units of a Blockchain. Each transaction represents the transfer of value or digital assets from one participant to another.

Each block includes the transactions, the previous block's hash value, a timestamp, and a nonce (a random number for verifying the hash). Due to the presence of a unique hash value, once generated, the information within each block cannot be altered. This ensures the network's immutability. In the first generation of Blockchain technology that handles cryptocurrencies, whenever a new transaction is created, it undergoes validation and verification through a consensus protocol carried out by the {\em miners}. Miners generate a new block of transactions after solving a mathematical puzzle called {\em Proof of Work}~(PoW) and then
propagate that block to the network. Other nodes in the network can validate the correctness of the generated block and only build upon it if both the transactions included in a new block and of the block itself are considered valid.

\par Ethereum\footnote{\url{https://ethereum.org}} has emerged as the second generation of Blockchain to
allow the building of complex distributed applications beyond the cryptocurrencies through the development of Smart Contracts. 
A smart contract is an executable code that automatically runs and enforces the terms of an agreement once the specified conditions are met~\cite{buterin2014next}. After being deployed on the Blockchain, the contract operates autonomously, and its code cannot be altered.
Usually, it is initiated by activating its constructor function via a transaction submitted to the network. This constructor function is then executed, and the resulting smart contract code is permanently stored on the Blockchain~\cite{khan2021Blockchain}. The execution of the smart contract is validated by a consensus mechanism called {\em Proof of Stakes} (PoS). Validator nodes who hold and ``stake'' a certain amount of cryptocurrency are chosen in a deterministic and pseudo-random way to create new blocks and validate transactions based on the amount of cryptocurrency they own and are willing to ``stake'' as collateral. Validators are incentivized by earning transaction fees.

The following Blockchain categories can be defined \cite{yang2020public}:

\begin{itemize}
    \item \textbf{Permissioned Blockchains} usually entails a set of participants who must obtain authorization to join the network, perform transactions, and validate blocks. Transactions are grouped, accessed, and verified by a designated group of nodes instead of anonymous miners. 
    \item \textbf{Permissionless Blockchains} allow anyone to join and participate without demanding prior authorization. Transaction verification relies on the work of many anonymous miners competing to solve a complex mathematical algorithm for that block of transactions via a trial-and-error approach.
    \item \textbf{Private Blockchains} are restricted to authorized participants. A single entity decides who can join the network and has full authority over the blockchain's management.
    \item \textbf{Public Blockchain} is an open and permissionless network accessible to anyone. Control is shared among all participants through consensus mechanisms. Since transactions are open to the public to verify, the risk of hacking and data manipulation is low even if information privacy can be menaced \cite{yang2020public}.
    \item \textbf{Hybrid Blockchain} integrates public and private Blockchain elements. 
    \item \textbf{Consortium Blockchains}, like Hybrid Blockchains, have private and public features. Access to the network is restricted to predetermined organizations or entities who jointly control the network. Decisions require consensus among the consortium participants.
\end{itemize}

\subsection{Federated Learning and Swarm Learning}

Both Federated Learning (FL) and Swarm Learning (SL) are decentralized approaches to machine learning that enable model training across distributed devices without the need to centrally share raw data. Since data is not transferred and centralized, these two methods have advantages regarding privacy preservation and network traffic reduction.

As for FL, the main actors of this protocol are $\mathcal{C}$ client devices~(or ``workers''), owning sensitive data and running local training on them; and a central server~(or ``aggregator''), that organizes the whole FL process aggregating the local updates. 
In particular, FL's goal is to train a global model $\mathbf{w}$ by uploading the weights of local models from workers
$\{\mathbf{w}^i|i \in \mathcal{C}\}$ to the parametric aggregator optimizing a loss function:

\begin{equation}
    \min\limits_{\mathbf{w}} l(\mathbf{w}) = \sum_{i=1}^n{\frac{s_i}{\mathcal{C}}L_i(\mathbf{w}^i)}
\end{equation}
\noindent
where $L_i(\mathbf{w^i})= \frac{1}{s_i}\sum_{j \in I_i}{l_j(\mathbf{w}^i, x_i)}$ is the loss function, $s_i$ is the local data size of the {\em i}-th worker, and $I_i$ identifies the set of data indices
with $|I_i|=s_i$, and $x_j$ is a data point.

\begin{figure}
    \centering
    \includegraphics[width=0.35\textwidth]{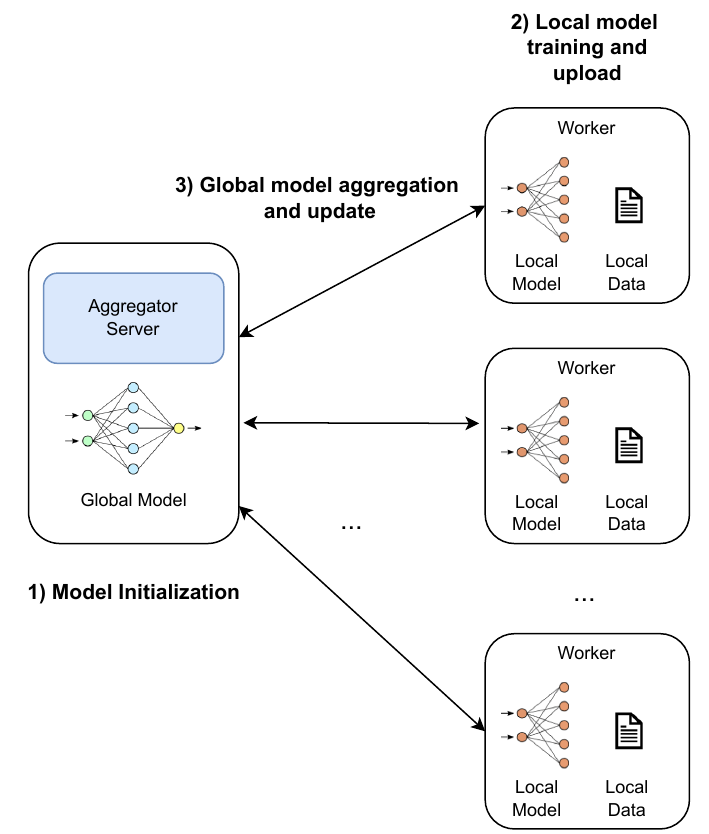}
    \caption{The Federated Learning workflow}\label{fig:workflowFL}
\end{figure}

As visible in Figure \ref{fig:workflowFL}, the basic FL workflow consists in the following phases~\cite{zhang2021survey}:

\begin{enumerate}
    \item \textbf{Model initialization}: phase in which the aggregator or server sets all the parameters useful for the global ML model $\mathbf{w}$ to their initial status. Moreover, this step also selects the random workers to be included in the process. 
    \item \textbf{Local model training and upload}: phase in which the workers execute local training using their private data after downloading the current global model. Then, each client computes the model parameter updates and sends them to the aggregator or server. The local training typically implicates multiple iterations of gradient descent, back-propagation, or other optimization methods to enhance the local model's performance. Specifically, at the {\em t}-iteration, each client updates the global model with the contributions coming from their datasets: $\mathbf{w}^i_t \leftarrow \mathbf{w}^i_t - \eta \frac{\partial L(\mathbf{w}_t,b)}{\partial \mathbf{w}^i_t}$ (where $\eta$ identify the learning rate and $b$ is local batch).
    \item \textbf{Global model aggregation and updation}: phase in which the central aggregator collects and aggregates the model parameter updates from all the workers, $\{\mathbf{w}^i|i \in \mathcal{C}\}$. The central server can employ different aggregation approaches like averaging, weighted averaging, or Secure Multi-party Computation~(SMC) to combine the received updates from each client.
\end{enumerate}

The security and fault tolerance of FL have been increasingly discussed because the central aggregator, keeping model parameters,
can be vulnerable to malicious attacks or system failures  \cite{nasr2019comprehensive,arazzi2023turning}. To address these problems, Swarm Learning (SL) has been recently introduced \cite{han2022demystifying}.

\begin{figure}
    \centering
    \includegraphics[width=0.45\textwidth]{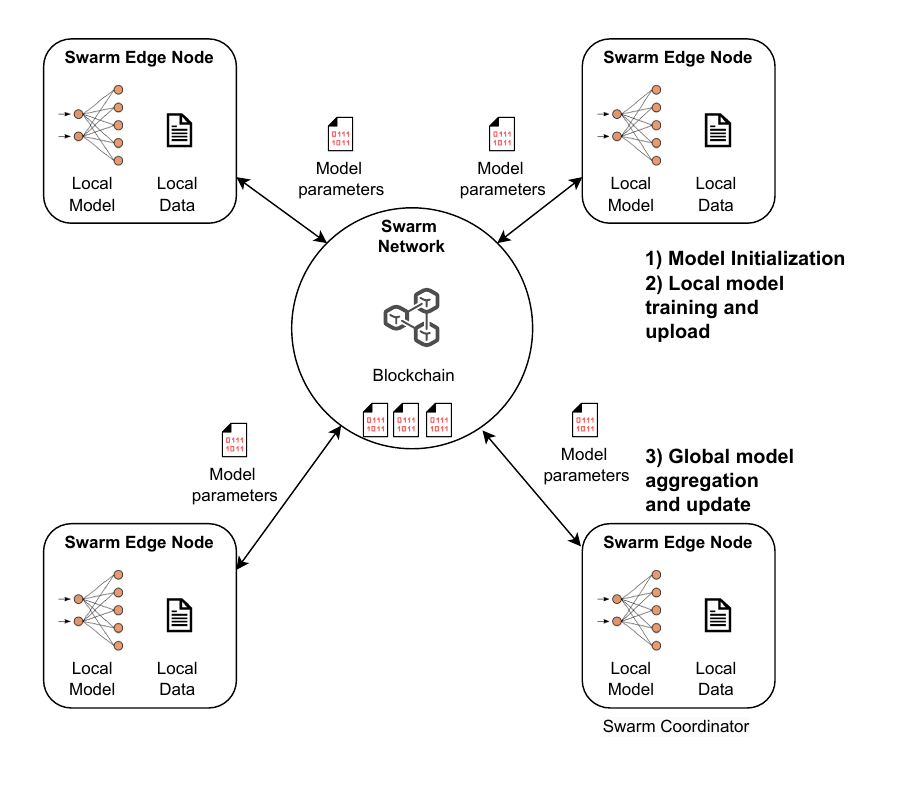}
    \caption{The Swarm Learning workflow}\label{fig:SLWorkflow} 
\end{figure}

As visible in Figure~\ref{fig:SLWorkflow}, SL exploits a Blockchain instead of a central aggregator server to securely onboard members and dynamically elect the leader. This allows for the performance of DL to be extremely decentralized. 

Moreover, it shares the model parameters via the Swarm Network and builds the models independently on private data at {\em Swarm Edge Nodes}, without the need for a central aggregator. In the workflow of SL, a new edge node enrolls via a Blockchain smart contract, obtains the model, and performs localized training until
a given interval. Then, local model parameters are exchanged between participants and combined to update
the global model before the next training round. 
A {\em Swarm coordinator} can be elected randomly\footnote{Due to deterministic characteristics of blockchains, randomization can be achieved through third-party services such as Chainlink available for public blockchain~\cite{kaleem2021demystifying}.} during each iteration and is responsible for maintaining metadata like the model state, training progress, and licenses without model
parameters.

\subsection{Zero-Knowledge Machine Learning}
Zero-Knowledge Proof or Zero-Knowledge Protocol~(ZKP, hereafter) is a cryptographic protocol, originally presented in~\cite{goldwasser1985knowledge}, that enables one party~(called the {\em prover}) to prove to another party~(called the {\em verifier}) that a piece of information is true 
without disclosing the actual details of that statement~(without revealing any information beyond the statement's validity).

The main features of the ZKP system include the following properties \cite{wu2014survey}:
{\em (i)} {\em Completeness}, which involves the fact that if the statement is correct, the verifier will always accept it; {\em (ii)} {\em Soundness}, which enforces the fact that if the statement is incorrect, the verifier will always reject it; {\em (iii)} {\em Zero Knowledge} means that no~(malicious) verifier can get any extra information from the proof, 
     except the correctness of the statement defined before.

These properties make ZKP widely used in  the 
context of Secure Multiparty Computation~(SMC), privacy and authentication. Zero Knowledge Machine Learning~(ZKML) is the application of ZKP on ML models in which parameters and operations are concealed from the verifier. The prover can demonstrate the computational correctness of the ML models without disclosing undesired information, promoting transparency and trust. The potential application of ZK to ML could determine that a particular piece of content is produced by applying a specific ML model to a given input~\cite{ganescu2024trust}. Interestingly, ZKML allows users to specify the desired information included in the proof, such as model parameters, input, output, or none.
 
\section{Proposed Approach}
\label{sec:approach}
In this section, we present a general overview of our approach. The framework aims at secure CTI sharing, allowing participants to train a Global Model~(GM, hereafter) collaboratively without revealing confidential information. Furthermore, our reputation approach provides an additional feature, indeed a set of {\em Validator} nodes is in charge of evaluating the quality of the local model contributions. The reputation mechanism detects low-quality models and prevents them from joining the aggregation. Moreover, the trustworthiness of {\em Validator} nodes is assessed by Zero-Knowledge Proof in our strategy. 

\begin{figure*}[ht]
	\centerline{
        \includegraphics[scale=0.55]{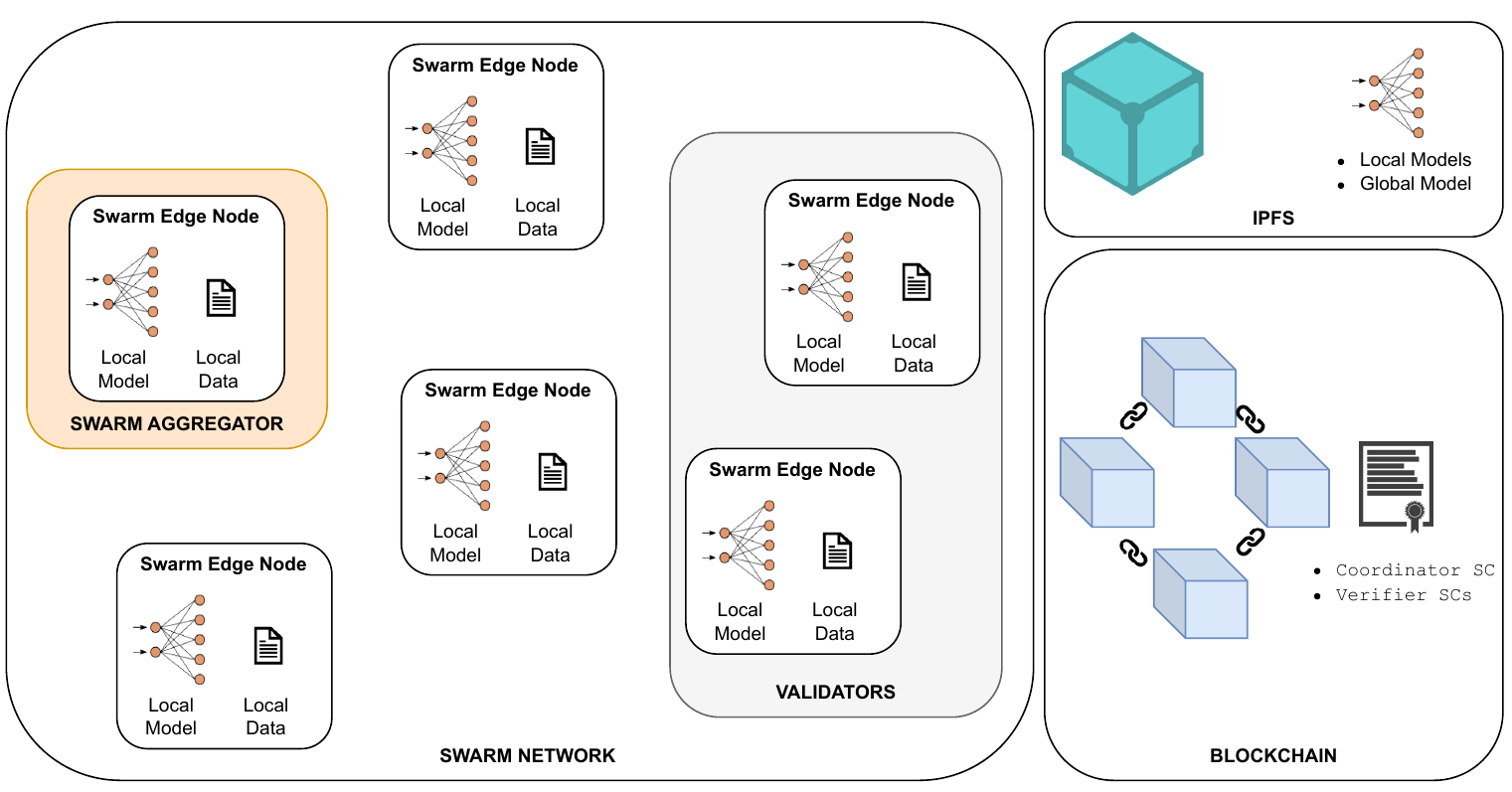}
    }
    \caption{The General SeCTIS Architecture} \label{fig:generalArchitecture}
\end{figure*}

A general architecture of our solution is reported in Figure \ref{fig:generalArchitecture} with 
the following actors:

\begin{itemize}
    \item \textbf{Swarm Edge Nodes}. These nodes represent different organizations that are the basic participants of our framework. They are both consumers and producers of shared CTI information, holding private local data and training the local model according to the SL mechanism. They are also called ``clients'' or ``workers'' of the SL model.
    \item \textbf{Validator Nodes}. They are chosen among the Swarm Nodes according to their computed reputation (see Section~\ref{sub:GMComputation} for details on the selection strategy). They rank all the local contributions based on reputation score for each iteration to obtain a GM with the best performance. Their trustworthiness is assessed via a mechanism based on ZKP.
    \item \textbf{Swarm Aggregator Node}. A randomly selected node is in charge of aggregating the best local models at each iteration.
     \item \textbf{IPFS}. This distributed File System stores the Global Model and the different Local Models for each iteration. 
    \item \textbf{Blockchain}. It is employed to provide a distributed ledger that stores the results of ZKP algorithms and transactions. Moreover, it allows the execution of several Smart Contracts, such as {\tt Coordinator SC} and {\tt Verifier SC}. {\tt Coordinator SC} contains the necessary functions to orchestrate the framework, like storing the verification contract addresses for all the local models at each iteration. It also maintains IPFS model addresses (for both the local and global models), computes trust and reputation scores, and keeps track of the iteration number.
    {\tt Verifier SC}, instead, receives the proof generated by the validator and produces a response based on its validity.
   
\end{itemize}

In practice, our solution comprises three main steps, namely: 

\begin{enumerate}
    \item \textbf{Local Models Training}. In this phase, the SL training iterations take place for each client.
    \item \textbf{Validators Verification}. In this phase, for each iteration, a ZKML algorithm is computed to verify the validators and assess their trustworthiness.
    \item \textbf{Global Model Aggregation}. In this step, the Global Model is aggregated by a selected node. This Aggregator ranks nodes according to a reputation score and aggregates the top-k model. 
\end{enumerate}

In the following sections, we detail each phase of our framework SeCTIS.

\subsection{Local Models Training}
\label{sub:localModelTraining}

This phase starts after the system setup and model initialization step, in which the different organizations register to the Blockchain and join the network.

For each iteration, the workers (i.e., the different organizations) download the Global Model from the IPFS. After that, local training with the CTI private data can be conducted for every client node. When the Local Model (LM) is computed, it is uploaded to the IPFS by the corresponding node. Moreover, the node submits the hash of the LM that identifies the model itself to the {\tt Coordinator SC} for the subsequent phases. Also, the {\tt Verifier SC} is deployed to the Blockchain, and its contract address is submitted to the {\tt Coordinator SC}. 

In summary, the Local Model Training phase comprises the following steps:
\begin{enumerate}
    \item Download the Global Model from IPFS. 
    \item Execute local training using their private CTI data.
    \item Upload the model into IPFS.
    \item Submit the IPFS 
    address (the hash) of their local models to {\tt Coordinator SC}.
    \item Deploy the {\tt Verifier SC} to the Blockchain. 
    \item Get the {\tt Verifier SC} address.
    \item Submit the {\tt Verifier SC} address to {\tt Coordinator SC}. 
\end{enumerate}

\begin{figure}[ht]
	\centerline{
        \includegraphics[scale=0.55]{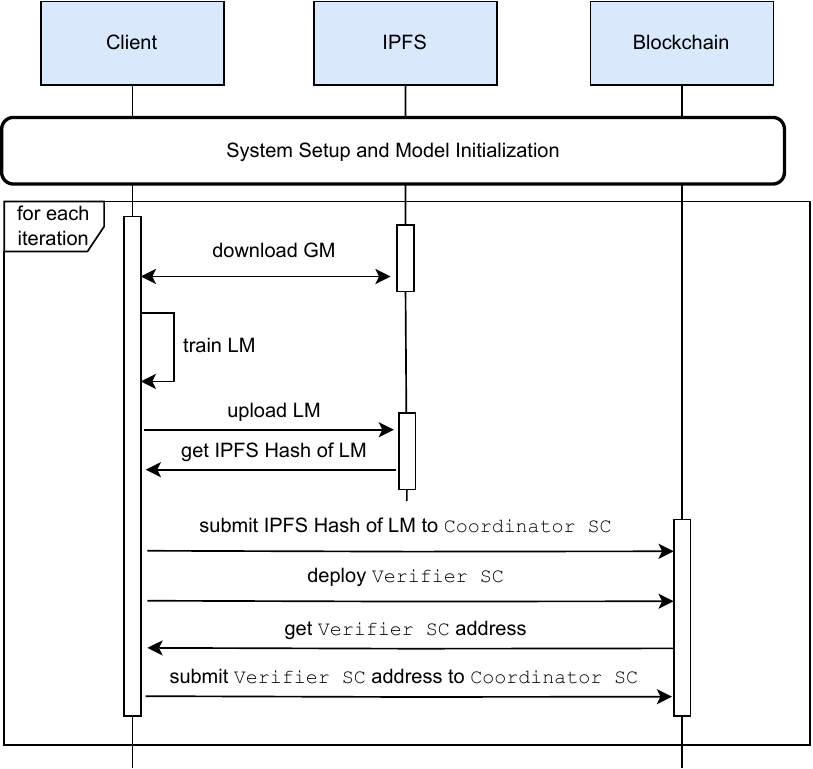}
    }
    \caption{Local Models Training Sequence Diagram} %
    \label{fig:LocalModelTraining}
\end{figure}

Figure \ref{fig:LocalModelTraining} shows a sequence diagram of all the steps performed during this first phase of the framework.

\subsection{Validators Verification} 
\label{sub:validatorsVerification}

In this phase, {\em Validator} nodes are in charge of testing all the LMs and producing proof of their trustworthiness. At each iteration of the framework, validators are chosen randomly among all the participant nodes. They have to query the {\tt Coordinator SC} and get the 
hashes of all the LMs. This is to be tested and identify them in the distributed filesystem and download them.
After that, they execute their test data on all the LMs to assess the quality of these models.
Since the validators can be malicious or malfunctioning and their test data can be corrupted, an immutable proof has to be run for each of them to assess the quality and trustworthiness of these nodes. In particular, each validator generates a Zero-Knowledge Proof containing: {\em(i)} a digest of the input data batch, and {\em(ii)} a digest of the model weight. 
Through this proof, we can assess that the validator's behavior is equal for all the clients. Specifically, this proof verifies that a given validator has tested all the models with the same data points in the same order. After the proof generation, for each tested model, the validator submits these proofs to associated {\tt Verifier SC}. Observe that for each iteration, there is a {\tt Verifier SC} for each participant node. 
In summary, during this phase of the framework, each {\em validator} node does the following steps for each iteration: 

\begin{enumerate}
    \item It queries the {\tt Coordinator SC} and gets the 
    hashes of all the LMs to be tested. 
    \item It downloads all the LMs from the IPFS.
    \item It executes its test data on LMs 
    \item It generates the ZKP. 
    \item It submits the proof along with its results (i.e., the outputs of the tested models) to the a {\tt Verifier SC} for each LM update. 
 \end{enumerate}

\begin{figure}[ht]
	\centerline{
        \includegraphics[scale=0.55]{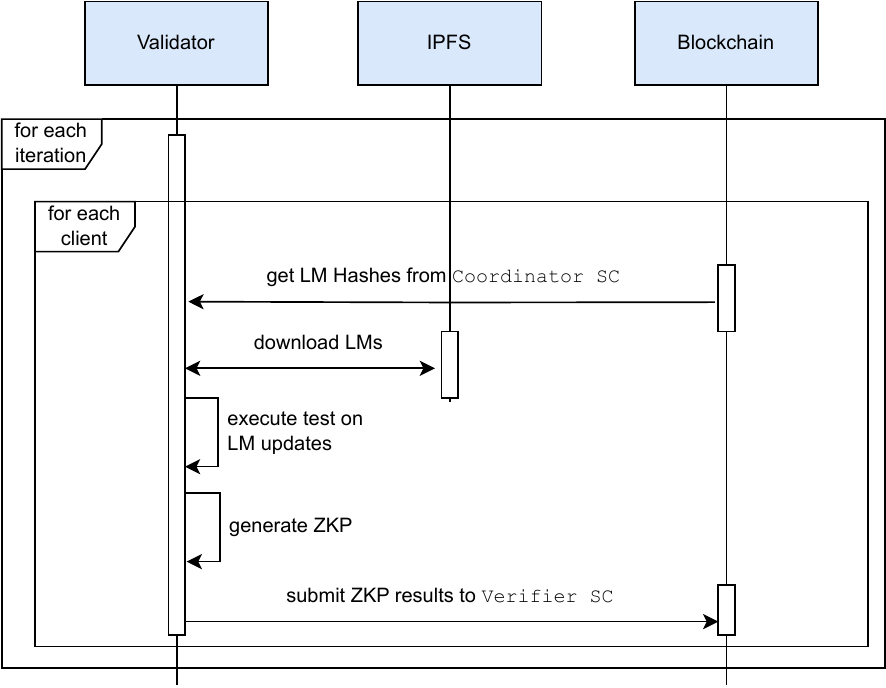}
    }
    \caption{Validators' Verification Sequence Diagram}
    \label{fig:validatorsVerification}
\end{figure}

Figure \ref{fig:validatorsVerification} shows a sequence diagram of all the steps performed in the second phase of SeCTIS.

\subsection{Global Model Computation}
\label{sub:GMComputation}
During the last phase of our framework, a GM is computed by aggregating the local model updates.
However, a data quality mechanism is applied to produce a GM that considers only the best local contributions.
To do this, as a first step, the {\tt Coordinator SC} randomly selects an Aggregator node among all the participants. This node has to compute the LMs' trust and reputation scores for all the validators. Then, it submits all these scores to the {\tt Coordinator SC}.
After that LMs are ranked and the top-$k$ models are aggregated. 
Finally, the aggregator publishes the results in the Blockchain.

In summary, the final stage of our framework consists of the following steps:

\begin{enumerate}
    \item An aggregator is randomly selected by the {\tt Coordinator SC}. 
    \item The Aggregator collects the outputs of validators for each LM. 
    \item The Aggregator computes the trust and reputation scores from the outputs that each validator has submitted.
    \item It updates the trust and reputation values in {\tt Coordinator SC}.
    \item A global rank of LM is computed.
    \item The top-k LMs are used to update the GM. 
    \item The Aggregator publishes the results in the Blockchain.
\end{enumerate}

\begin{figure}[ht]
	\centerline{
        \includegraphics[scale=0.55]{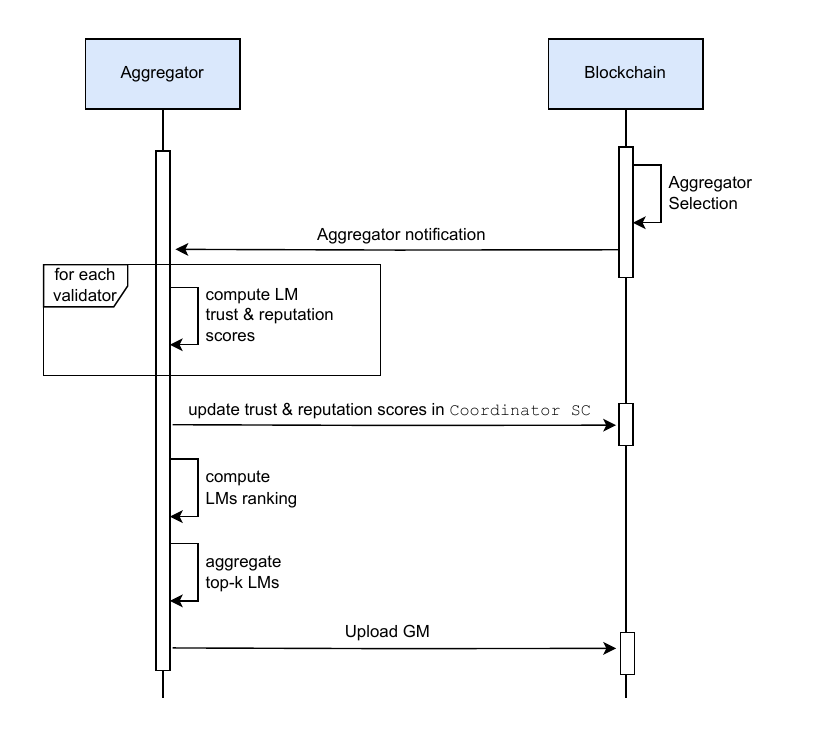}
    }
    \caption{Global Model Computation Sequence Diagram}
    \label{fig:GMComputation}
\end{figure}

Figure \ref{fig:GMComputation} shows a sequence diagram of all the steps performed for the Aggregator selection and computation of the GM.

In the following, we define our trust and reputation model to assess the trustworthiness of nodes. These metrics are useful to select the {\em validator} nodes for the subsequent iteration. Indeed, If a participant's reputation is damaged over a set number of interactions, the organization they represent may be suspended or even removed from the system. Hence, it can no longer be selected as a validator or participate in the framework. The node reputation is computed through the model trust scores for each iteration and is defined as follows.

Given the set of local models $\{m_1, m_2, \cdots, m_z \}$, let $m_i$ be the $i_{th}$ local model. 
For each validator $j$, test results are deployed into the Blockchain in the form of a $n-tuple$ containing the following information: 

\begin{enumerate}
    \item a digest of the employed test set $TS_j$ with size $|TS_j| = s$,
    \item a digest of the verified model $\mathcal{H}(m_i)$,
    \item the output of the model for each data point in $TS_j$.
\end{enumerate}

The output of the $i_{th}$ model is a vector of probabilities $\mathcal{P}_i^d[p_1, p_2, \dots, p_n]$ for each data point $d$ of $TS_j$, where $n$ is the number of possible classes. This vector contains the probabilities that $d$ belongs to each considered class (or label).
The result can be represented as follows:

\begin{center}
$<\mathcal{H}(TS_j), \mathcal{H}(m_i), \mathcal{M}_{n \times s}>$
\end{center}

\noindent
where $\mathcal{H}$ is the hash function to compute the digest, and $\mathcal{M}_{n \times s}$ is the matrix containing all the probabilities vectors returned by the $i_{th}$ model for the data points of $TS_j$.

To compute the trust score, we start by defining the average error $P_{ij}^{k}$, where $i$ is the $i_{th}$ model, $j$ is the $j_{th}$ validator and $k$ is the iteration, according to the following equation:

\begin{equation}
    P_{ij}^{k}= \frac{\sum_{d \in TS_j}{\lVert\mathcal{P}_i^d-\mathcal{P}_c^d\rVert}}{|TS_j|} 
\end{equation}

\noindent
where $\mathcal{P}_c^d$ is the centroid of the different output vectors produced by all the local models for the data point $d$. In practice, we consider each probability vector as coordinates of a point in an Euclidean $n$-space. Hence, given a point $d$ of the test set $TS_j$ of $j$, we locate all the outputs produced by the available local models on $d$ in the Euclidean space and identify the most central one as the reference centroid for $d$. Finally, $\lVert\mathcal{P}_i^d-\mathcal{P}_c^d\rVert$ denotes 
the distance between the output of the $i_{th}$ model for the data point $d$ and the related centroid. Higher values of $P_{ij}^{k}$ mean that the results of the $i_{th}$ model differ greatly from the average.

Now we define $P_{i}^{k}$ as the average value of $P_{ij}^{k}$ for each validators, for the $k_{th}$ iteration and for the $i_{th}$ model, as:

\begin{equation}
    P_{i}^{k}= \frac{\sum_{j \in V}{P_{ij}}}{|V|}
\end{equation}

\noindent where $V$ is the set of all validators, and 
$|V|$ is its size.

The trust value for the $i_{th}$ model at the $k_{th}$ iteration can be computed as:

\begin{equation}
T_i^{k}= 1-P_i^{k}
\end{equation}

Finally, the reputation score for the $i_{th}$ model at the $k_{th}$ iteration is represented by the following equation:

\begin{equation}
    R_i^{k}= (1- \alpha) R_i^{k-1} + \alpha T_i^k
\end{equation}

\noindent
where $\alpha$ is a weight parameter ranging from [0,1]. The lower the parameter $\alpha$ the higher the importance of past reputation scores for the given node.
 
\section{Attack Model}
\label{sec:attackModel}
In our attack model, we focus on possible attacks to the mechanism of assurance of data and model quality rather than addressing common threats like poisoning attacks, backdoors, or denial-of-service (DoS) in Swarm Learning, as discussed in~\cite{chen2023backdoor,bector2023poisoning}. 

We can identify two main phases in which data can be altered in our framework, namely: {\em(i)} during the data collection and {\em(ii)} during the labeling phases.
As for the first case, the different organizations usually gather data from IoT sensors or custom applications. These sources can be compromised or malfunctioning, and data can present errors, be incomplete, or be false. Moreover, the labeling phase, since it has to be carried out manually by some experts, can produce unintentional errors and noisy labels~\cite{song2022learning}. In our attack model, the only type of malicious attack to model and data quality we consider is {\em label flipping} attack. The {\em label flipping} attack is a form of data poisoning where an adversary intentionally mislabels data. 
This threat aims to 
corrupting the model's training data and potentially leading to degraded performance or incorrect predictions \cite{abrishami2022classification} (see Section \ref{sub:reputation_results1} for further detail on this attack).

Moreover, the {\em validator} nodes pose potential threats. 
There is a risk of collusion between certain validators and Byzantine clients \cite{guerraoui2024byzantine} - clients containing noisy labels that can behave differently. This collusion enhances the reputation of the Byzantine clients while diminishing that of benign clients. In this attack model, a malicious validator selects a set of verification data tailored to a specific colluding malicious client, as explained in Section~\ref{sub:reputation_results1}. 

To fortify our framework against this form of attacks, we incorporate both the Zero-Knowledge Proof~(ZKP) and a reputation mechanism designed to foster consensus among validators. These approaches ensure that the behavior of validators remains consistent across all clients. Additionally, this approach ensures that each validator tests all models using the same data points and in the same sequence, thereby maintaining the integrity of the validation process and mitigating the risk of collusion.

Finally, by combining ZPK, Smart Contracts, and Blockchain technologies, we can guarantee that SeCTIS is secure and can assess the trustworthiness of all the participants of our framework. Hence, classical attacks on trust and reputation systems~(such as Slandering, Whitewashing, or Sybil attacks \cite{hoffman2009survey}) are implicitly solved by design.

\begin{figure*}[t]
    \centering
    \subfigure[10\% Malicious Client]
    {
    \includegraphics[width=0.31\textwidth]{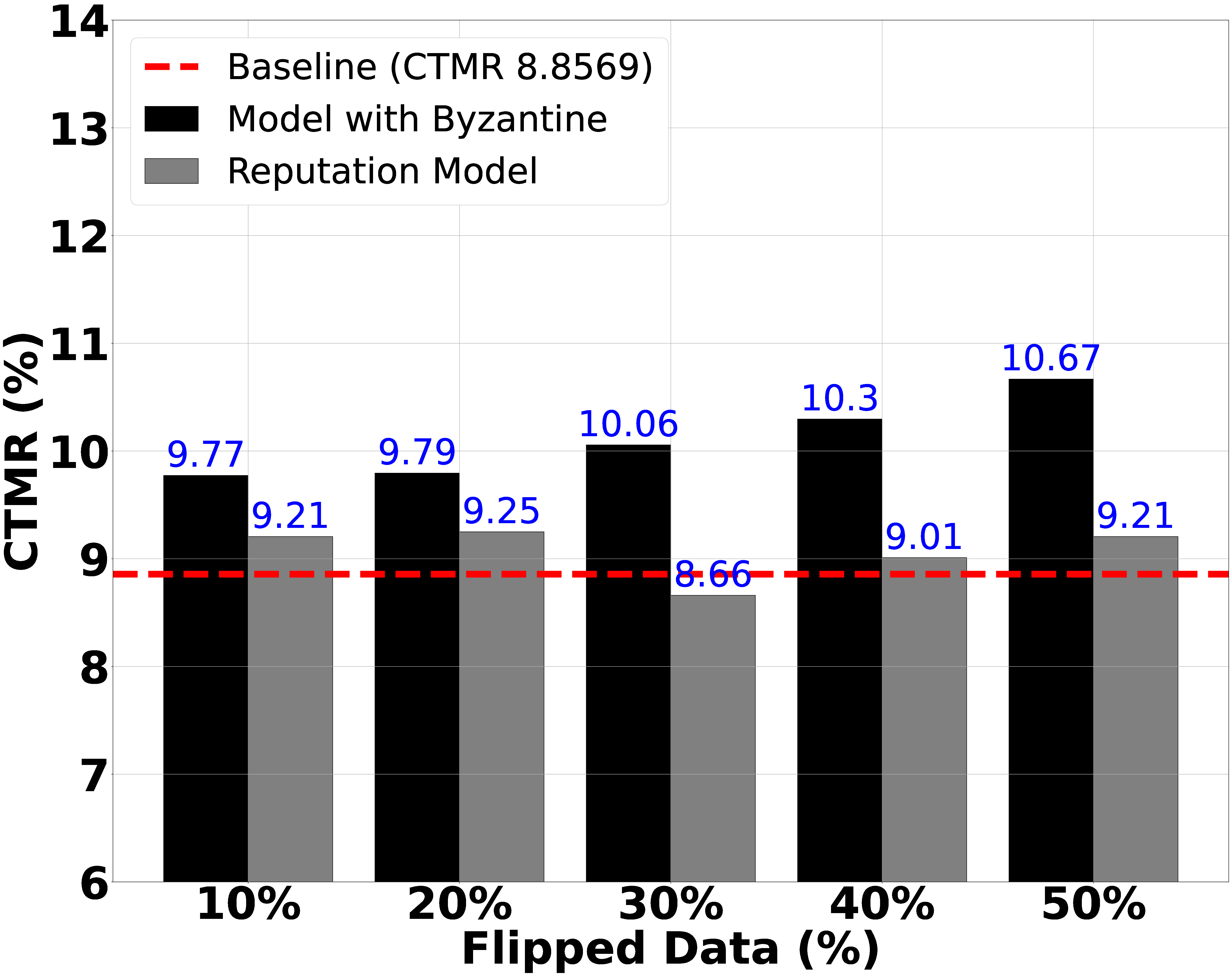}
    \label{fig:1_client_mcr_reputation}
    } 
    \subfigure[20\% Malicious Client]{
    \includegraphics[width=0.31\textwidth]{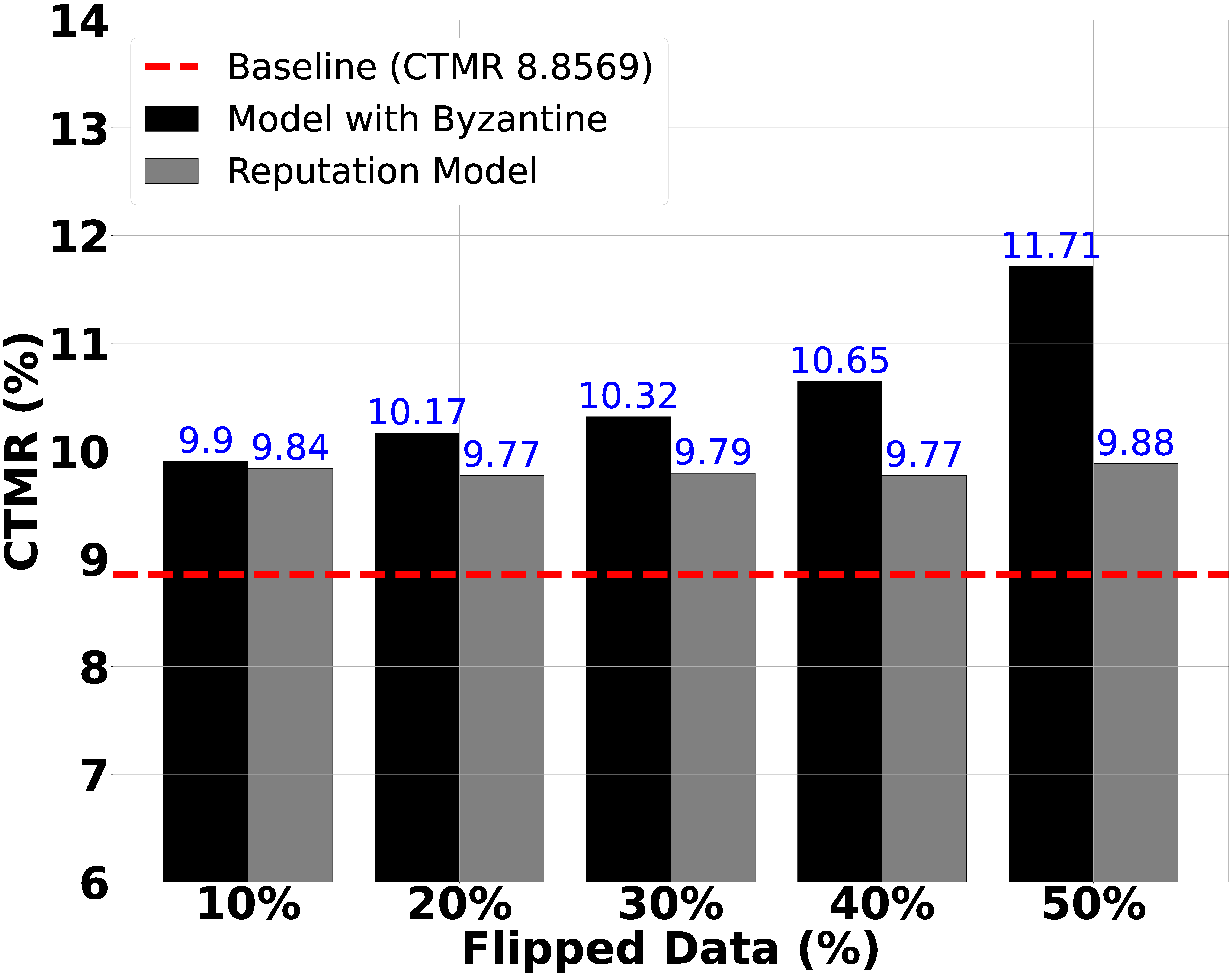}
    \label{fig:2_client_mcr_reputation}
    } 
    \subfigure[30\% Malicious Client]
    {
    \includegraphics[width=0.31\textwidth]{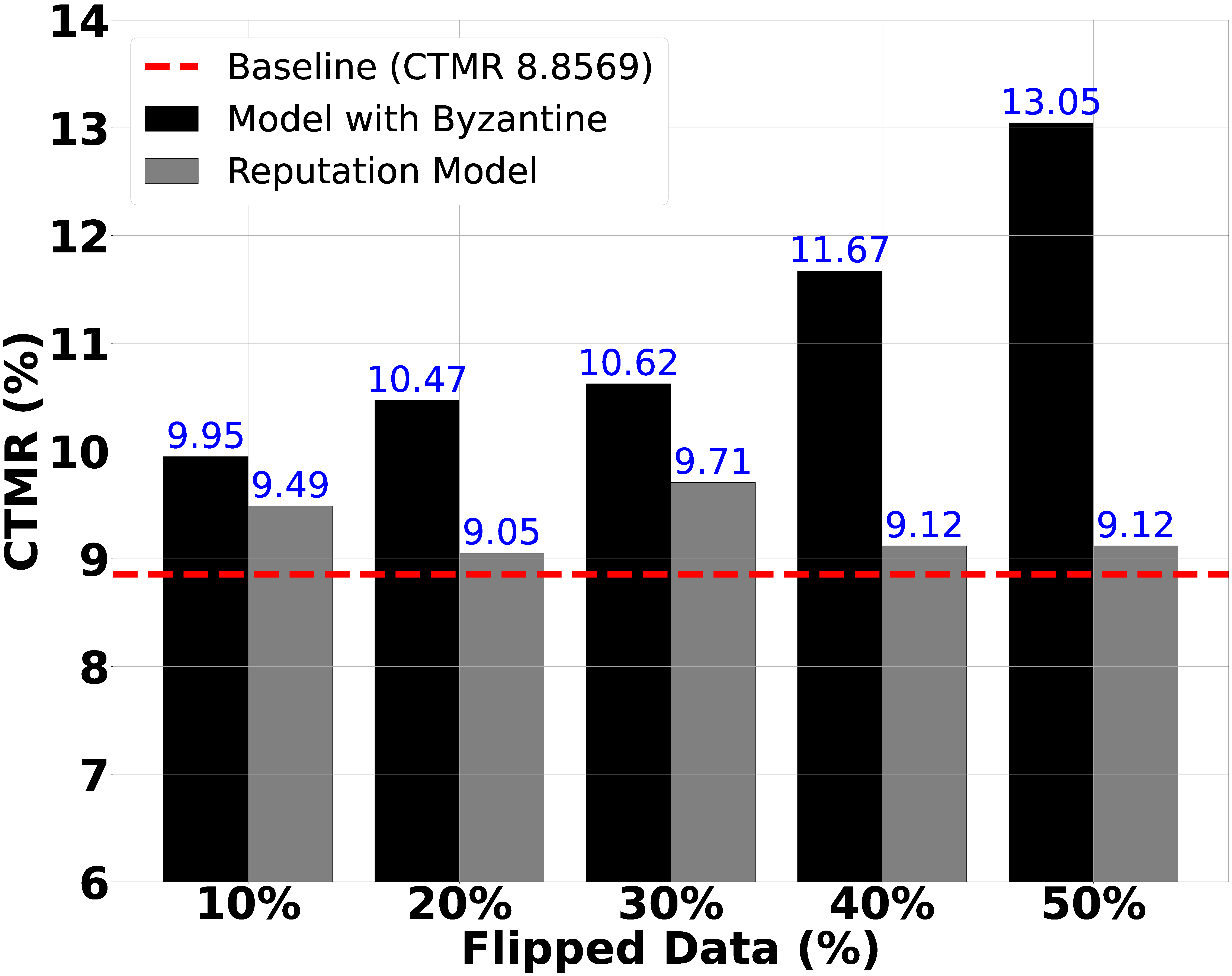}
    \label{fig:3_client_mcr_reputation}
    }    
    \caption{Misclassification Rate} 
    \label{fig:mcr_reputation}
\end{figure*}
\begin{figure*}[t]
    \centering
    \subfigure[10\% Malicious Client]
    {
    \includegraphics[width=0.31\textwidth]{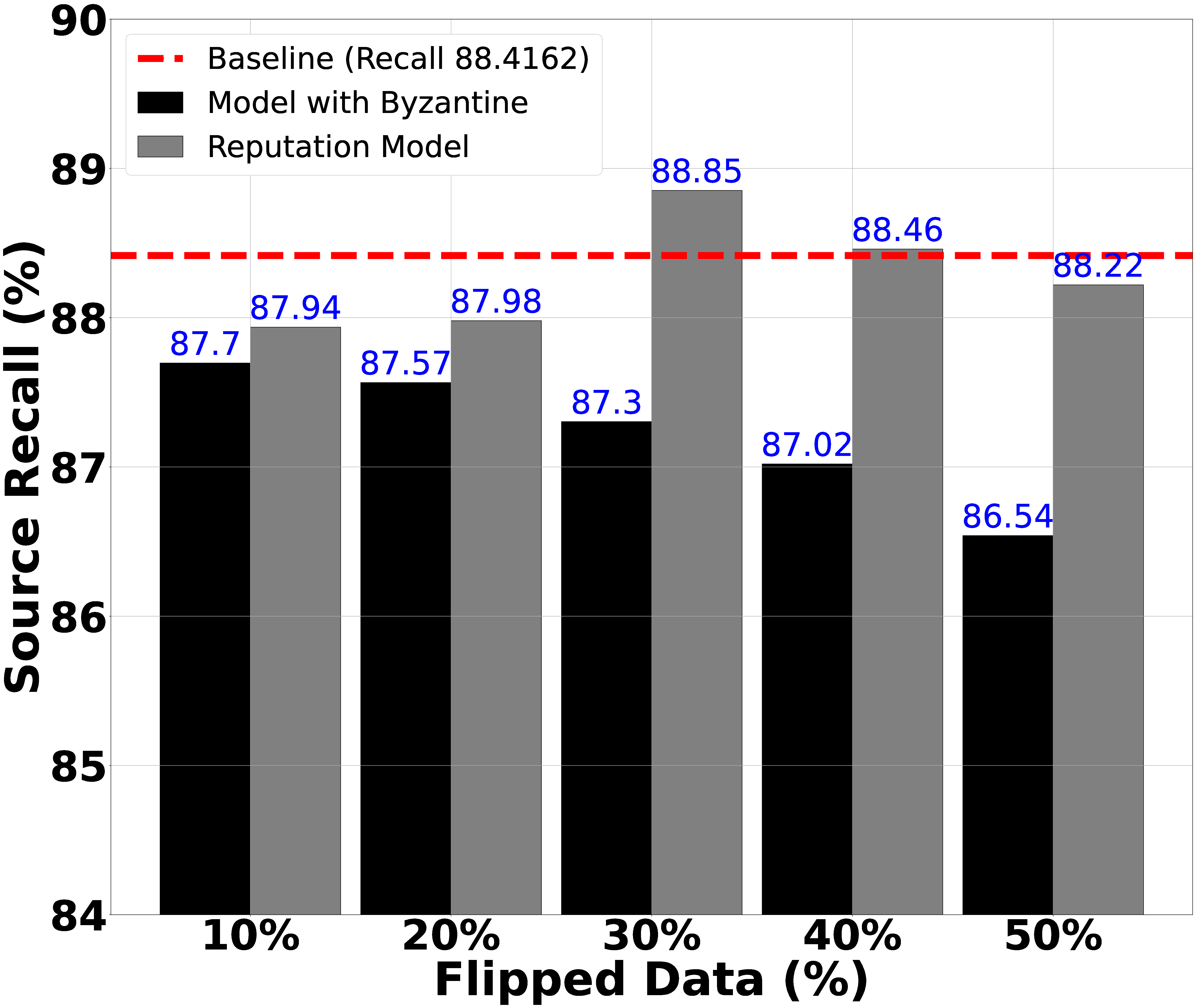}
    } 
    \subfigure[20\% Malicious Client]{
    \includegraphics[width=0.31\textwidth]{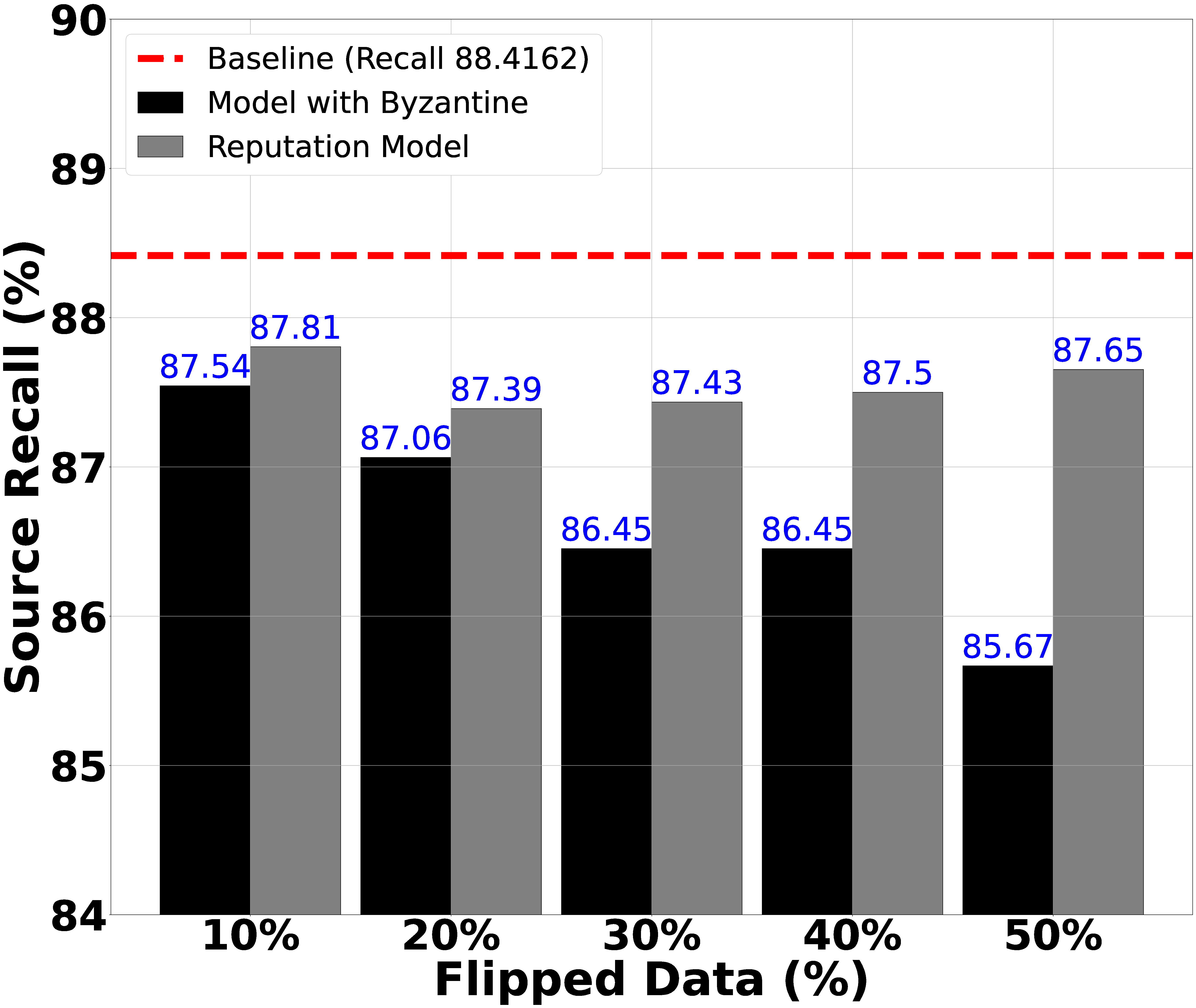}
    } 
    \subfigure[30\% Malicious Client]
    {
    \includegraphics[width=0.31\textwidth]{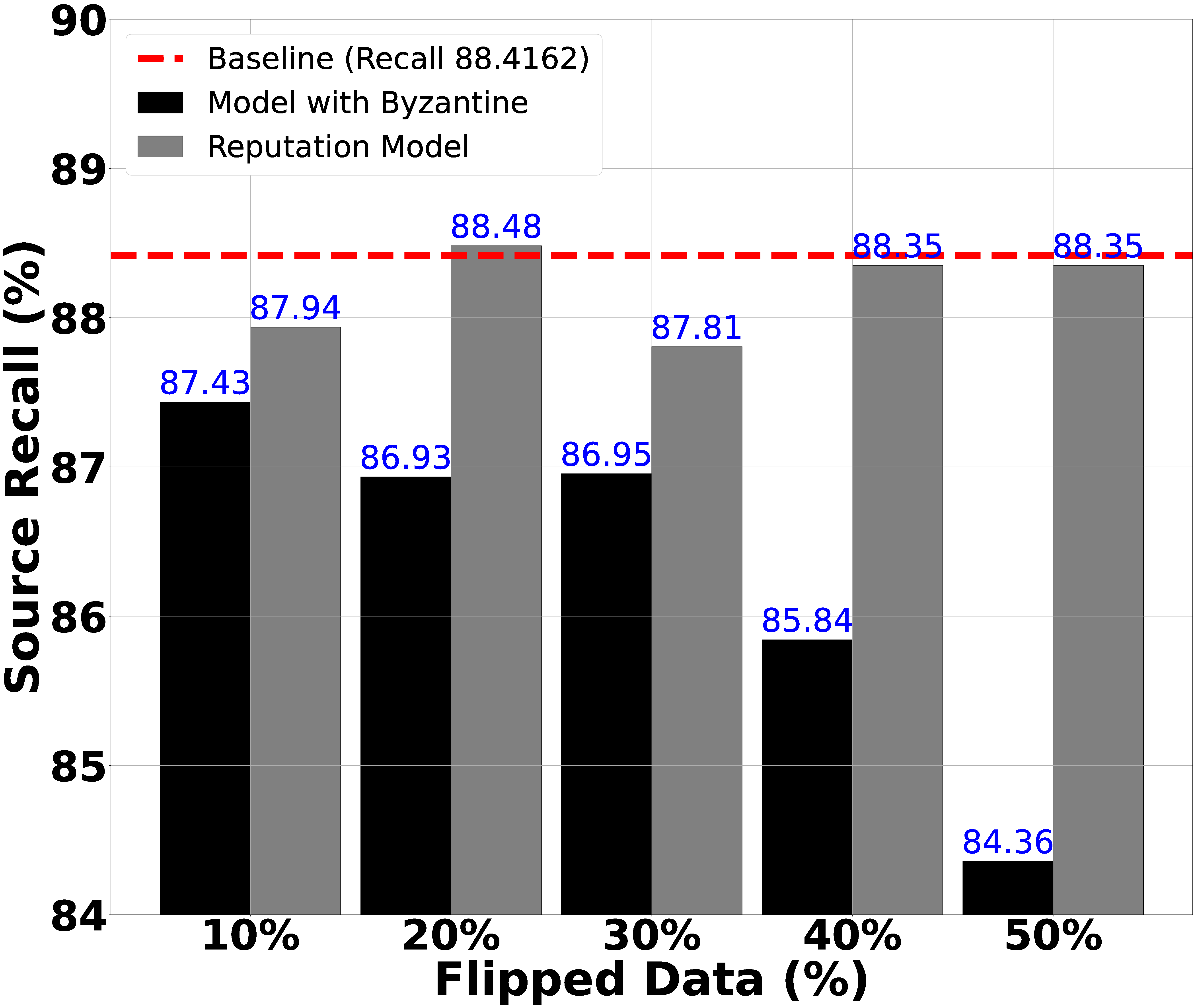}
    }    
    \caption{Attack Success Rate} 
    \label{fig:source_recall_reputation}
\end{figure*}

\section{Experimental Results}
\label{sec:experiment}
In this section, we discuss the experiments carried out to assess the performance of our framework.
Specifically, in Section~\ref{sub:datasetdesc}, we describe the dataset, the evaluation metrics, and the environment used for our experiments. Section~\ref{sub:reputation_results1} is dedicated to analyzing the findings and performance of our reputation approach.

\subsection{Testbed Description}
\label{sub:datasetdesc}
In our experimentation, we have employed a testbed comprising elements of Swarm Learning, integrating both Federated Learning and Blockchain technologies.

\subsubsection{Dataset} Darknet traffic frequently encompasses communications linked to malicious activities such as botnet, command and control, malware distribution, and phishing. Analyzing this traffic can help organizations identify patterns and signatures of such malicious activities and proactively defend them. Therefore, we conducted our experiments utilizing the publicly accessible CIC-Darknet 2020 dataset by the Canadian Institute of Cybersecurity~\cite{habibi2020didarknet}. This dataset includes traffic from various categories: Non-Tor, Non-VPN, Tor, and VPN. The dataset contains 93,356 Non-Tor entries, 23,863 Non-VPN entries, 1,392 Tor entries, and 22,919 VPN entries. To conduct SL and evaluate the model, we partitioned the traffic data into an $80:20$ ratio, with $80\%$ allocated for collaborative training and the remaining $20\%$ for testing. We carried out our experiments under an Independent and Identically Distributed~(IID) data distribution, i.e., the training data is evenly divided among each client. The testing data is solely employed for model evaluation purposes.

\subsubsection{Evaluation Metrics} We employed various evaluation metrics to assess the impact of the label-flipping attack and the performance of the proposed framework. These metrics include:
\begin{itemize}
    \item \textbf{Model F1-score}. Due to the imbalanced nature of our dataset, we adopted the F1-score as a metric to evaluate the model's performance. The F1-score considers both precision $\mathcal{P}$ and recall $\mathcal{R}$, and it is computed by Equation~\ref{eq:f1}.
    
    \begin{equation}
        F1\text{-score} = \frac{2 \times \mathcal{P} \times \mathcal{R}}{\mathcal{P} + \mathcal{R}}
        \label{eq:f1}
    \end{equation}
  
where $\mathcal{P}$ is the ratio of true positive predictions to the total predicted positives and $\mathcal{R}$ is the ratio of true positive predictions to the total actual positives.

\item \textbf{Class Transition Misclassification Rate}. The Class Transition Misclassification Rate~(CTMR) measures the percentage of instances from the source class, $s$, that are incorrectly classified as the target, $t$, class. CTMR is calculated using the Equation~\ref{eq:ctmr}. A higher CTMR indicates a greater proportion of instances were misclassified.
\begin{equation}
    \text{CTMR} = \frac{\sum_{j=1}^{N_{s}} [(y_j == s) \land (\hat{y}_j == t)]}{N_s}
    \label{eq:ctmr}
\end{equation}
where 
$N_{s}$ is the total number of instances with the class label $s$, $y_j$ is the true label of instance $j$, and $\hat{y}_j$ is the predicted label of instance $j$.
\item \textbf{Source Recall}. Source Recall measures the ability of the model to correctly identify instances of a particular class $s$. The decrease in recall for a specific class, $s$, indicates that the attack has manipulated the model to misclassify instances of that class. Equation~\ref{eq:source_recall} determines the recall of particular class $s$.
\begin{equation}
    \mathcal{R}_{s} = \frac{\mathcal{X}}{\mathcal{X} + \mathcal{\hat{X}}}
    \label{eq:source_recall}
\end{equation}

where $\mathcal{X}= \sum_{j=1}^{N_{s}} [(y_j == s) \land (\hat{y}_j == s)]$; $\mathcal{\hat{X}}= \sum_{j=1}^{N_{s}} [(y_j == s) \land (\hat{y}_j \neq s)]$; $N_{s}$ is the total number of instances belonging to class $s$, $y_j$ is the true label of instance $j$, and $\hat{y}_j$ is the predicted label of instance $j$.
\end{itemize}

\subsubsection{Environment Setup} The experiments were conducted on a Windows 11 Pro system featuring an Intel Core i9 processor, $32$ GB of RAM, and an NVIDIA Quadro P2000 with 5 GB of GDDR5X memory. We implemented the SeCTIS using the \textit{PyTorch} framework. Additionally, the visualization of results was facilitated by employing the \textit{Matplotlib} library.

\subsubsection{Model Settings} We implemented a Deep Neural Network~(DNN) comprising two hidden layers and one output layer. The first hidden layer consists of $64$ neurons, followed by a layer with $32$ neurons. These hidden layers employ the Rectified Linear Unit (ReLU) activation function to introduce non-linearity. The output layer, instead, is composed of neurons corresponding to the four classes, with each neuron representing the likelihood of the input belonging to a specific class.

We conducted Global Model training for $50$ rounds, where in each round, the organizations train their models for $5$ local epochs with a batch size of $32$. Clients utilized the learning rate of $0.01$ with Stochastic Gradient Descent (SGD) and Cross-entropy loss function for training their models.
\begin{figure*}[t]
    \centering
    \subfigure[10\% Flipped Data]
    {
    \includegraphics[width=0.16\textwidth]{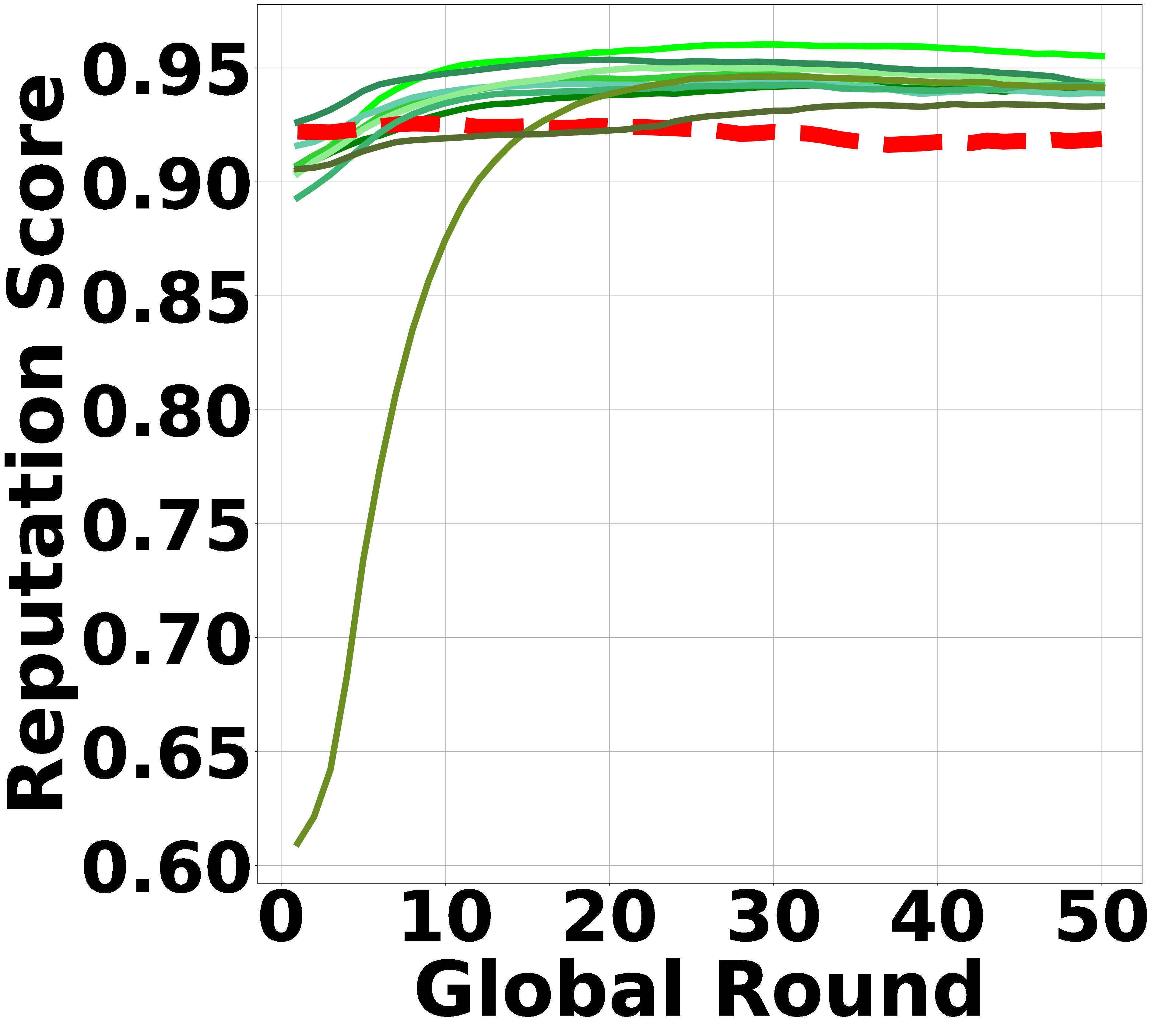}
    \label{fig:1_Client_reputation_10}
    } 
    \subfigure[20\% Flipped Data]{
    \includegraphics[width=0.16\textwidth]{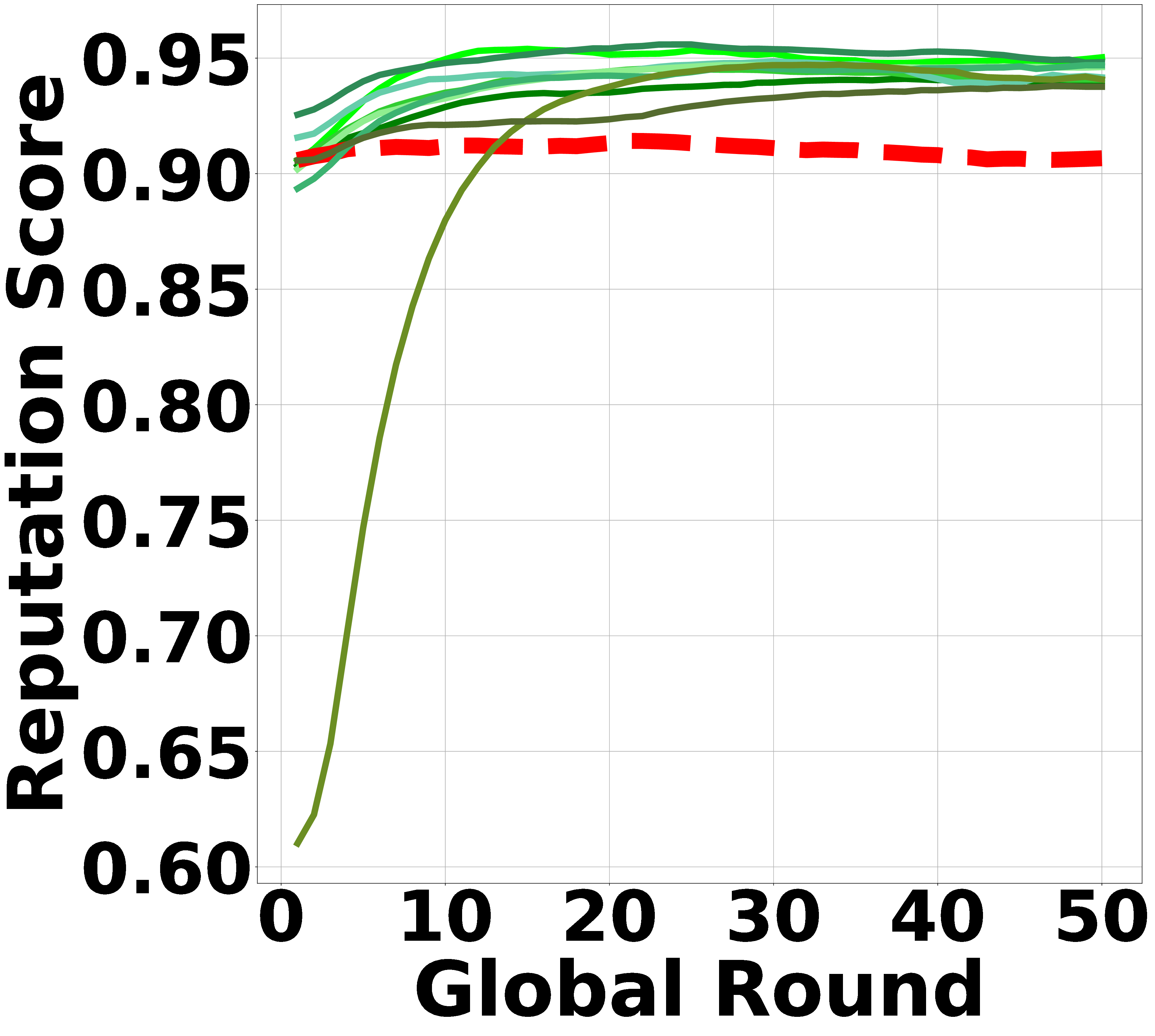}
    \label{fig:1_Client_reputation_20}
    } 
    \subfigure[30\% Flipped Data]
    {
    \includegraphics[width=0.16\textwidth]{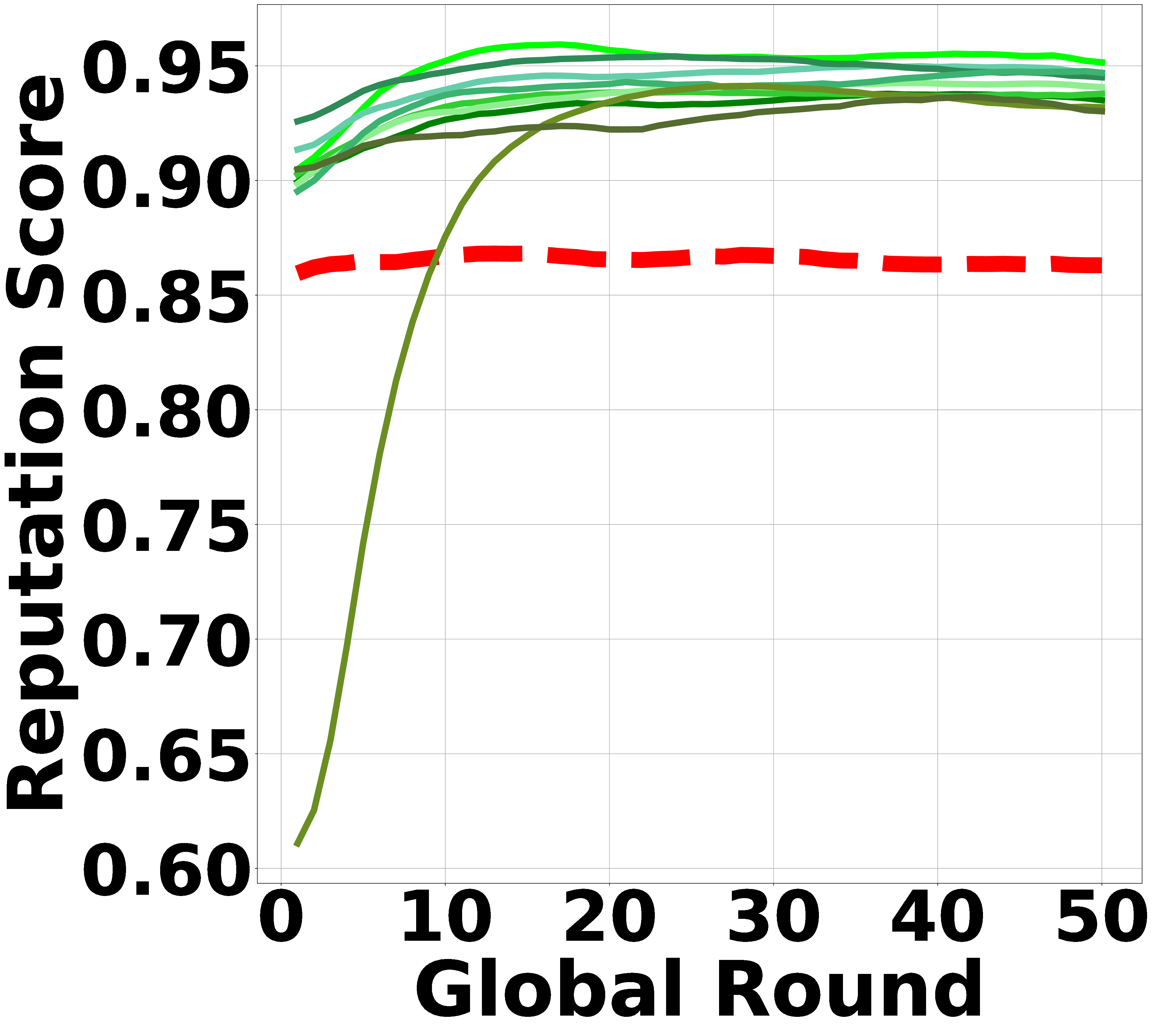}
    \label{fig:1_Client_reputation_30}
    } 
    \subfigure[40\% Flipped Data]
    {
    \includegraphics[width=0.16\textwidth]{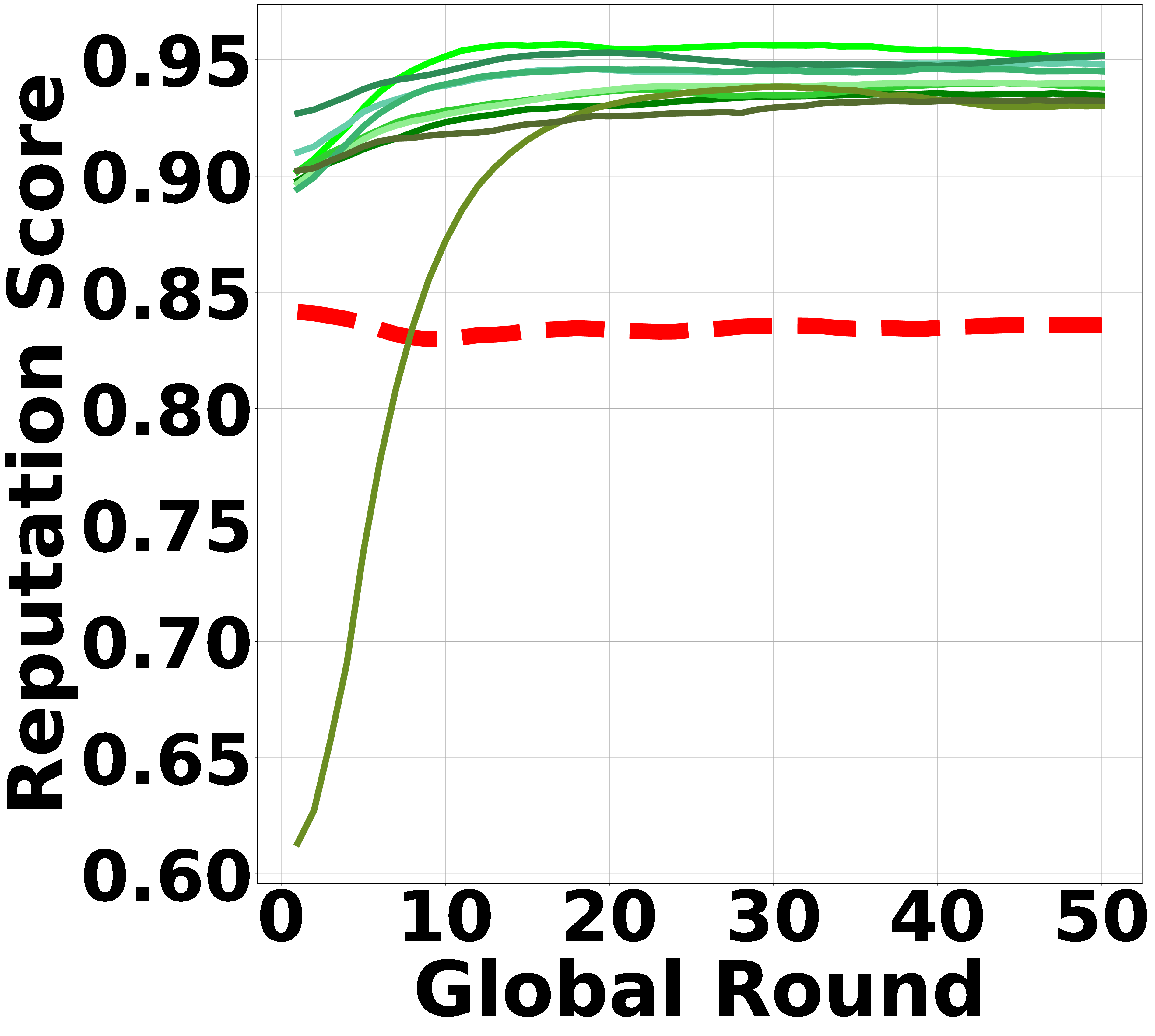}
    \label{fig:1_Client_reputation_40}
    } 
    \subfigure[50\% Flipped Data]
    {
    \includegraphics[width=0.16\textwidth]{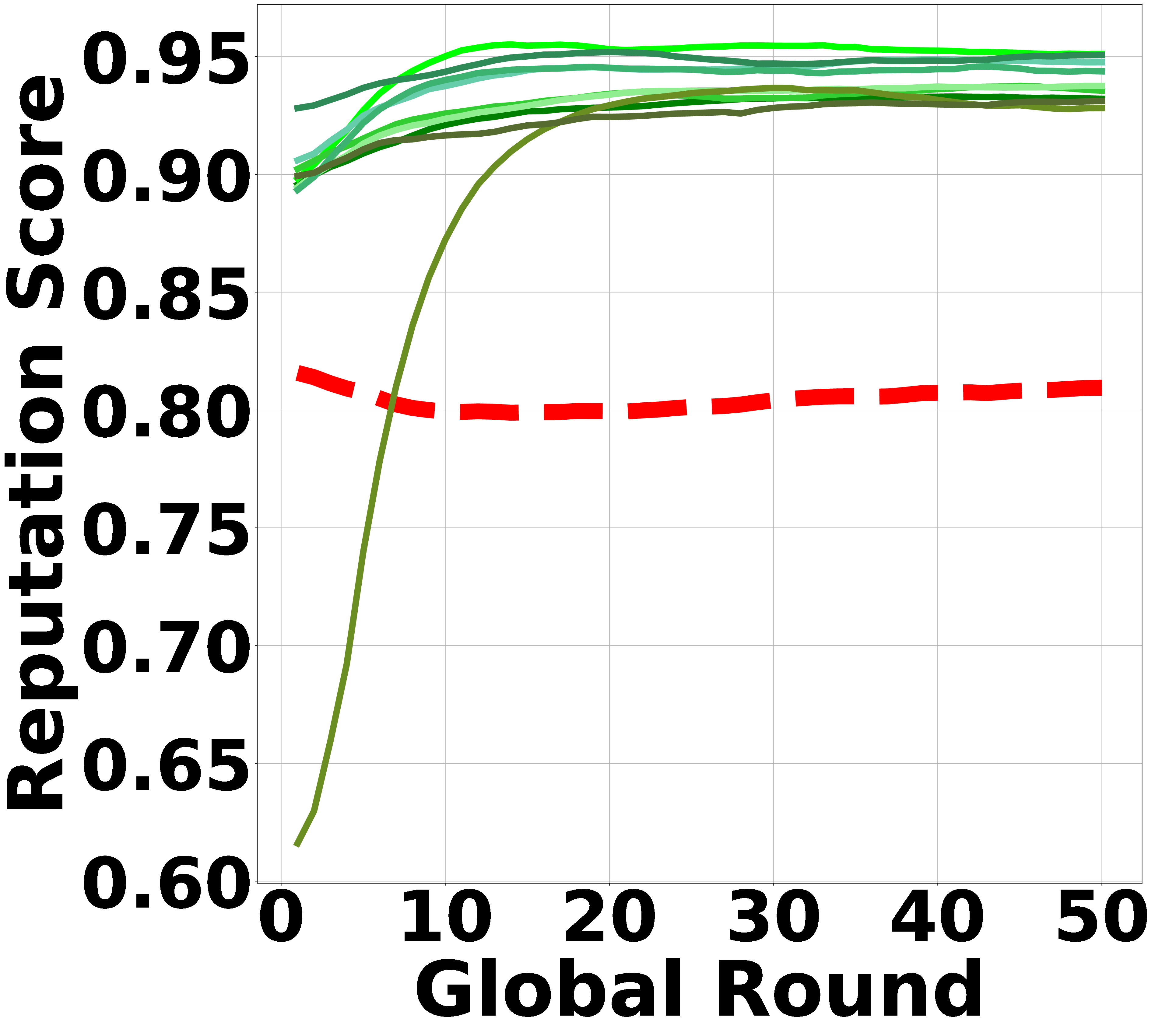}
    \label{fig:1_Client_reputation_50}
    } 
    \subfigure
    {
    \includegraphics[width=0.08\textwidth]{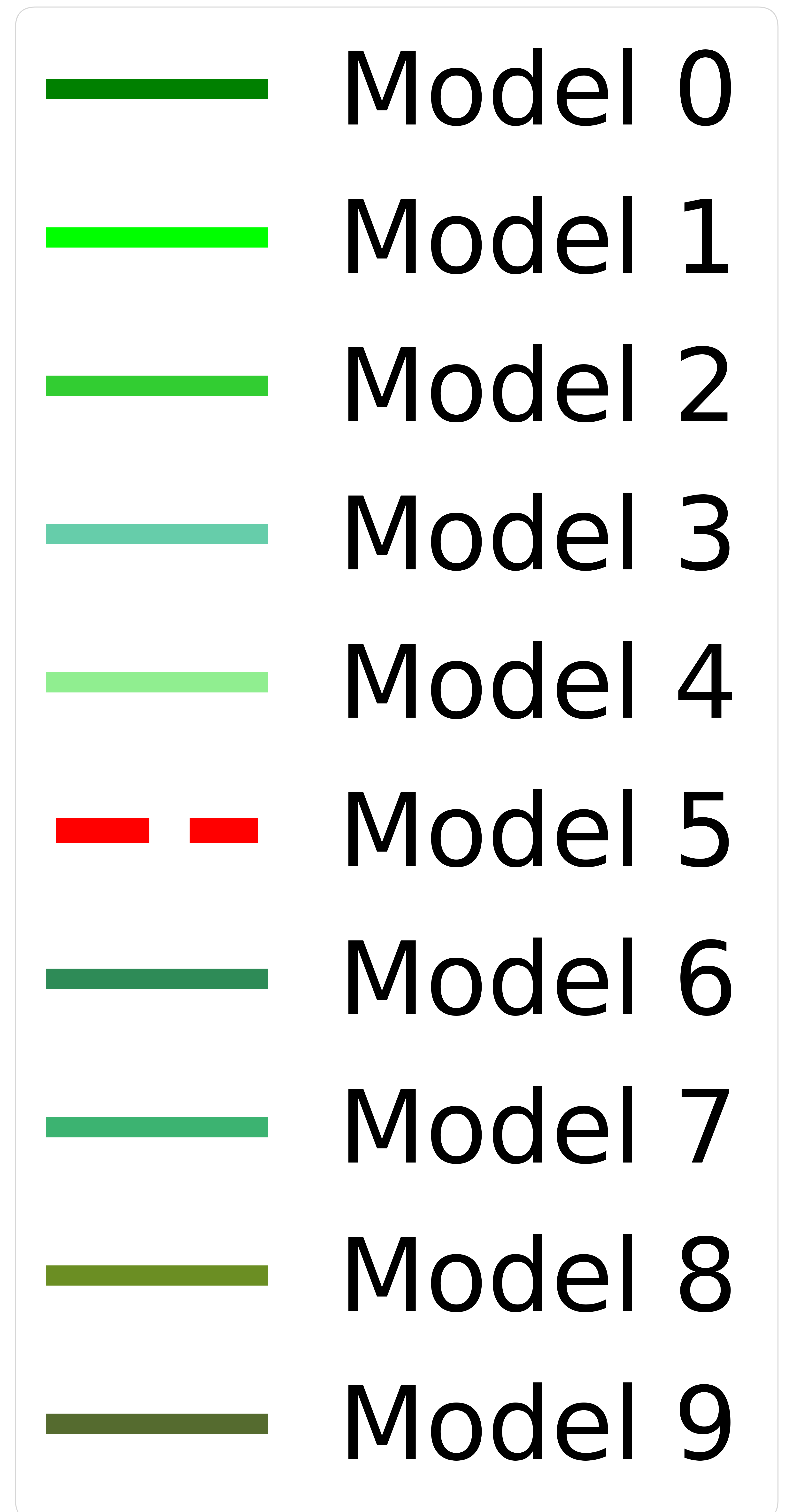}
    } 
   
    \caption{Reputation scores of the model with 10\% Byzantine clients. Byzantine clients(Model 5), highlighted in red, contrast with benign participants represented in green.} 
    \label{fig:1_Client_reputation}
\end{figure*}
\begin{figure*}[t]
    \centering
    \subfigure[10\% Flipped Data]
    {
    \includegraphics[width=0.16\textwidth]{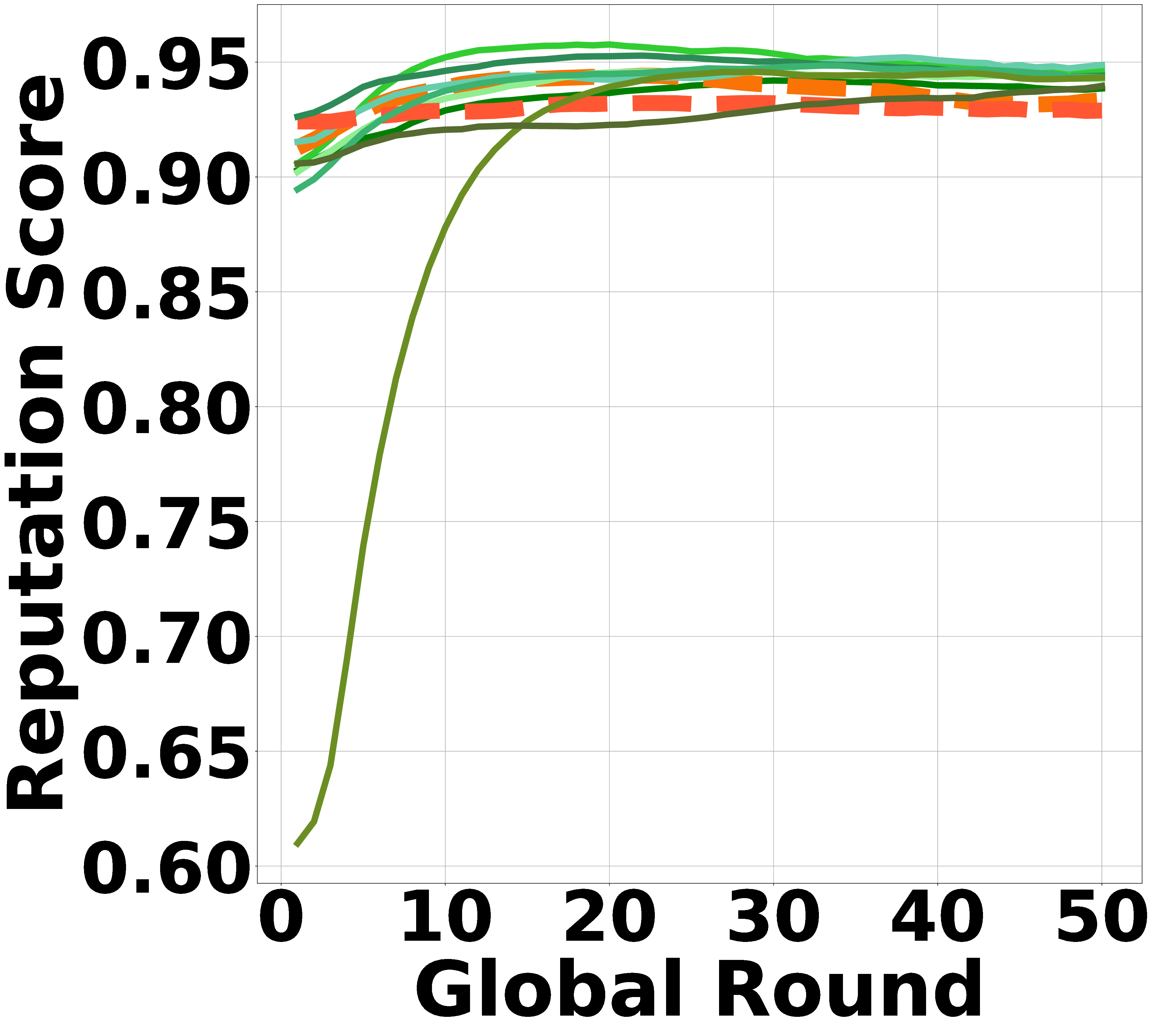}
    \label{fig:2_Client_reputation_10}
    } 
    \subfigure[20\% Flipped Data]{
    \includegraphics[width=0.16\textwidth]{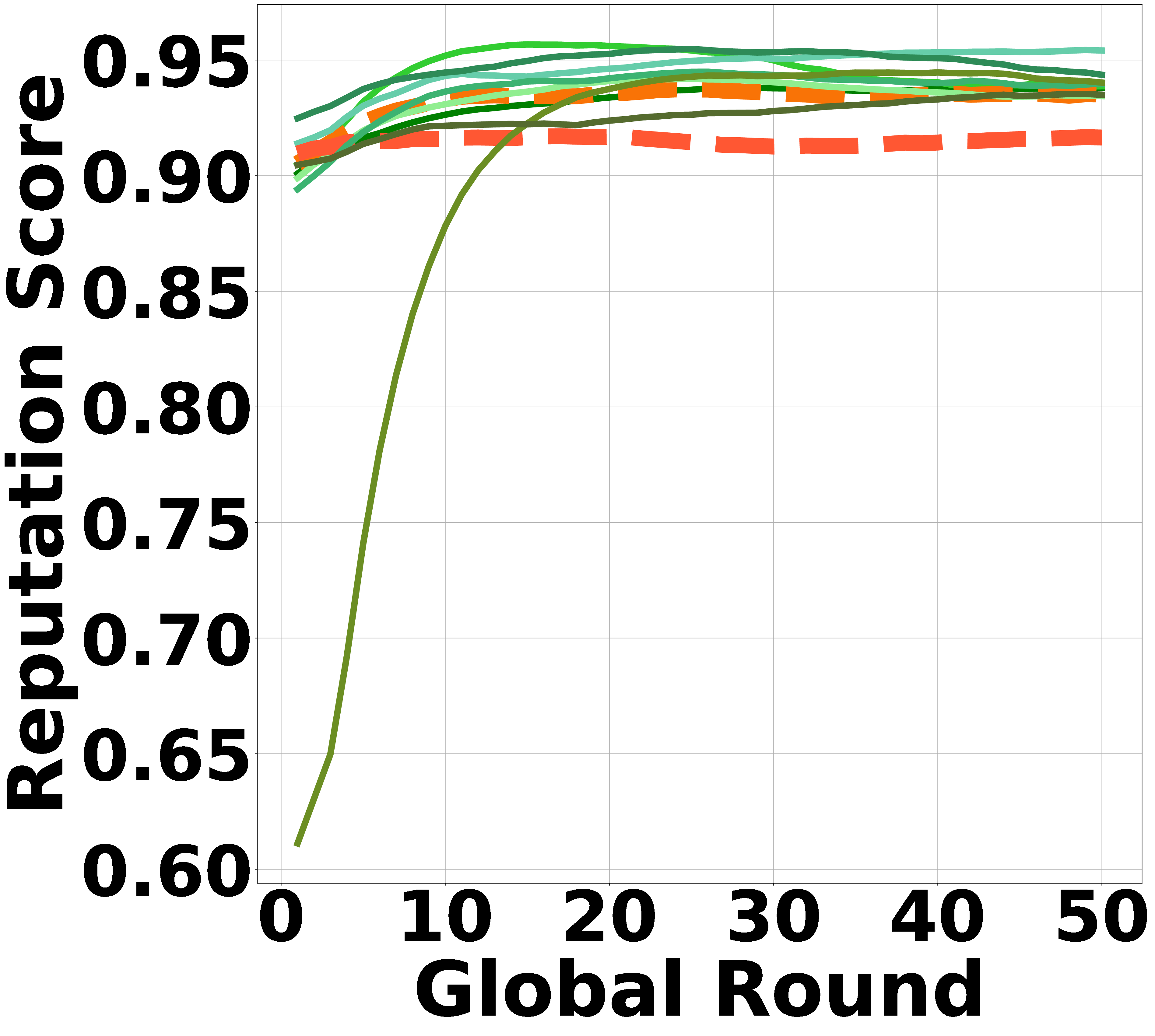}
    \label{fig:2_Client_reputation_20}
    } 
    \subfigure[30\% Flipped Data]
    {
    \includegraphics[width=0.16\textwidth]{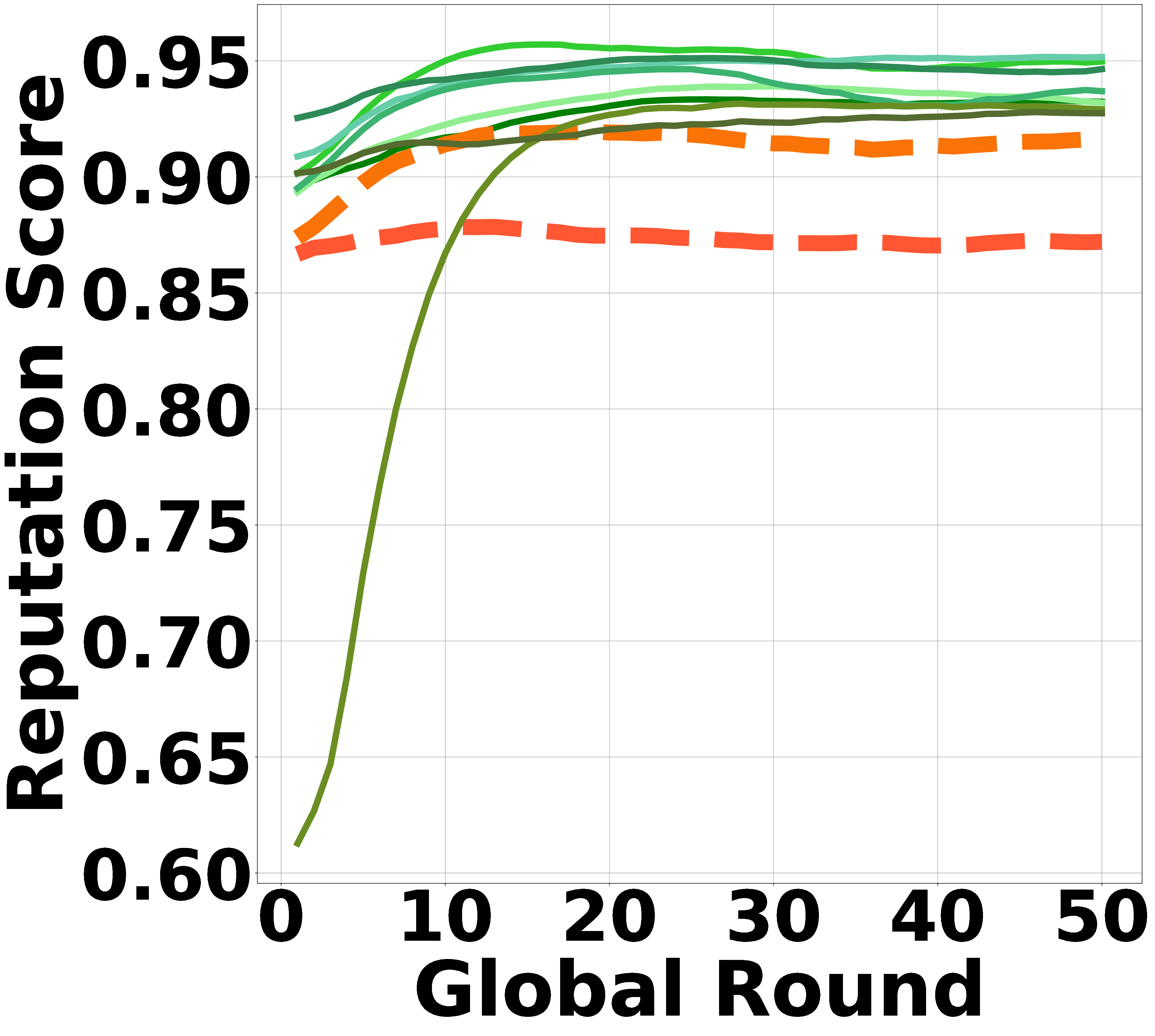}
    \label{fig:2_Client_reputation_30}
    } 
    \subfigure[40\% Flipped Data]
    {
    \includegraphics[width=0.16\textwidth]{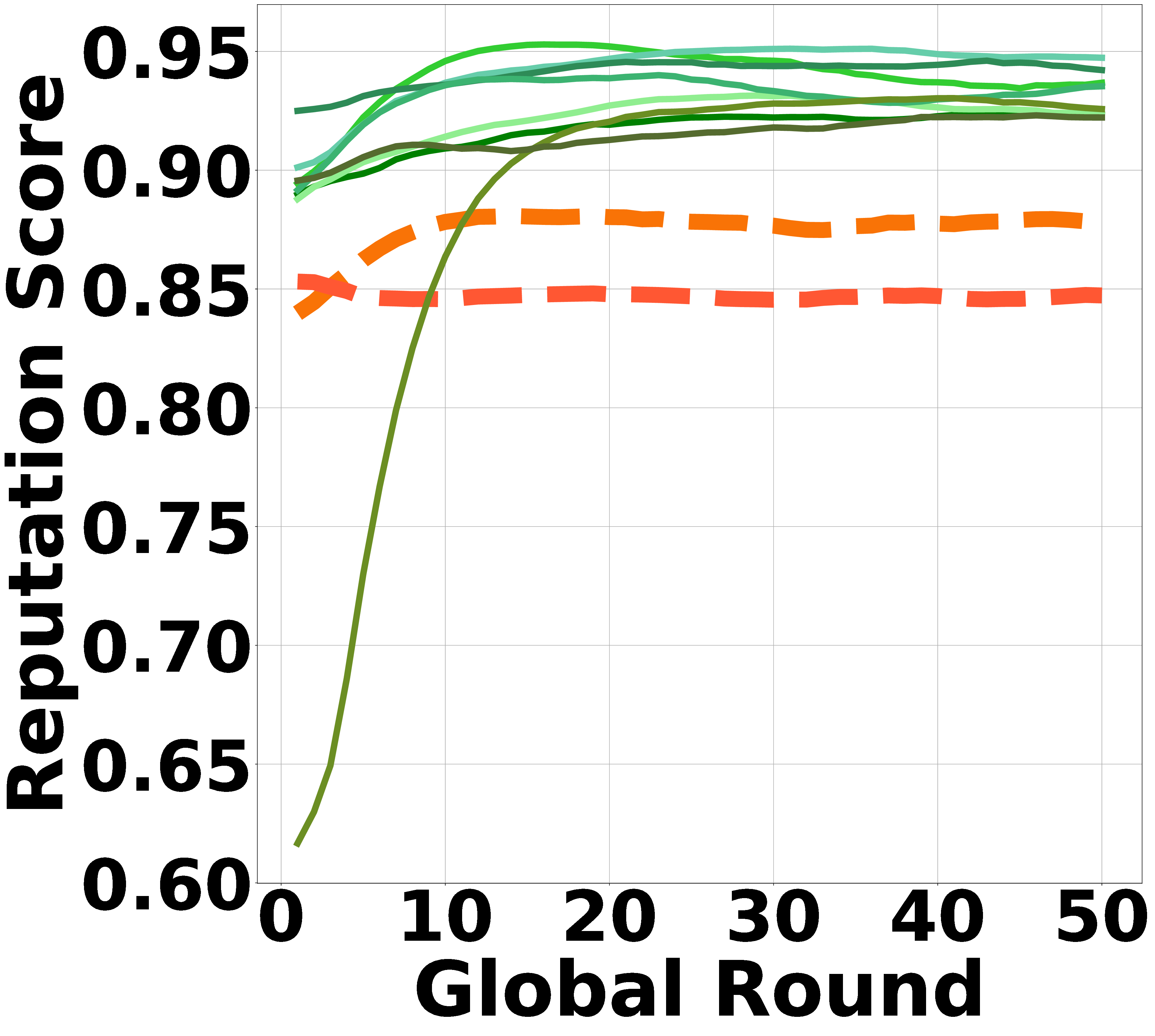}
    \label{fig:2_Client_reputation_40}
    } 
    \subfigure[50\% Flipped Data]
    {
    \includegraphics[width=0.16\textwidth]{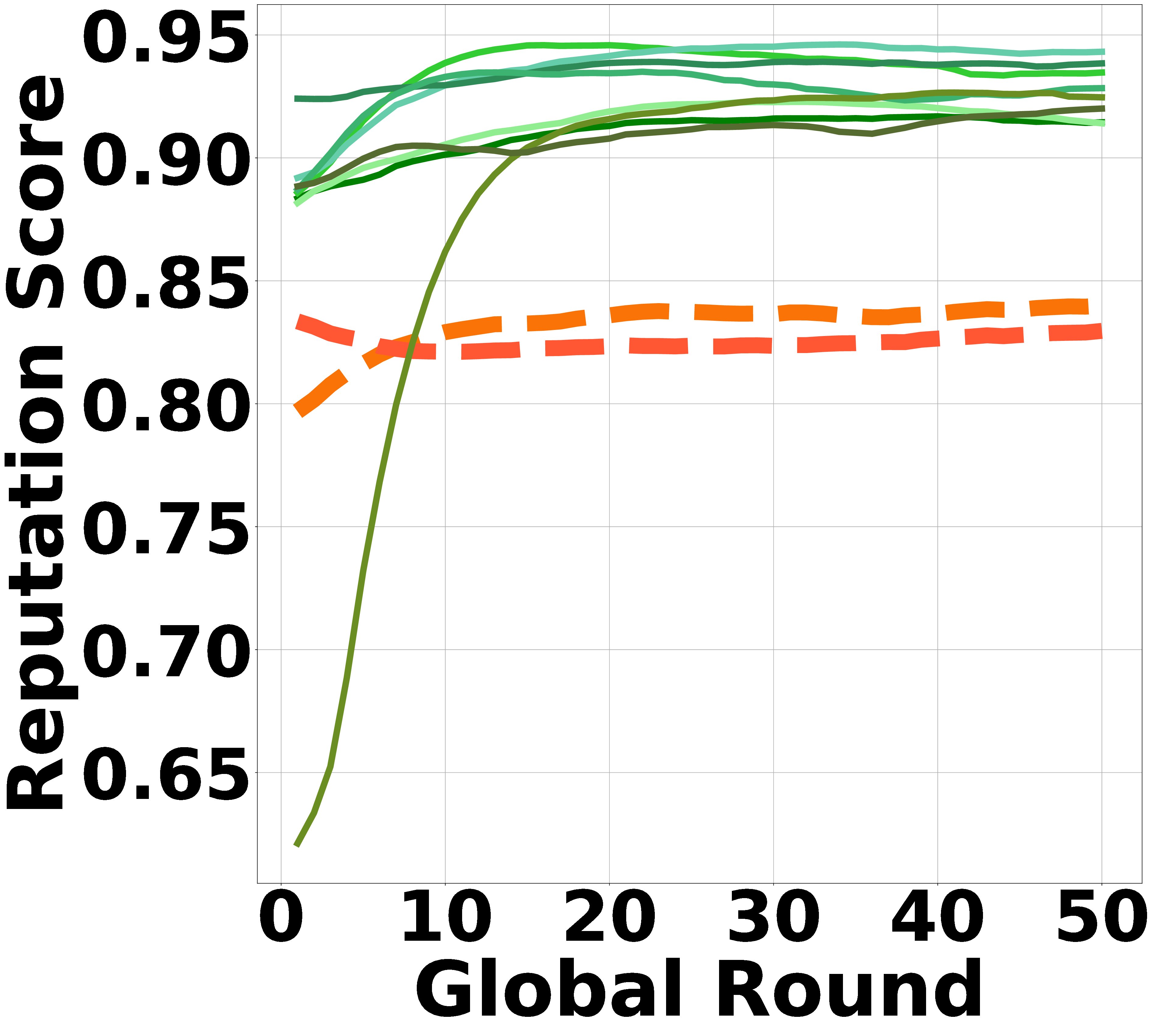}
    \label{fig:2_Client_reputation_50}
    } 
    \subfigure
    {
    \includegraphics[width=0.08\textwidth]{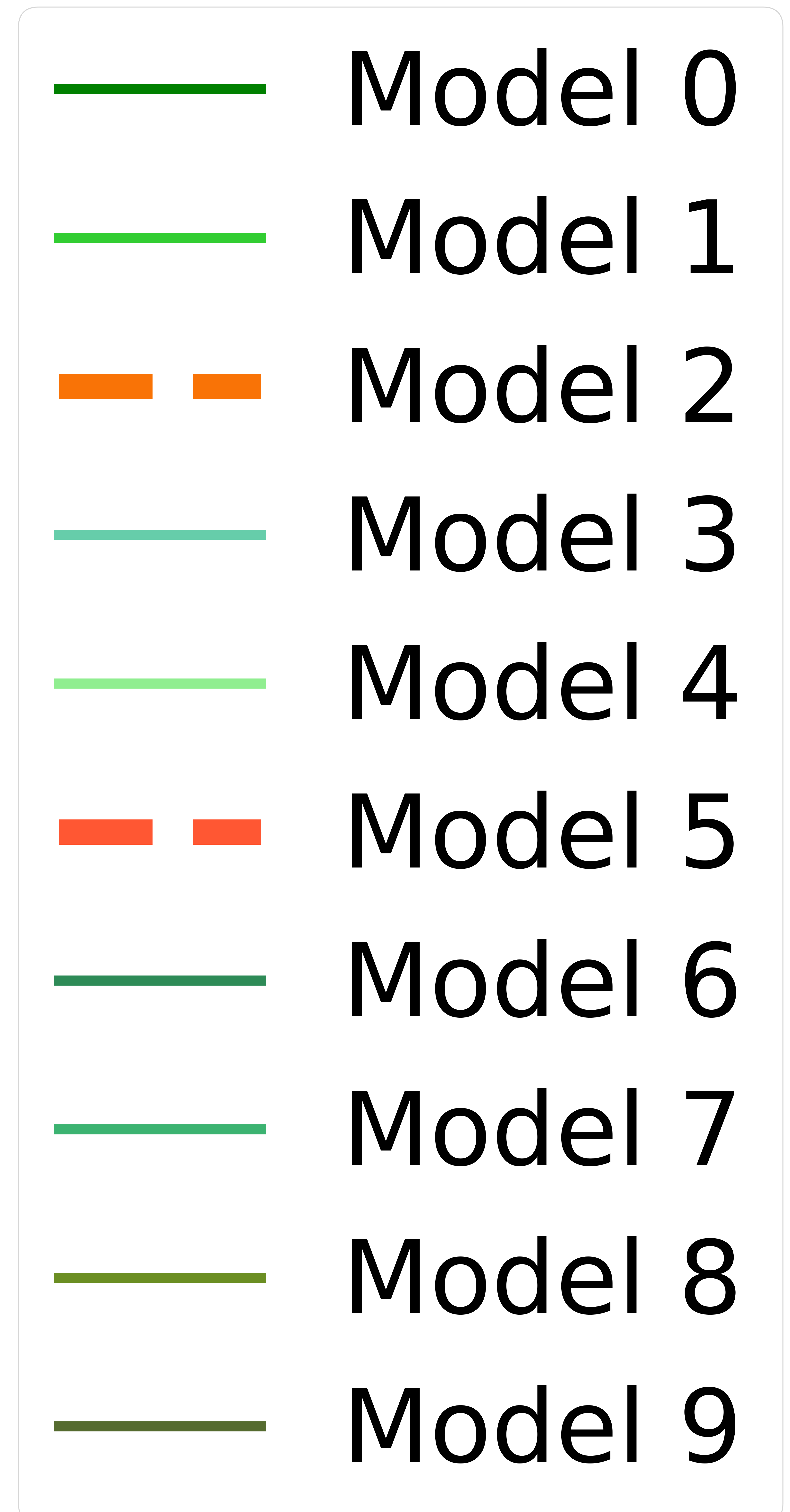}
    } 
   
    \caption{Reputation scores throughout rounds of swarm learning involving 20\% Byzantine clients. Two Byzantine models(Model 2 and 5) are emphasized in red variant colors, and legitimate participants are highlighted in green.} 
    \label{fig:2_Client_reputation}
\end{figure*}
\begin{figure*}[t]
    \centering
    \subfigure[10\% Flipped Data]
    {
    \includegraphics[width=0.16\textwidth]{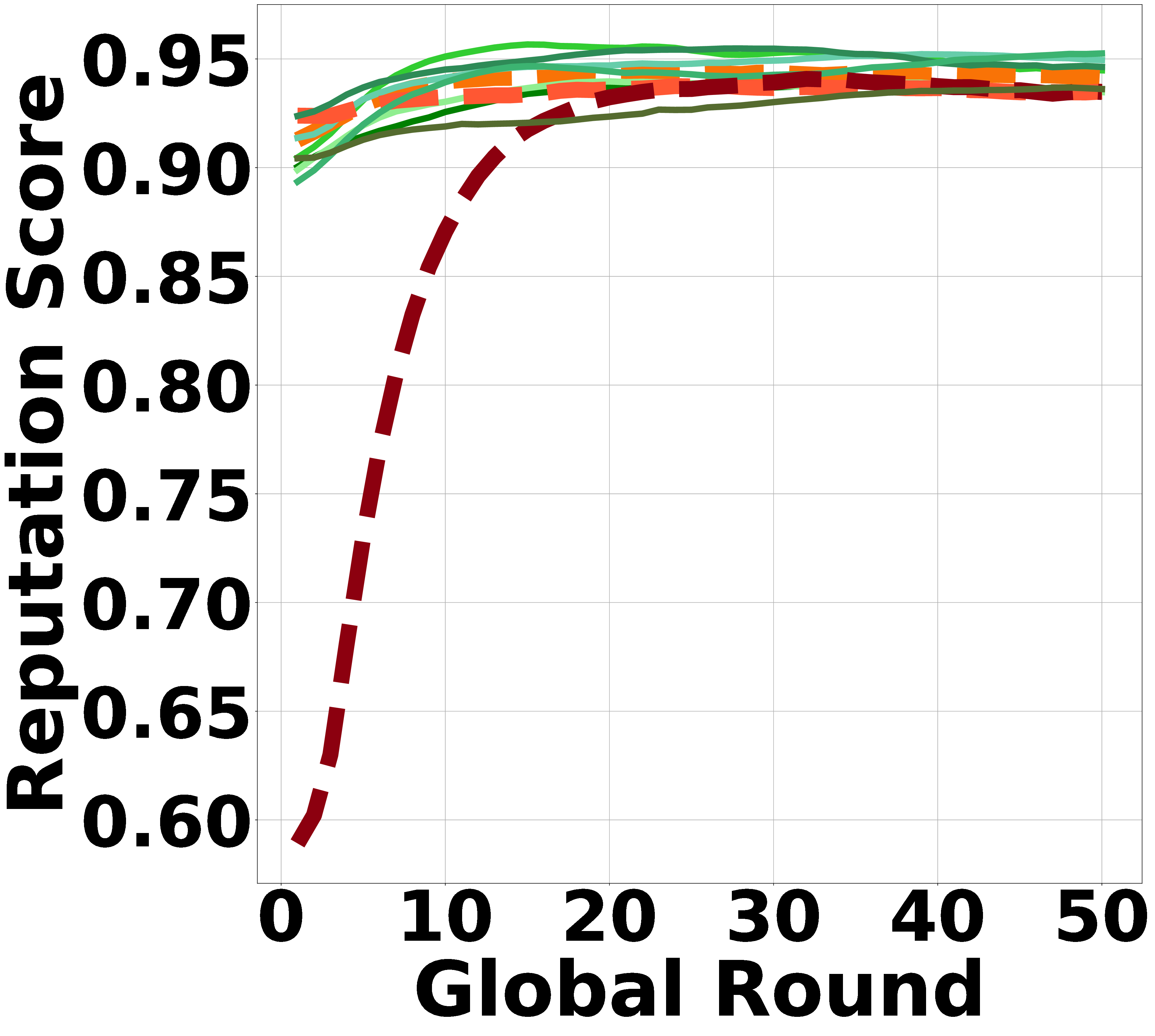}
    \label{fig:3_Client_reputation_10}
    } 
    \subfigure[20\% Flipped Data]{
    \includegraphics[width=0.16\textwidth]{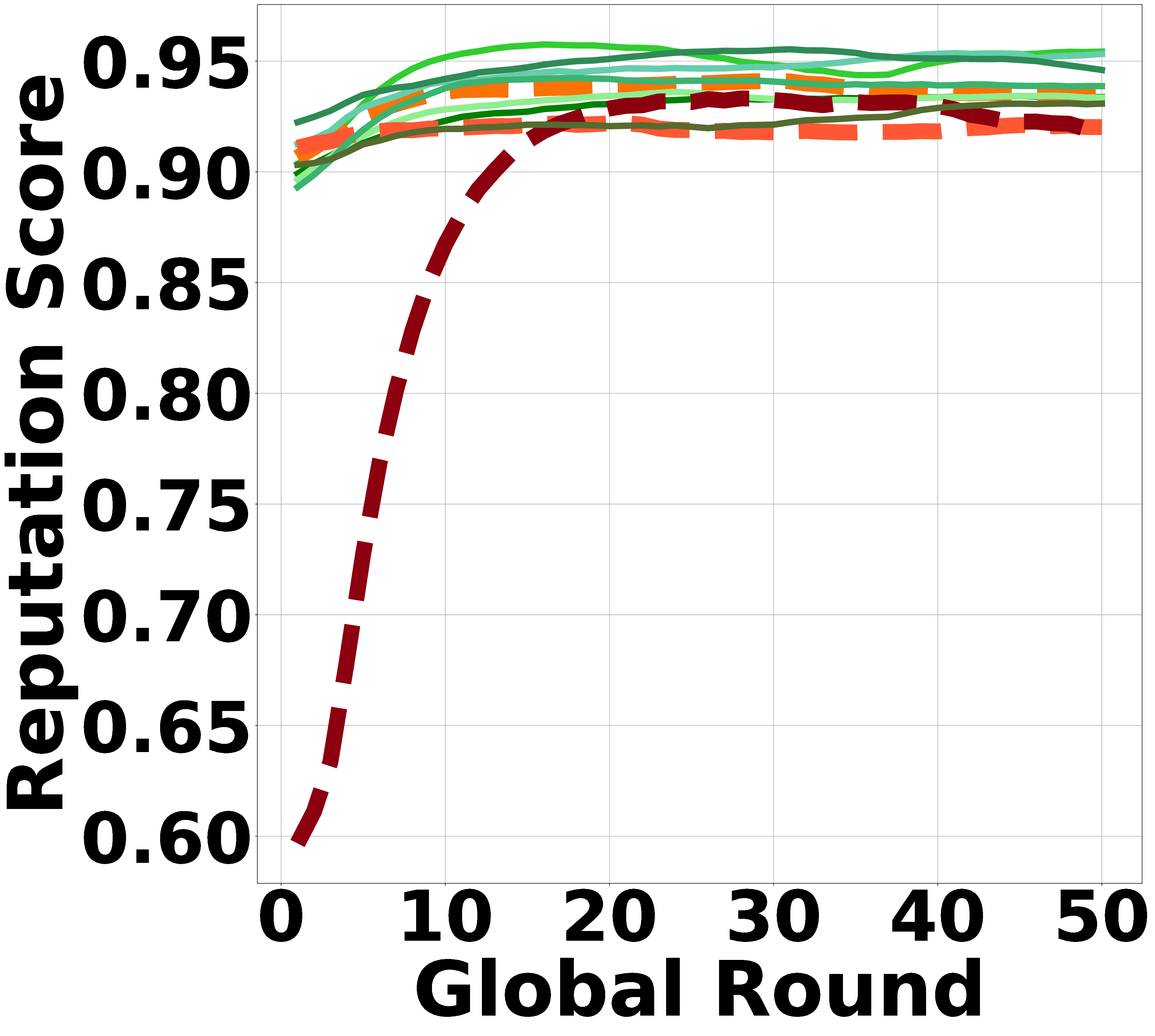}
    \label{fig:3_Client_reputation_20}
    } 
    \subfigure[30\% Flipped Data]
    {
    \includegraphics[width=0.16\textwidth]{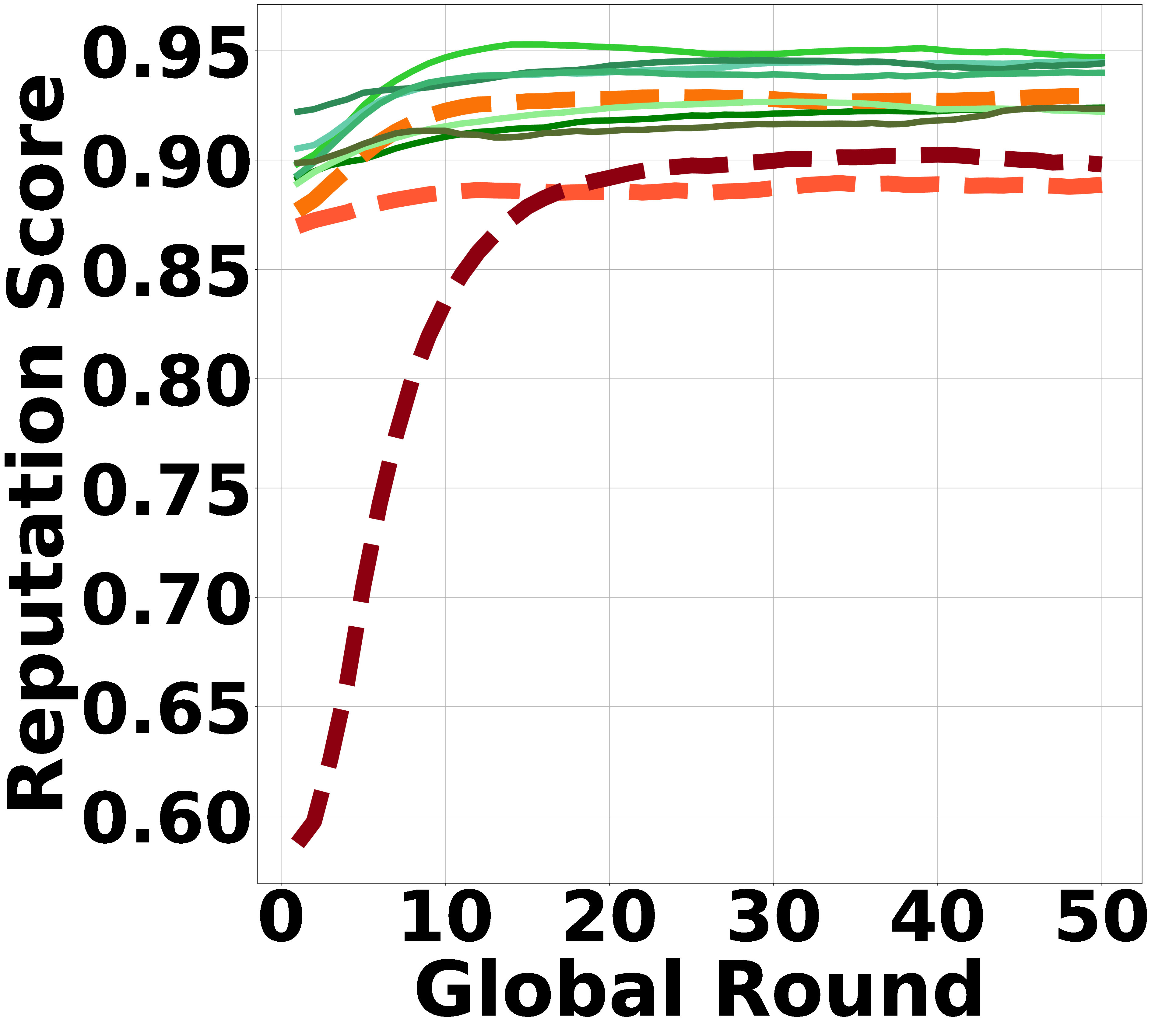}
    \label{fig:3_Client_reputation_30}
    } 
    \subfigure[40\% Flipped Data]
    {
    \includegraphics[width=0.16\textwidth]{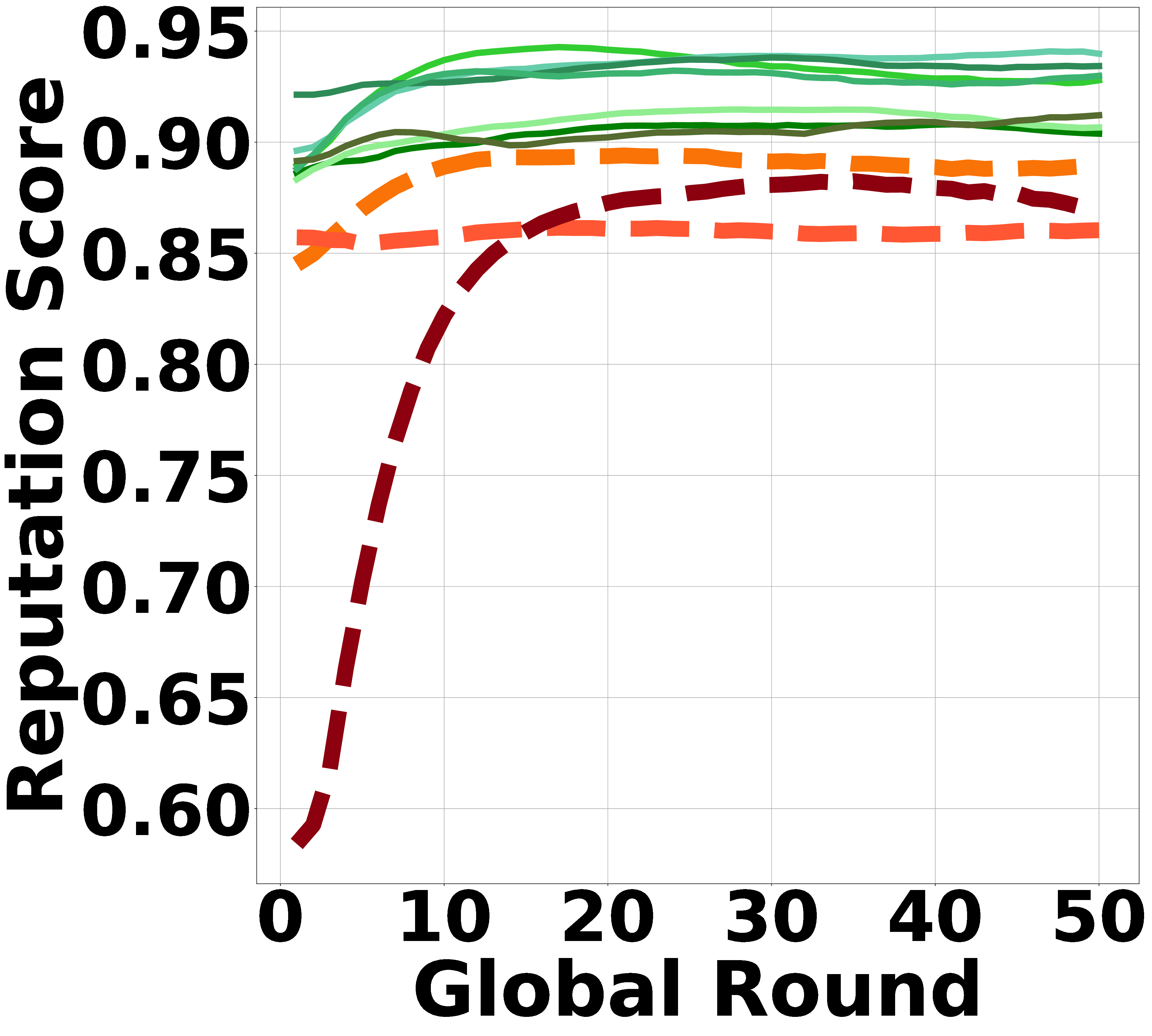}
    \label{fig:3_Client_reputation_40}
    } 
    \subfigure[50\% Flipped Data]
    {
    \includegraphics[width=0.16\textwidth]{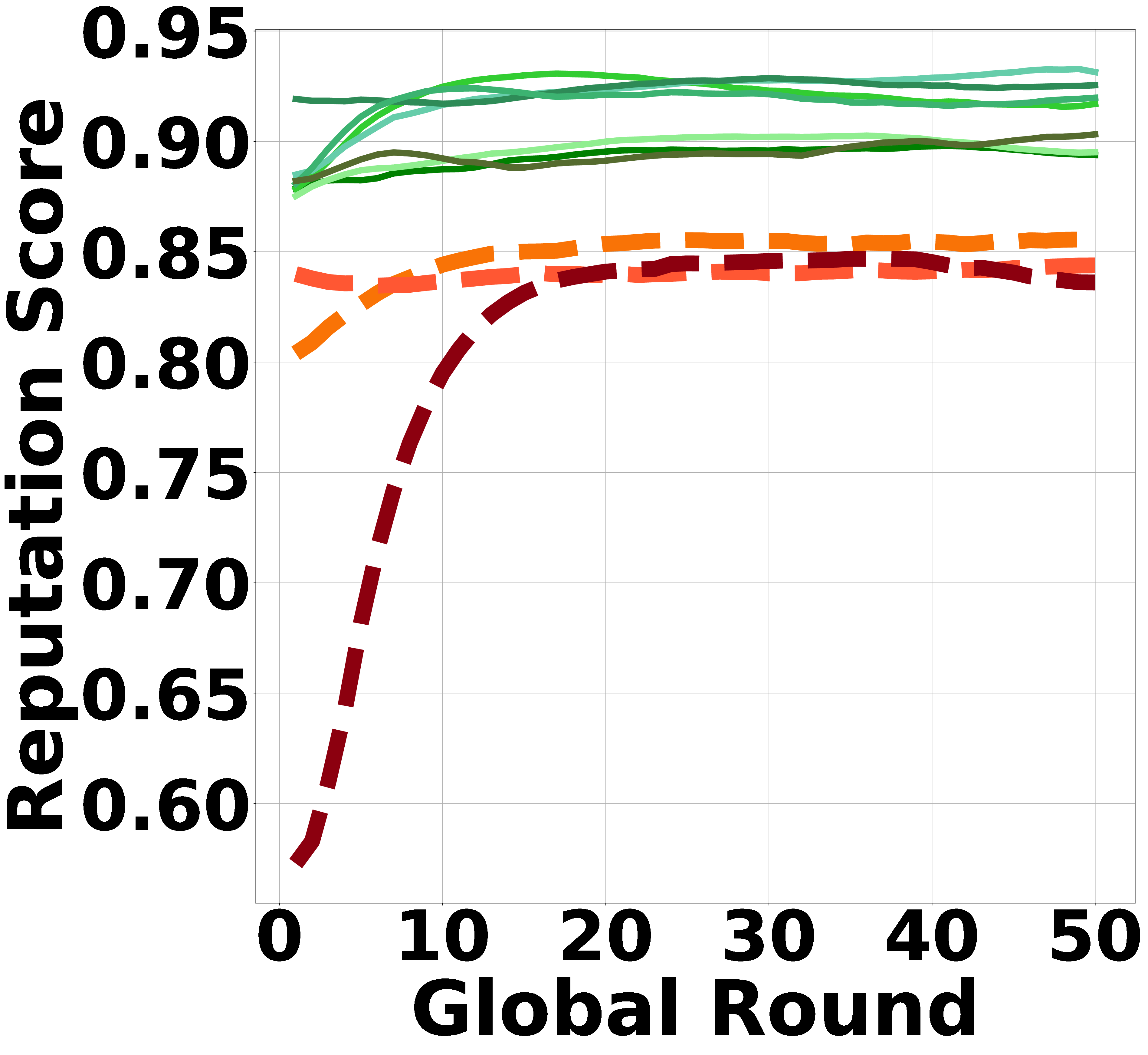}
    \label{fig:3_Client_reputation_50}
    } 
    \subfigure
    {
    \includegraphics[width=0.08\textwidth]{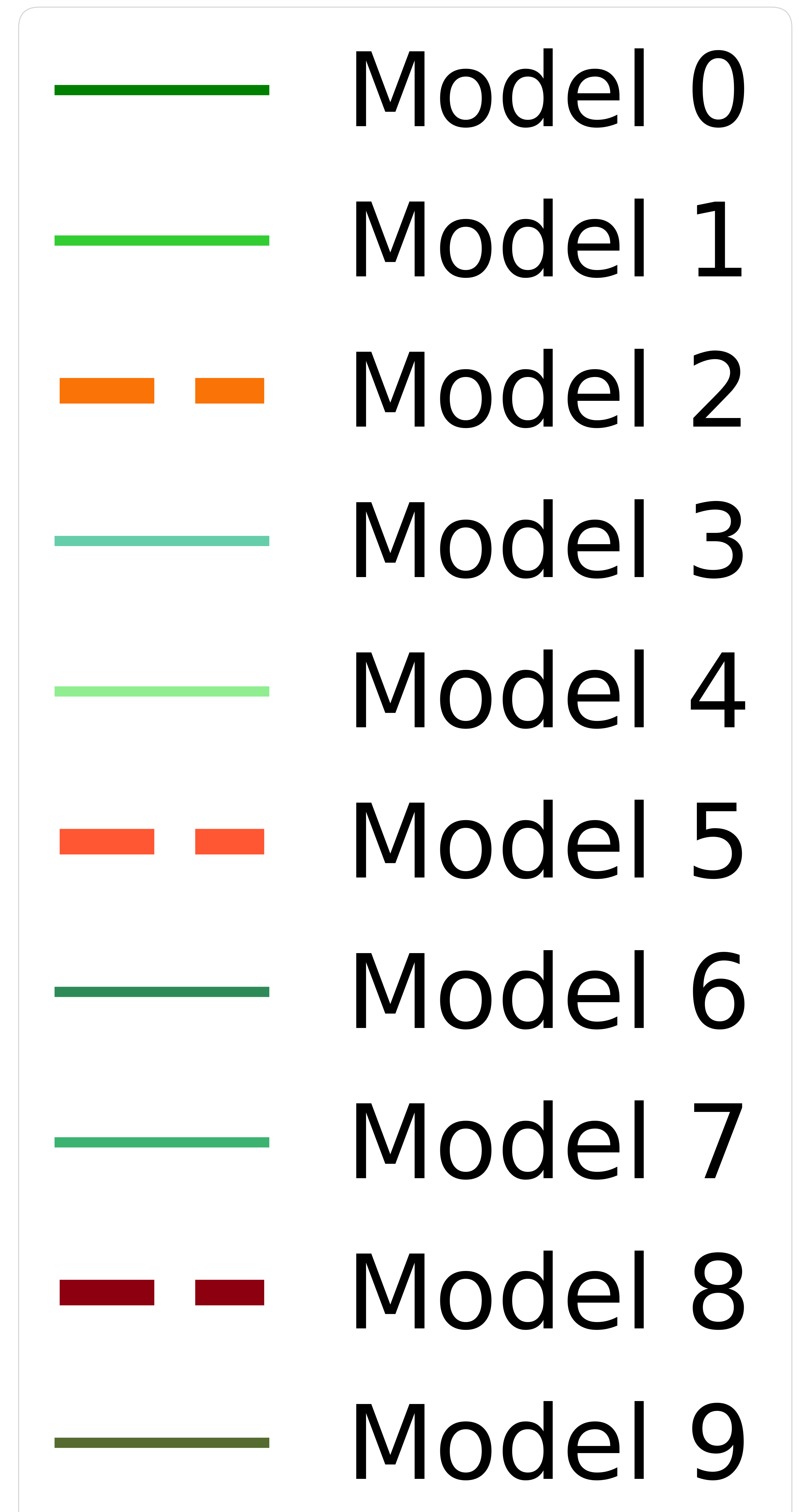}
    } 
   
    \caption{Reputation scores across communication rounds with a presence of 30\% Byzantine clients. Three Byzantine models(Models 2, 5, and 8) are vivid in red, while benign models are represented in green.} 
    \label{fig:3_Client_reputation}
\end{figure*}

\subsection{Experimental Analysis of Reputation Model}
\label{sub:reputation_results1}
The data collection and labeling phases of CTI generation within the organization may unintentionally produce noisy labels. This may potentially obstruct the collaborative learning process. As stated in Section~\ref{sec:attackModel}, we randomly flipped labels to simulate instances of incorrect labels. These simulations involved $t\%$ of the $N$ participating organization acting as \textit{Byzantine client}.
Additionally, to analyze how the noisy labels affect the model's performance, we conducted experiments where $x\%$ of samples from specific classes were changed to another class.
\par
 We initially investigated the CTI model using a Swarm Learning setup with network traffic data involving 10 participating organizations. Our SL model for network traffic classification yielded an F1-score of $0.928$. Subsequently, we conducted experiments with $10\%$, $20\%$, and $30\%$ of the total participating organizations as Byzantine clients. Since we have to assess whether our reputation model could identify low-quality models, we designated Byzantine nodes as fixed clients. For the scenario with $t = 10\%$, Client 5 was designated as the Byzantine client. In the case of $20\%$, Clients 2 and 5 were chosen, while for the $30\%$ setup, Clients 2, 5, and 8 were selected. For each scenario, we also conducted experiments with different percentages of label flipping from VPN class to NonVPN, where $x$ ranges from $10\%$ to $50\%$.
\par
We analyzed the misclassification rate on the test set under two cases: {\em(i)} when flipping is performed and {\em(ii)} when the SL model is implemented with the reputation technique. Figure~\ref{fig:mcr_reputation} illustrates the misclassification rates across three different scenarios: with one Byzantine client, two Byzantine clients, and three Byzantine clients. In Figure~\ref{fig:1_client_mcr_reputation}, the horizontal dashed line represents the SL model's misclassification rate (with a value of $8.8569$) without any Byzantine clients. This indicates that approximately $8.9\%$ of the VPN data is misclassified to the NonVPN class within the baseline model. However, the misclassification rate increases for the model containing one Byzantine client. Also, the misclassification rate rises proportionally with the increased percentage of incorrect labels. Specifically, when $10\%$ of VPN samples are changed to the NonVPN class, the misclassification rate increases to $9.77\%$. When $50\%$ of VPN class samples are flipped, the misclassification rate reaches $10.67\%$ in the test set. The results for two and three Byzantine clients follow the same pattern as those for one Byzantine client. Figures~\ref{fig:2_client_mcr_reputation}, \ref{fig:3_client_mcr_reputation} depict a clear increase in the misclassification rate as the number of Byzantine clients increases. For instance, when three Byzantine clients are present in the SL training round, and $50\%$ of samples are changed to another class, the misclassification rate reaches $13.05\%$. 

To avoid the inclusion of low-quality models in aggregation, we implemented SL and computed a reputation score for each model using validators, as explained in Section~\ref{sub:validatorsVerification}. Within this framework, we examined two scenarios. First, we assessed the computation of the reputation score under the assumption that all validators are honest within the SL system, even when confronted with noisy labels. Subsequently, we examined the performance of the reputation model when faced with dishonest validators.
 
\subsubsection{Scenario 1: Reputation model with noisy labels and honest validators}

In this, all the validators are honest, and each validator uses the same data to evaluate different models, but different validators use different validation data. Figure~\ref{fig:1_Client_reputation} illustrates the reputation scores of 10 participating organizations in each round of our GM training. In this Figure, Model 5, which is highlighted by a dashed line, represents the Byzantine client, while the legitimate models are depicted in solid lines. Also, the five different cases, corresponding to varying percentages of label flipping, are presented in the subfigures. 
In the initial rounds of SL training, the reputation scores are slightly lower than in subsequent rounds. Specifically, the reputation score of Model 8 is notably low in the first few rounds. However, following the reception of the aggregated model, Model 8 shows improvement, with its reputation score steadily increasing. As the rounds of SL progress, the models converge, and the reputation scores appear to stabilize. This indicates minimal changes compared to previous rounds. Furthermore, it is evident that the reputation scores of Byzantine clients are consistently lower than those of other clients. With an increase in the number of incorrect labels, there is a noticeable divergence in the reputation scores between legitimate clients and Byzantine clients. Particularly, when 30\%, 40\%, and 50\% of samples are flipped, a clear distinction emerges, showing the quality of legitimate clients. Similarly, the scenario involving two Byzantine clients shows a similar pattern to that of a single Byzantine client. From Figure~\ref{fig:2_Client_reputation}, it is evident that both Byzantine models (Model 2 and Model 5) have lower reputation scores than others. Also, when 50\% labels of samples are altered, the reputation scores of Byzantine clients range between 0.80 to 0.85, whereas others range from 0.90 to 0.95. 

SL with three Byzantine clients also exhibits low reputation scores for those clients. Figure~\ref{fig:3_Client_reputation} depicts the reputation scores of each model across multiple GM training rounds containing three Byzantine clients. In this case, Models 2, 5, and 8 are identified as Byzantine, with their reputation scores notably lower than others. The reputation score of Model 8 varies depending on whether it behaves as Byzantine or not. Specifically, with 50\% of samples flipped, the reputation score of Model 8 drops to approximately 0.85, whereas it is above 0.90 when not considered as a Byzantine node. Byzantine clients will exhibit lower reputation scores than others for 10\% and 20\% flipping. However, the difference is insignificant due to the limited number of label alterations. This indicates that in scenarios where organizations with a higher proportion of incorrect labels exhibit comparatively lower reputation scores. So, based on the reputation score, we opted to include only 70\% of the total participating organizations with high reputation scores for aggregation.
\par
The misclassification rate of SL with reputation, compared to the model without reputation scheme, is illustrated in Figure~\ref{fig:mcr_reputation}. Figures~\ref{fig:1_client_mcr_reputation},~\ref{fig:2_client_mcr_reputation}, and~\ref{fig:3_client_mcr_reputation} clearly demonstrate a reduction in the misclassification rate when low-quality models are excluded based on the reputation score. For instance, with three Byzantine clients and 50\% of label flipping, the misclassification rate reduced to 9.12\% from 13.05\%. As the reputation model excludes low-quality models from aggregation, our SeCTIS framework reduces misclassification rates. Achieving the same results as FL without Byzantine clients becomes challenging. However, in some cases, the misclassification rate reaches the baseline.
\par
We also examined the recall of the VPN class under two conditions: when the reputation scheme is not utilized and when SL with reputation is applied, as depicted in Figure~\ref{fig:source_recall_reputation}. In this figure, the recall of the baseline model is denoted by a dashed line, with a value of 88.4162. The presence of Byzantine clients during collaborative learning will diminish the source recall; for instance, with SL involving three Byzantine clients and 50\% of label alterations, the source recall drops to 84.36\%. However, with aggregation based on the reputation scheme, the recall consistently improves across all scenarios, closely approaching the performance of the baseline model. Specifically, when three Byzantine clients are present, and 50\% of the labels are flipped, our SeCTIS framework enables a significant increase in the recall rate to 88.35\%.

Our experimental findings show that integrating a reputation scheme into SL effectively reduces the impact of Byzantine clients, leading to lower misclassification rates and improved recall rates. Furthermore, this integration helps to identify and exclude low-quality models during the GM aggregation. Thus enhancing the overall performance and resilience of collaborative learning.
\begin{figure*}[t]
    \centering
    \subfigure[10\% Flipped Data]
    {
    \includegraphics[width=0.16\textwidth]{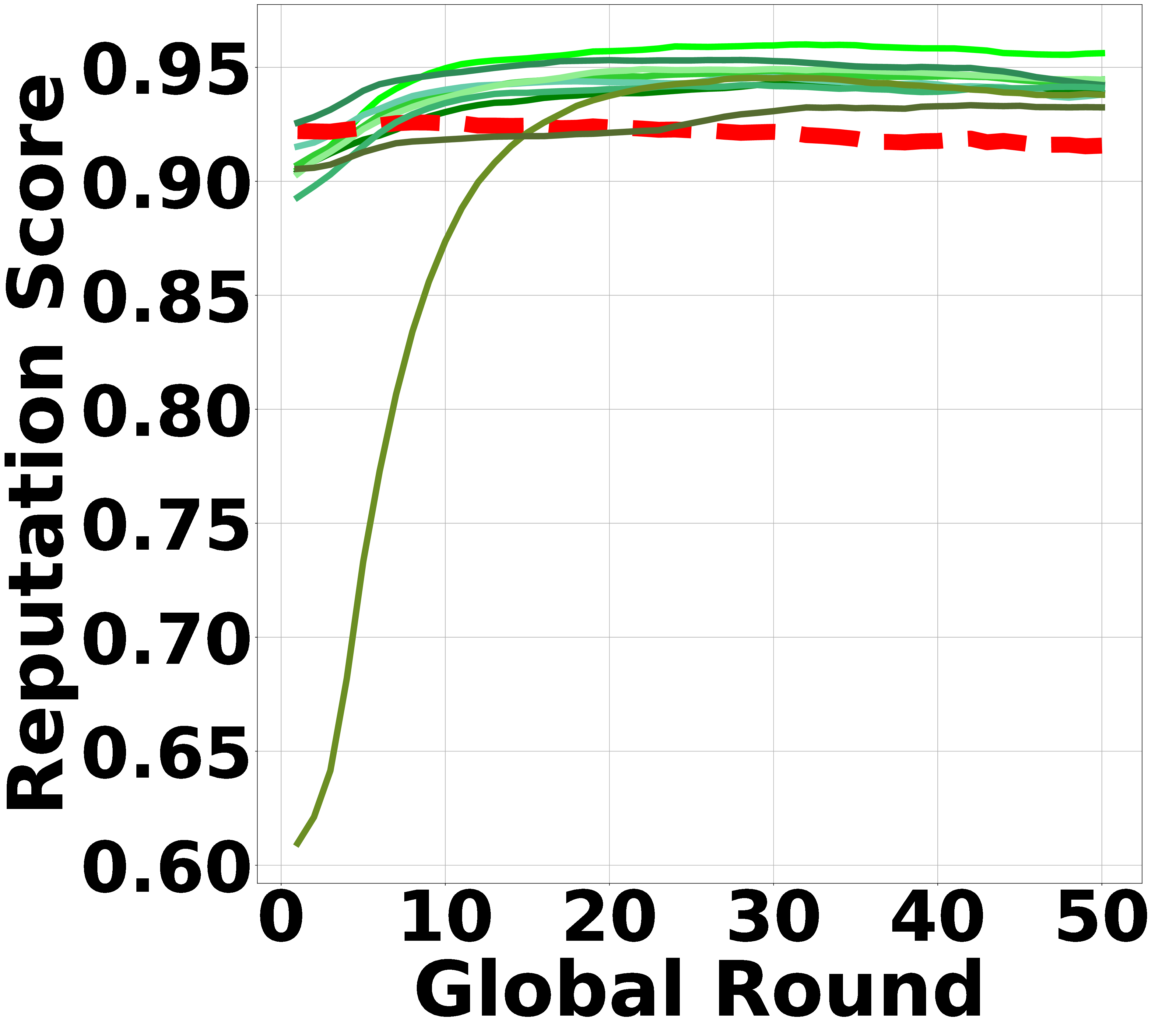}
    \label{fig:Dishonest_1_Client_reputation_10}
    } 
    \subfigure[20\% Flipped Data]{
    \includegraphics[width=0.16\textwidth]{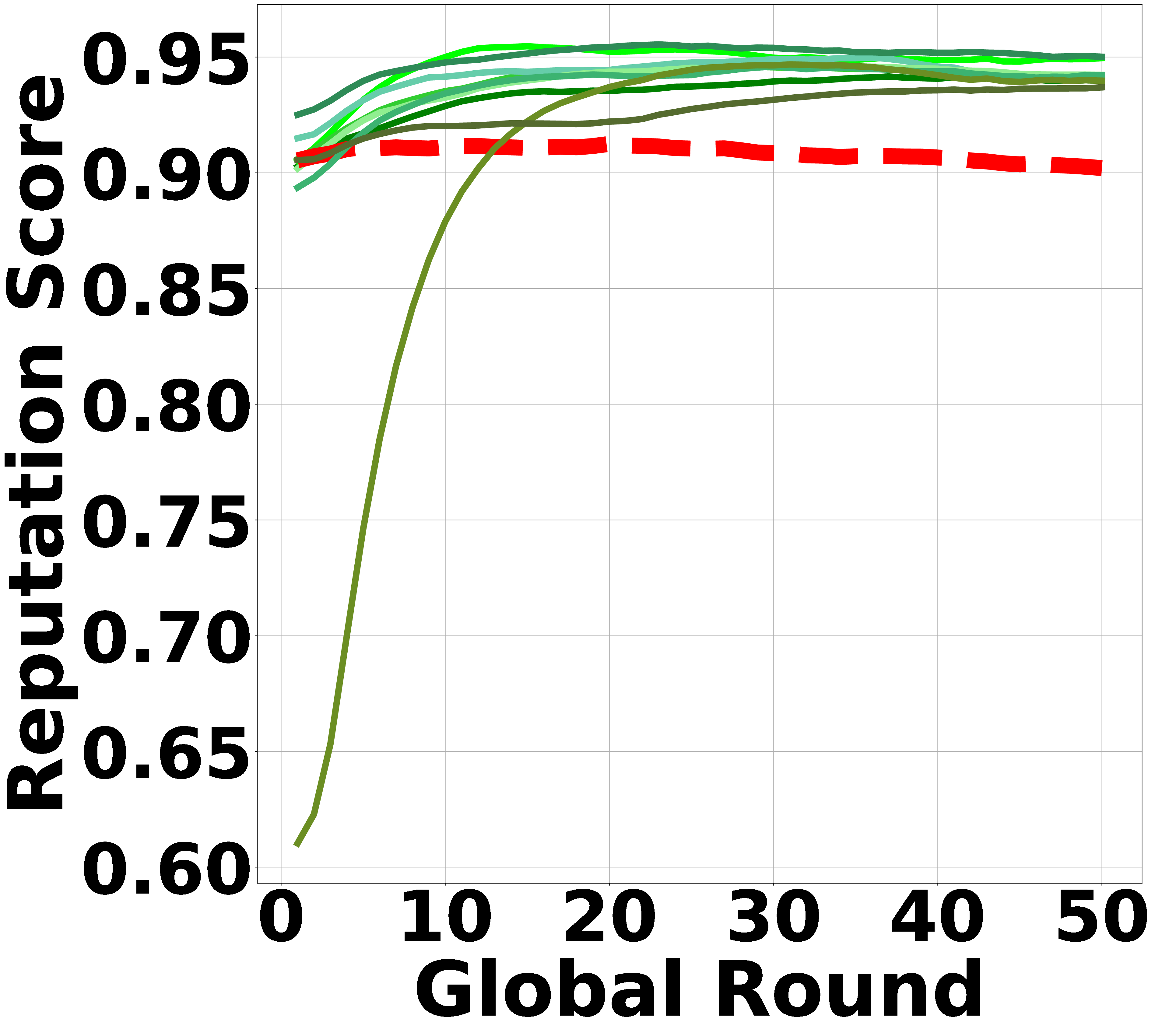}
    \label{fig:Dishonest_1_Client_reputation_20}
    } 
    \subfigure[30\% Flipped Data]
    {
    \includegraphics[width=0.16\textwidth]{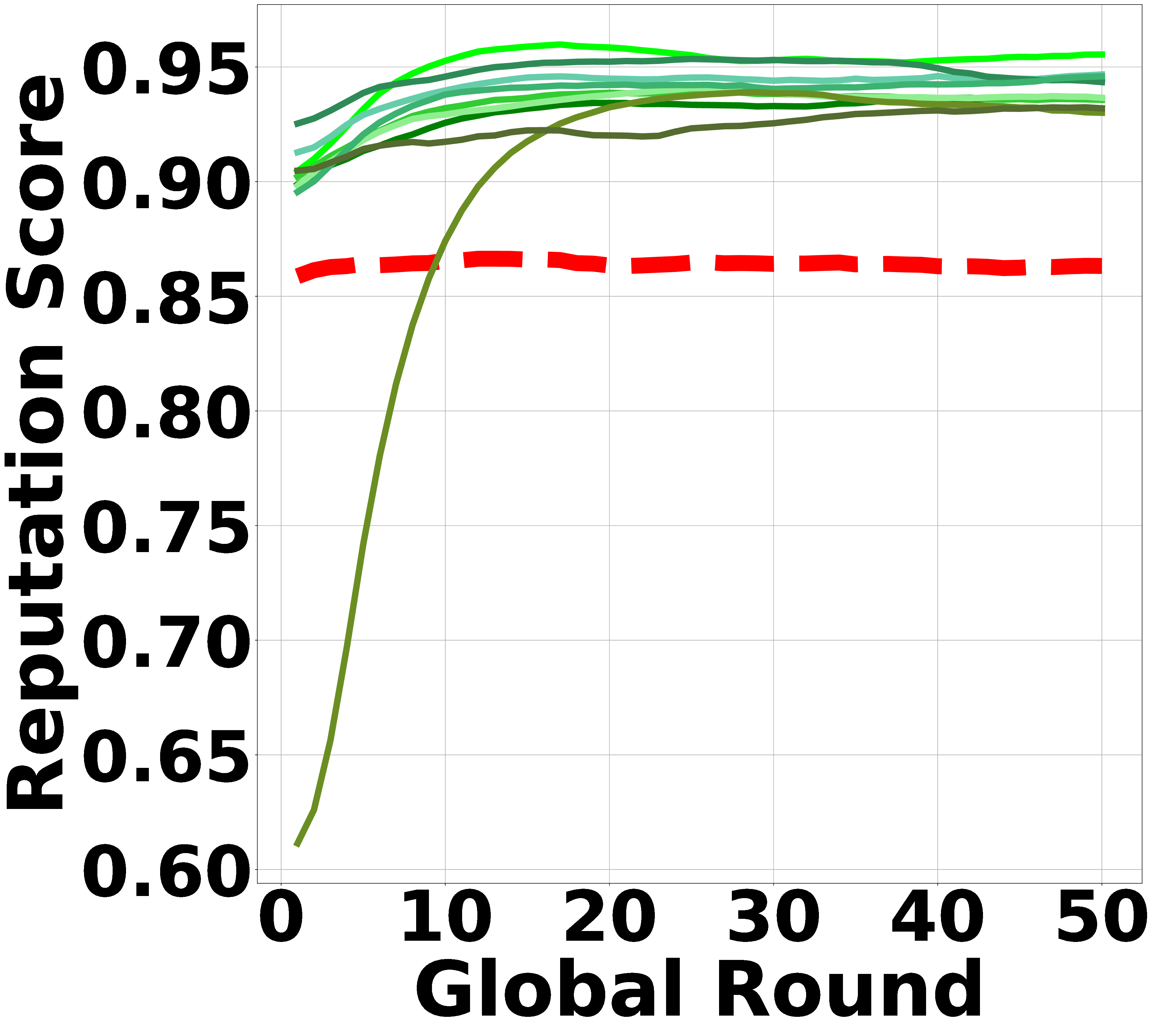}
    \label{fig:Dishonest_1_Client_reputation_30}
    } 
    \subfigure[40\% Flipped Data]
    {
    \includegraphics[width=0.16\textwidth]{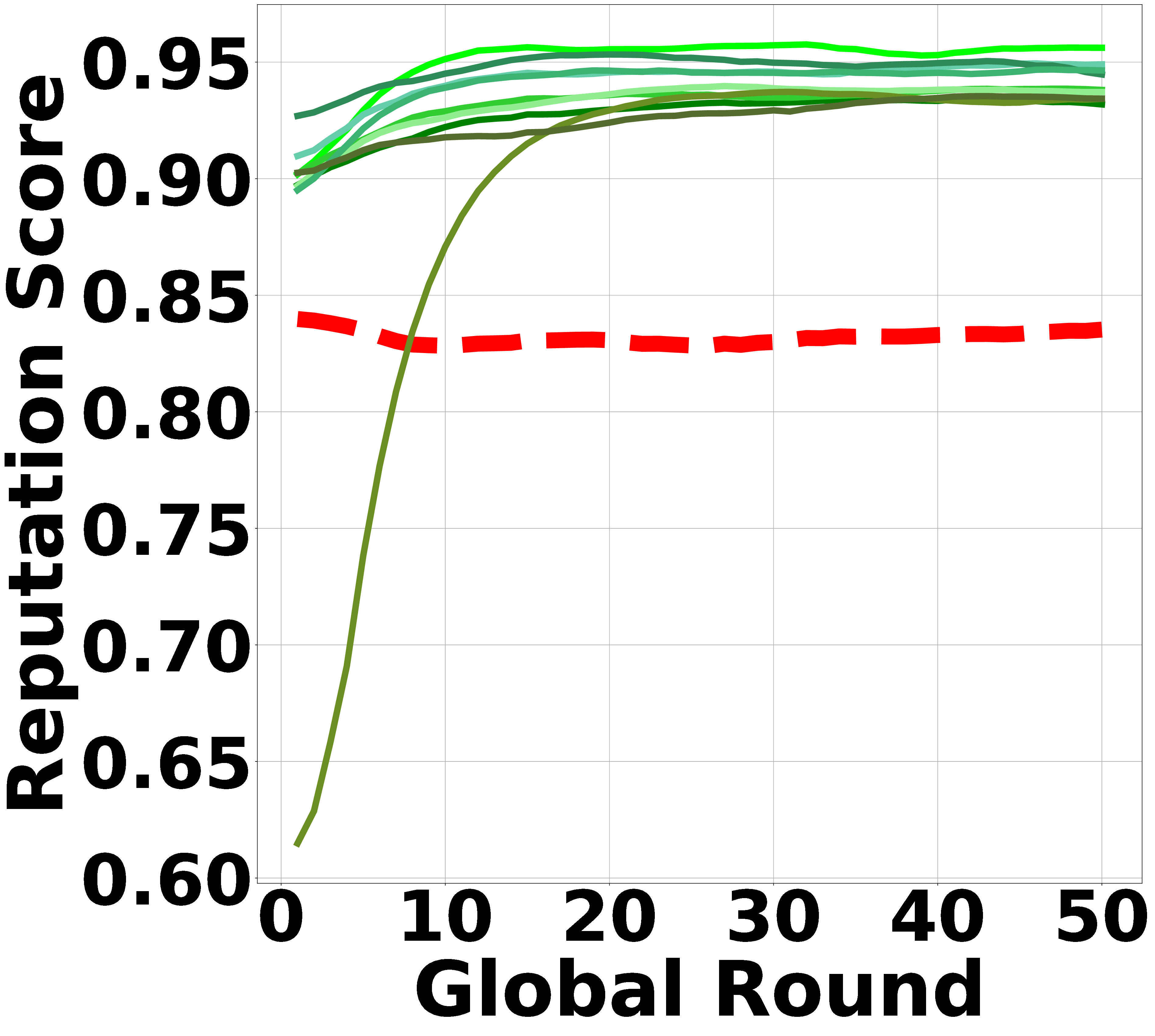}
    \label{fig:Dishonest_1_Client_reputation_40}
    } 
    \subfigure[50\% Flipped Data]
    {
    \includegraphics[width=0.16\textwidth]{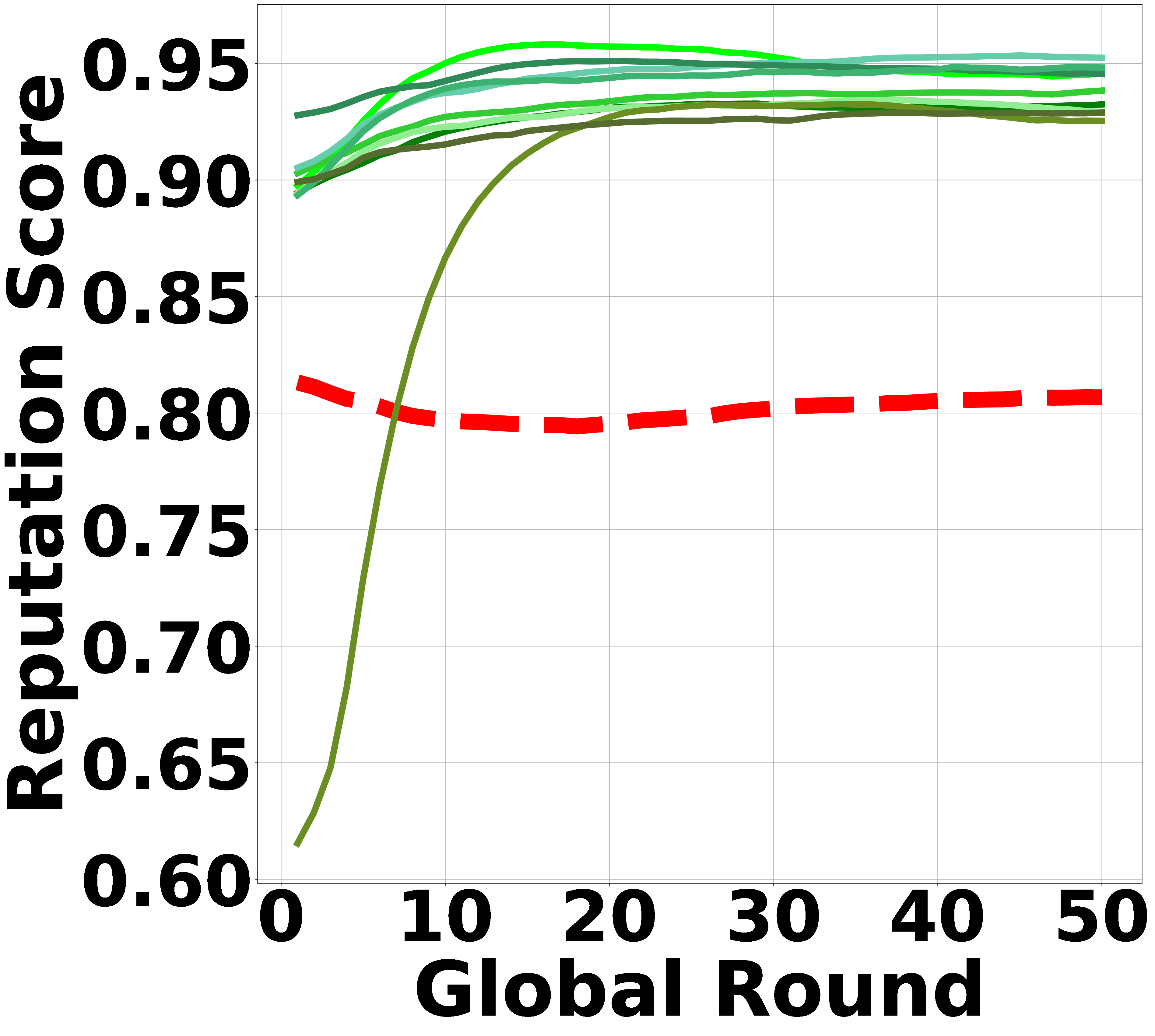}
    \label{fig:Dishonest_1_Client_reputation_50}
    } 
    \subfigure
    {
    \includegraphics[width=0.08\textwidth]{1_Client_legend}
    } 
   
    \caption{Reputation scores of the model with $10\%$ Byzantine clients. Byzantine clients (Model 5), highlighted in red, contrast with benign participants represented in green.} 
    \label{fig:dishonest_1_Client_reputation}
\end{figure*}
\begin{figure*}[t]
    \centering
    \subfigure[10\% Flipped Data]
    {
    \includegraphics[width=0.16\textwidth]{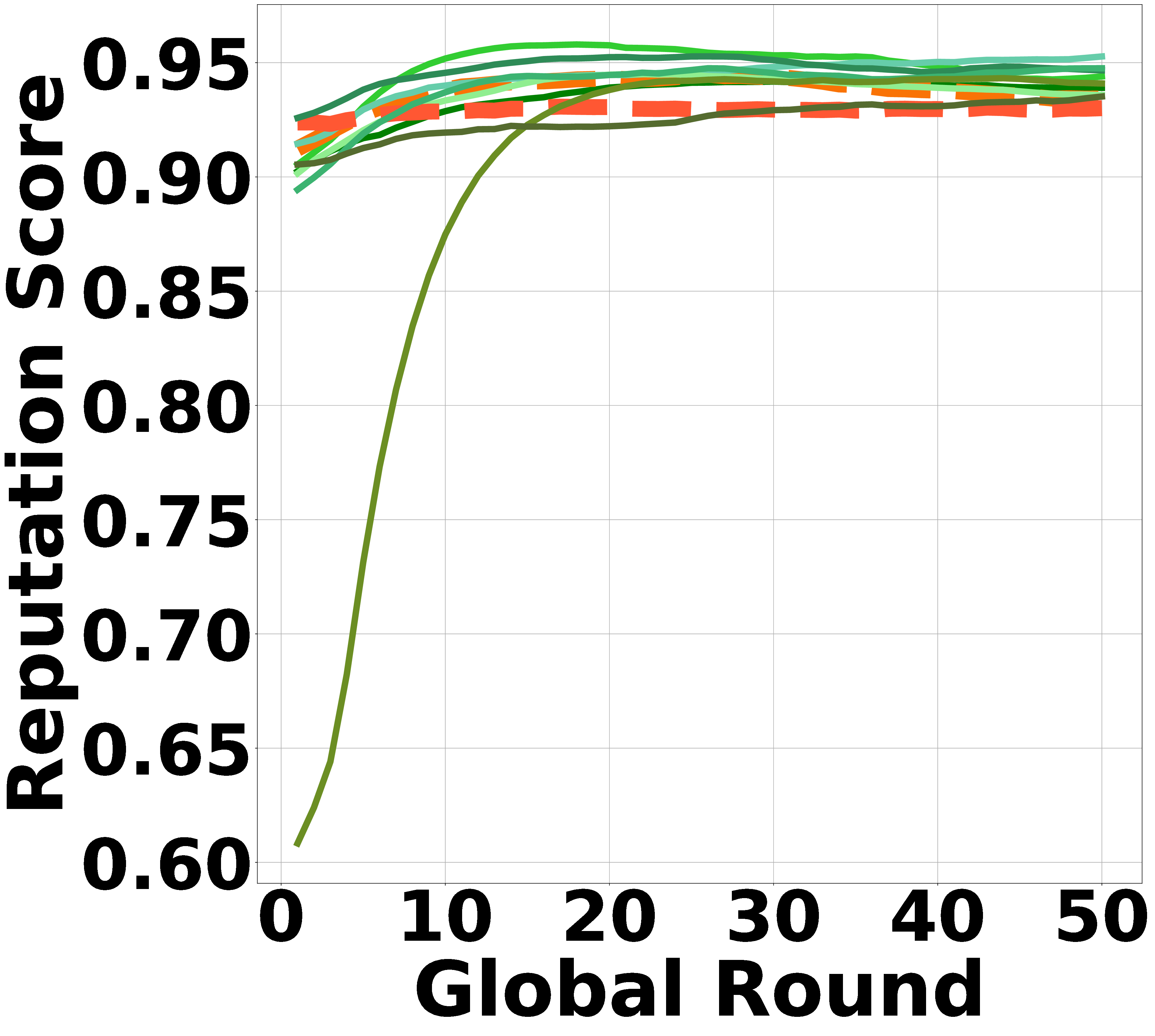}
    \label{fig:Dishonest_2_Client_reputation_10}
    } 
    \subfigure[20\% Flipped Data]{
    \includegraphics[width=0.16\textwidth]{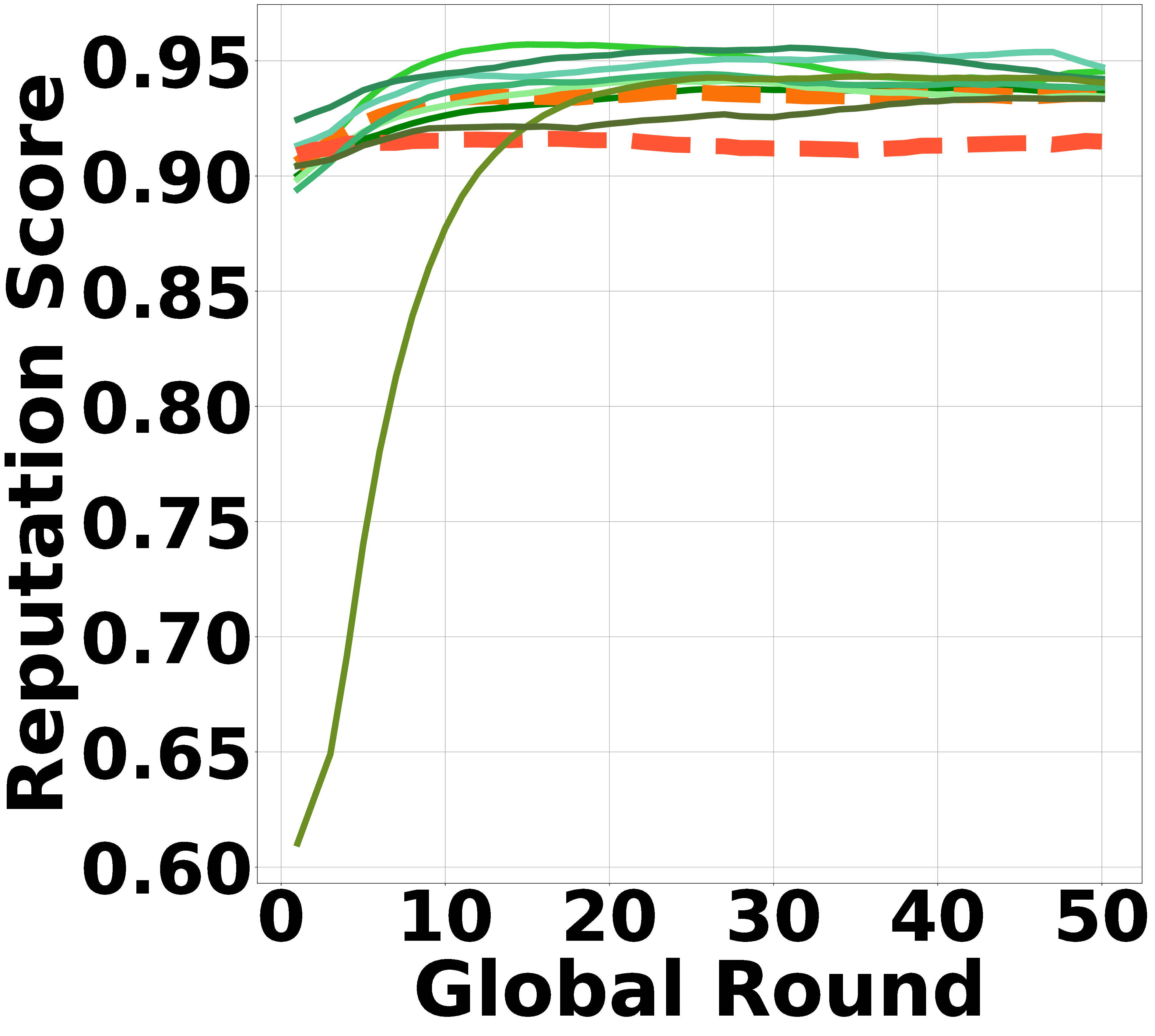}
    \label{fig:Dishonest_2_Client_reputation_20}
    } 
    \subfigure[30\% Flipped Data]
    {
    \includegraphics[width=0.16\textwidth]{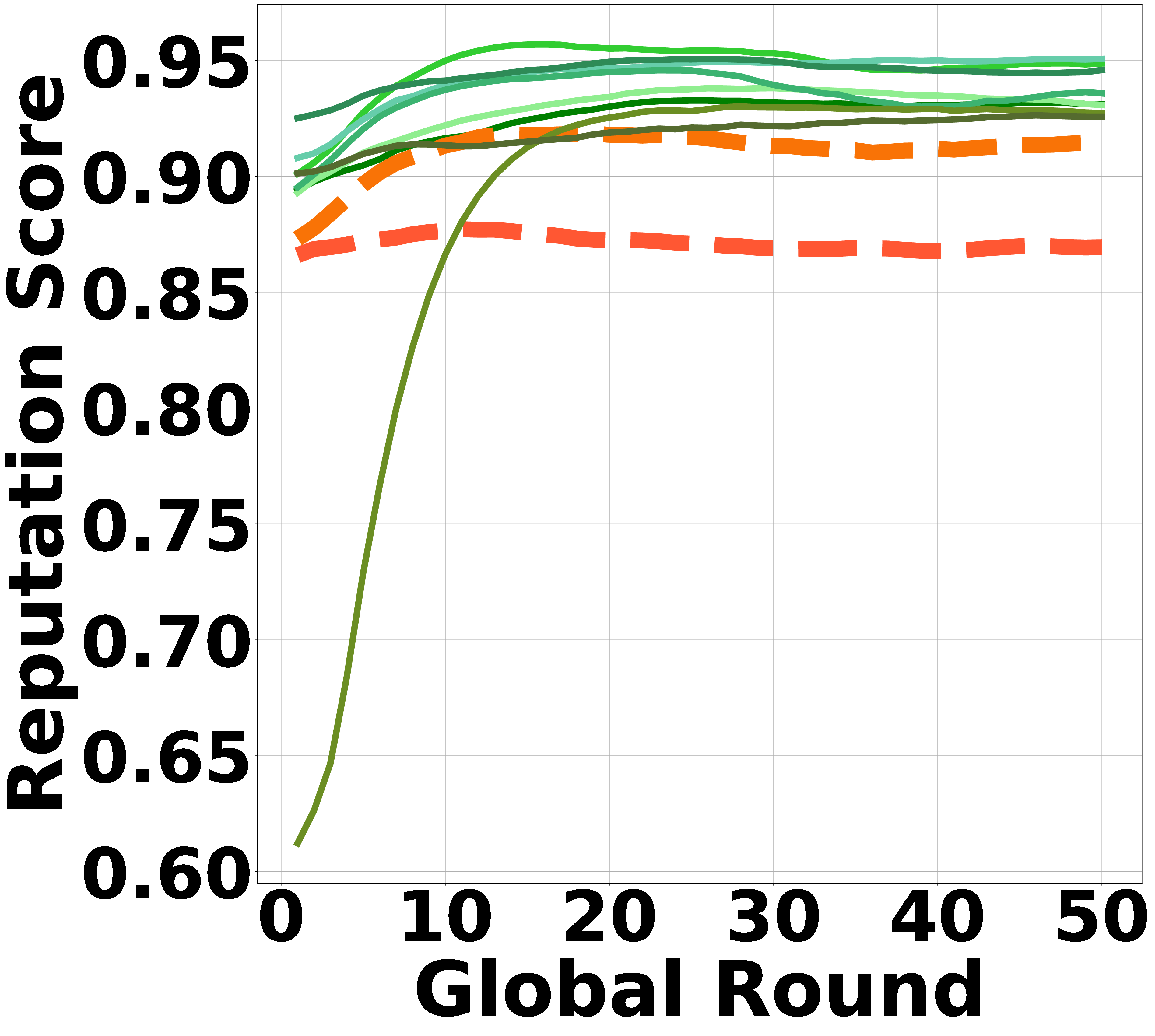}
    \label{fig:Dishonest_2_Client_reputation_30}
    } 
    \subfigure[40\% Flipped Data]
    {
    \includegraphics[width=0.16\textwidth]{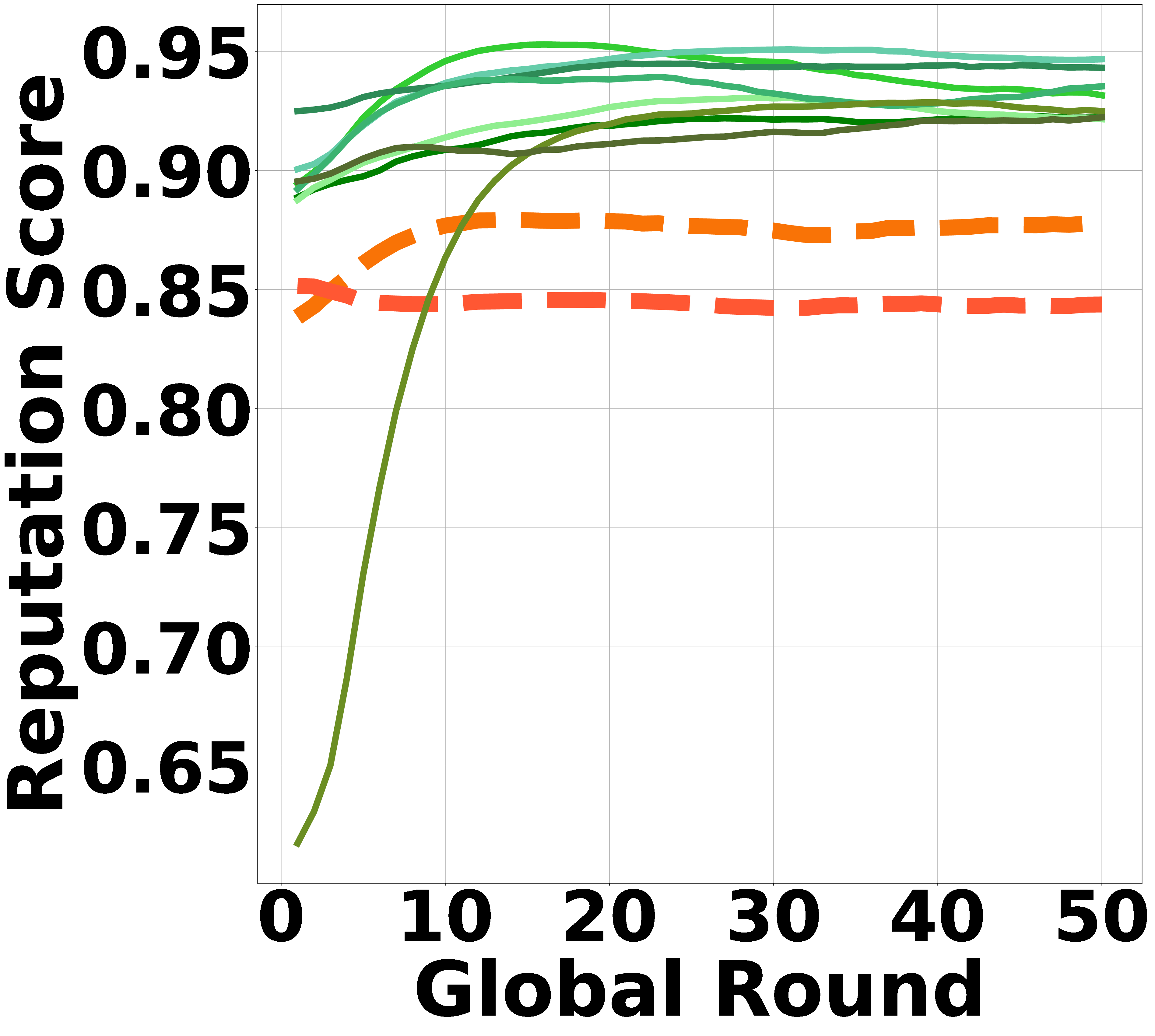}
    \label{fig:Dishonest_2_Client_reputation_40}
    } 
    \subfigure[50\% Flipped Data]
    {
    \includegraphics[width=0.16\textwidth]{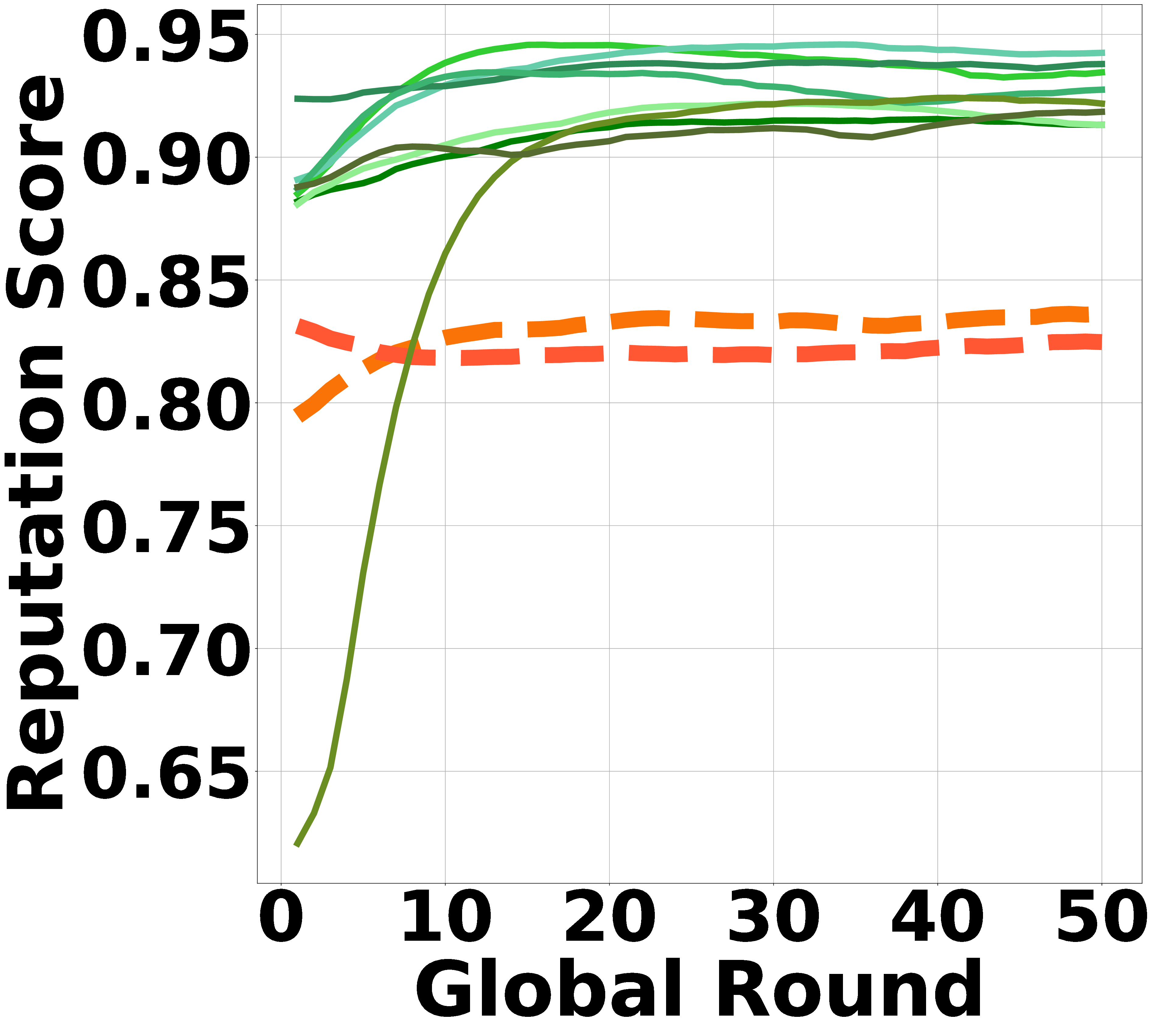}
    \label{fig:Dishonest_2_Client_reputation_50}
    } 
    \subfigure
    {
    \includegraphics[width=0.08\textwidth]{2_Client_legend}
    } 
   
    \caption{Reputation scores throughout rounds of swarm learning involving $20\%$ Byzantine clients. Two Byzantine models (Model 2 and 5) are emphasized in red variant colors, and legitimate participants are highlighted in green.} 
    \label{fig:dishonest_2_Client_reputation}
\end{figure*}
\begin{figure*}[t]
    \centering
    \subfigure[10\% Flipped Data]
    {
    \includegraphics[width=0.16\textwidth]{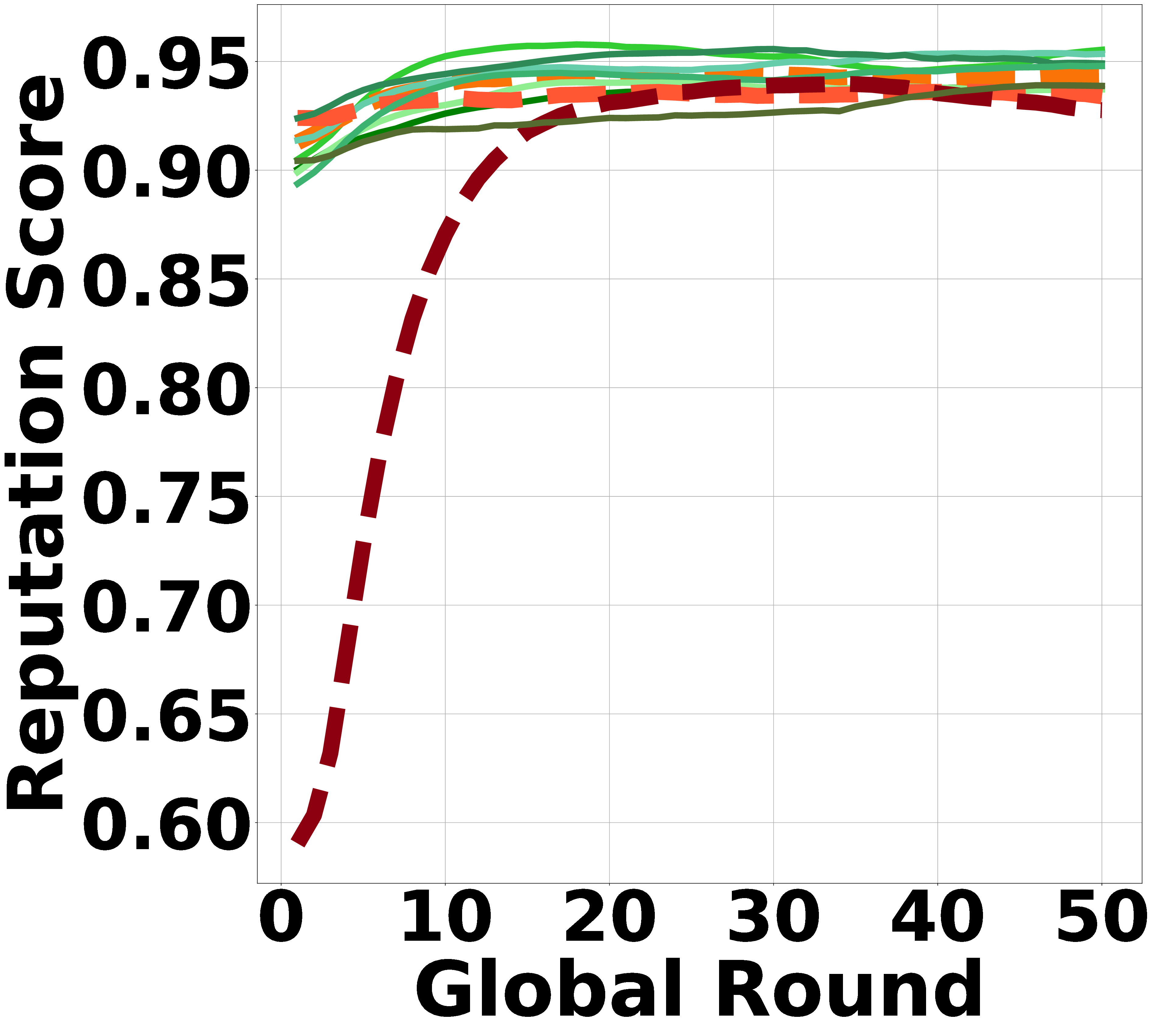}
    \label{fig:Dishonest_3_Client_reputation_10}
    } 
    \subfigure[20\% Flipped Data]{
    \includegraphics[width=0.16\textwidth]{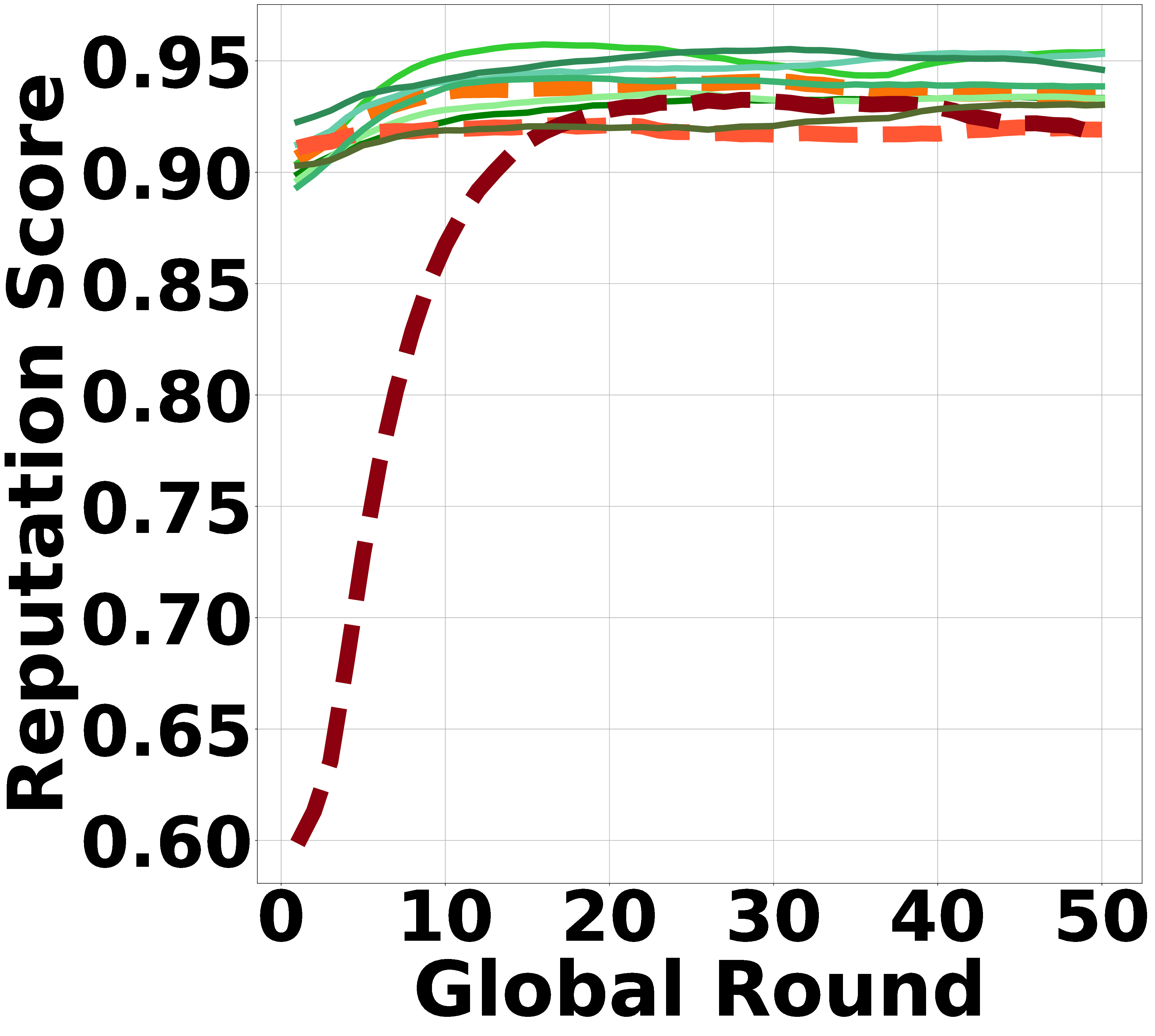}
    \label{fig:Dishonest_3_Client_reputation_20}
    } 
    \subfigure[30\% Flipped Data]
    {
    \includegraphics[width=0.16\textwidth]{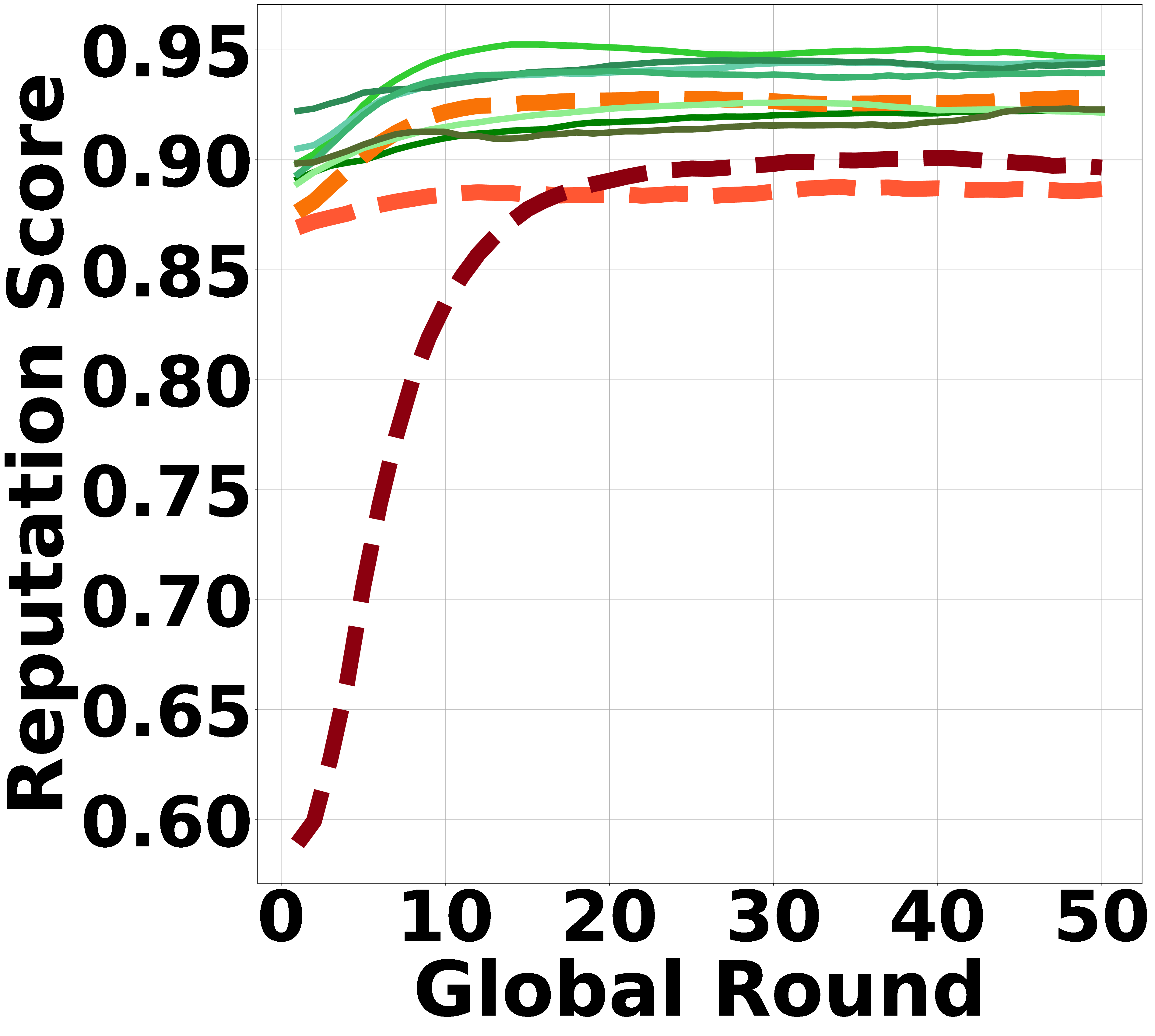}
    \label{fig:Dishonest_3_Client_reputation_30}
    } 
    \subfigure[40\% Flipped Data]
    {
    \includegraphics[width=0.16\textwidth]{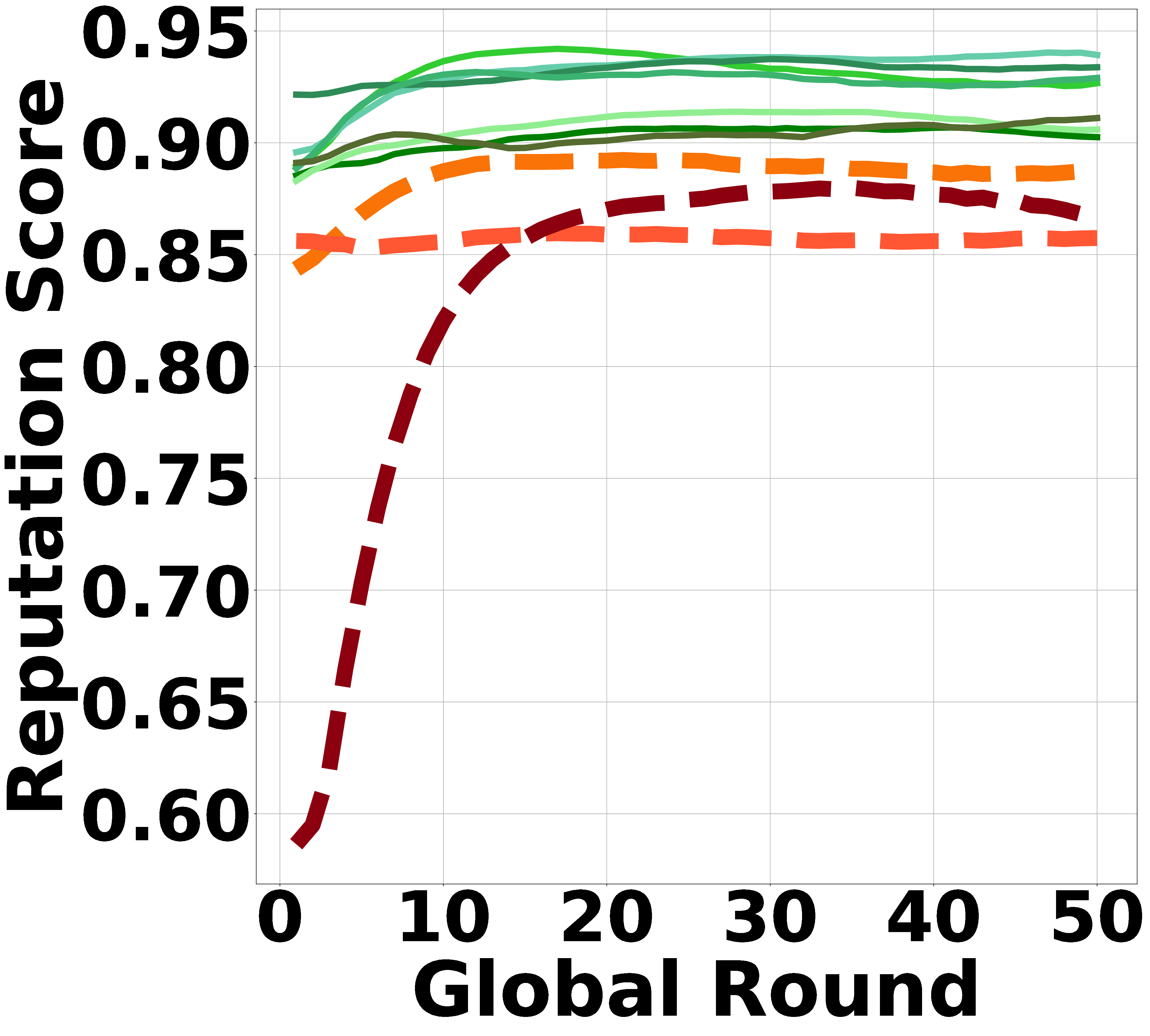}
    \label{fig:Dishonest_3_Client_reputation_40}
    } 
    \subfigure[50\% Flipped Data]
    {
    \includegraphics[width=0.16\textwidth]{Reputation_3_Client_50_percentage_3_1}
    \label{fig:Dishonest_3_Client_reputation_50}
    } 
    \subfigure
    {
    \includegraphics[width=0.08\textwidth]{3_Client_legend}
    } 
   
    \caption{Reputation scores across communication rounds with a presence of $30\%$ Byzantine clients. Three Byzantine models (Models 2, 5, and 8) are vivid in red, while benign models are represented in green.} 
    \label{fig:dishonest_3_Client_reputation}
\end{figure*}
\subsubsection{Scenario 2: Reputation model with noisy labels and faulty validators}
We introduced faulty validators that collude with Byzantine clients to evaluate our framework with dishonest validators. Our implementation of a consensus mechanism is designed to ensure fault tolerance and robustness in the face of these challenges. This mechanism operates effectively as long as a majority of the nodes remain honest and operational. For a system with $N$ nodes with $f$ faulty nodes, the system ensures correctness if at least $2f+1$ nodes are honest and operational. 

In our framework, we evaluated scenarios with a total of 10 validators, indicating that this configuration can withstand up to three faulty validators. For this, we considered scenarios with three faulty validators colluding with Byzantine clients. We simulated the situation with one Byzantine client and three faulty validators, specifically Validators 0, 1, and 5, colluding with Byzantine Client 5 and  
 holding Client 5's noisy data. This setup aims to improve the reputation score of Model 5 artificially. However, since the majority of validators are honest, the overall system can still identify Model 5 as low quality. The reputation scores calculated in the presence of faulty validators are illustrated in Figure~\ref{fig:dishonest_1_Client_reputation}. When comparing Figures\ref{fig:1_Client_reputation} and \ref{fig:dishonest_1_Client_reputation}, there is no significant difference in the overall reputation scores of the clients. This consistency is achieved with three faulty validators as the system reaches a consensus on the correct state without exceeding the tolerated threshold.

Similarly, we simulated scenarios with two Byzantine clients and three Byzantine clients. With two Byzantine clients, Validators 0, 2, and 5 were faulty; Validators 0 and 5 colluded with Byzantine Client 5, while Validator 2 colluded with Byzantine Client 2. Validators 2, 5, and 8 were faulty in the three Byzantine clients scenario and colluded with Byzantine Client 2, Byzantine Client 5, and Byzantine Client 8, respectively. Our framework exhibits the same behavior when there are two or three Byzantine clients and three faulty validators. However, there are marginal differences in the reputation scores when comparing scenarios with no faulty validators to those with up to three validators. Our framework demonstrates the same pattern as when all validators are honest, as shown in Figures \ref{fig:2_Client_reputation}, \ref{fig:dishonest_2_Client_reputation} and \ref{fig:3_Client_reputation}, \ref{fig:dishonest_3_Client_reputation}. In all cases, our reputation model discards the Byzantine clients, even if the system includes faulty validators. The experimental results show that the framework can handle up to three faulty validators effectively, ensuring reliable consensus and maintaining the system's integrity. 

In our evaluation, our framework maintains safety and liveliness, ensuring system integrity and responsiveness. Safety mechanisms guarantee correctness and consistency, even when faced with adversarial actions. Despite challenges posed by Byzantine clients and colluding validators, the majority of honest validators consistently identified and rejected malicious data injections. For example, in the simulated scenario featuring one Byzantine client and three faulty validators, attempts to manipulate the reputation score of Model 5 were thwarted by the vigilance of honest validators. 


\section{Security Analysis of Validator Operations}
\label{security-analysis-of-zkp-operations}
We employed the EZKL library in our framework to implement and manage ZKPs~\cite{ezkl-docs}. EZKL is specifically designed to facilitate efficient and secure proof generation and verification, making it an integral component in ensuring the privacy and integrity of our computational processes. This section details how the underlying mechanisms upheld the security guarantees using EZKL. In Subsections A, B, and C, we will provide a high-level overview of the workflow EZKL in our solution. 

\begin{table}
\scriptsize
\centering
\caption{List of Terms used in ZKP concept} \label{tab:ZkConceptSymbols}
\begin{tabular}{|l|p{5.3cm}|}
\hline
    Elliptic Curve Point & A point on an elliptic curve \\\hline
    Commitment & Cryptographic binding of data to a fixed value \\\hline
    KZG-commitments & A polynomial commitment scheme used in ZKPs \cite{kate2010constant} \\\hline
    Poseidon Hash & A cryptographic hash function used in ZKPs \\\hline
    Witness & Data used to generate a zero-knowledge proof \\\hline
    SRS & Structured Reference String \cite{ef2023kzg} \\\hline
    Circuit & A set of equality constraints for ZKP computations \\\hline
    EZKL & Python library for zero-knowledge proof systems \cite{ezkl-docs} \\\hline
    HALO2 & A recursive Zero-Knowledge Proof protocol \cite{zcash-halo2} \\\hline
   
    ZKSNARK & A Zero Knowledge Proof System \cite{ben2019succinct} \\\hline
    
\end{tabular}
\end{table}


\subsection{Setup Phase}

 This is the initial phase which configures Zero Knowledge solution. The objective of this phase is the generation of a key pair, $<proving\ key,\ verification\ key>$, 
 and setting the circuit as done in~\cite{south2024verifiable}. The key pair and circuit settings are used in the proof generation and, subsequently, in the verification phases, which are handled by the validators and the { \tt Verifier SC}, respectively. For this objective,  based on the aforementioned key pair and circuit setting, the federated learning clients build an Ethereum Virtual Machine (EVM) verifier and deploy it to the blockchain. In our solution, we inherit the functionalities of the EZKL framework and use them to carry out the steps above. 
 

The input to the setup phase includes:
\begin{itemize}
    \item The local model in the ONNX format.
    \item The Structured Reference String (SRS) contains all the information to generate and verify the proofs \cite{maller2019sonic,ef2023kzg}. In our framework, we use a global SRS for all ZKP operations across all learner nodes. The SRS is stored in IPFS and its address is available in the {\tt Coordinator SC}.
\end{itemize}

The outputs of the setup phase are:
\begin{itemize}
    \item The { \em proving key} used to generate the proof. This key can either be provided to the validators by a trusted third party or built through a Secure Multiparty Computation performed by the involved organizations \cite{south2024verifiable}.
    \item The { \em verification key}. { \tt Verifier SC} takes the proof submitted by the validator as input and uses this { \em verification key} to asses whether the proof is valid or not. The { \em verification key} is pre-embedded in { \tt Verifier SC} smart contract on the blockchain.
    \item The circuit settings. The circuit is created based on the EZKL framework. Circuit settings include tolerance, input scale, and lookup range. Tolerance defines the acceptable error margin in computations, while the input scale specifies the scaling factor for input values. Lookup range setting the range of values for lookup operations. By changing these settings, we can fine-tune the performance and accuracy of cryptographic computations.
\end{itemize}


\subsection{Commitment Creation}
\label{commitments-using-hash-elliptic-curves}
Validators generate proofs using HALO2 and KZG commitments. These commitments include the input data hash, model hash, and public output. The hash is created using the \textit{Poseidon hash function}, with elliptic curve commitments facilitated by EZKL.

The Structured Reference String (SRS) defined during the Setup Phase includes a generator \( g \) and its powers based on a secret scalar \( s \). The SRS ensures the security of the commitments. In our solution, we define the following commitments:

\subsubsection{Input Data Commitment}
This represents the input data \( x \) to the ONNX model used by the validator. The dataset Poseidon hash is used to build the commitment as follows:

\begin{equation}\label{eq:commitX}
C(PH(x)) = g^{PH(x)}
\end{equation}

\subsubsection{Model Commitment}
Similarly, the commitment for the ONNX model, denoted as \( M \), is defined as follows:

\begin{equation}\label{eq:commitM}
C(PH(M)) = g^{PH(M)}
\end{equation}

The model hash maps the commitment to the target ONNX model.

\subsubsection{Public Output Commitment}
The public output \( y \), generated by running the model  \( M \) on its input data \( x \), is committed as follows:

\begin{equation}\label{eq:commitY}
C(y) = g^y
\end{equation}

These commitments bind the data to specific values while keeping them hidden, except for the public model results.

In our solution, a dedicated verifier contract is in charge of checking and detecting the correctness of the proofs submitted by validators. 
If an attacker attempts to alter the input data, model, or public output, the resulting commitments \( C(PH(x')) \), \( C(PH(M')) \), and \( C(y') \) will be different from the original \( C(PH(x)) \), \( C(PH(M)) \), and \( C(y) \). Due to the properties of the hash function and the elliptic curve commitments, any deviation will be detected by the verifier:


$$
g^{PH(x')} \neq g^{PH(x)},\ g^{PH(M')} \neq g^{PH(M)},\ g^{y'} \neq g^{y}
$$

Also, to forge a valid proof using different points, an attacker would need to find consistent polynomial evaluations for points \( g^{PH(x')} \), \( g^{PH(M')} \), and \( g^{y'} \) that correspond to valid inputs. This is equivalent to solving the polynomial evaluation problem at secret points defined by the SRS, which is computationally infeasible. This ensures that any change in the inputs or attempts to generate fake proofs will result in a detectable change in the results, maintaining the binding property. The security of our commitments relies on the computational infeasibility of deriving the secret \( s \) from the SRS. Additionally, the Poseidon hash function ensures collision-resistance and preimage-resistance, further securing the integrity of the commitments.

 
\subsection{Proof Generation and Verification}

After the generation of the commitments as explained in Section \ref{commitments-using-hash-elliptic-curves}, the proof can be generated and verified as follows.

\subsubsection{Proof Generation}
The generation of the proof leverages the {\em proving key} for the validator generated during the setup phase.
As mandated in the HALO2 ZK-SNARK proof system, the prover (i.e., the validator in our scenario) uses the  {\em proving key} to generate a zero-knowledge proof, namely \( \pi \), which includes the commitments\footnote{We do not report the details about the proof generation as they are part of the basic HALO2 ZK-SNARK framework and, hence, not defined in our solution.}. The proof demonstrates that the prover uses \( PH(x) \), \( PH(M) \), and obtains \( y \) as output.

\subsubsection{Verification of Proof}  
To validate the proof \( \pi \) returned by a validator, {\tt Verifier SC} uses the {\em verification key} to ensure the proof \( \pi \) correctly proves knowledge of \( PH(x) \), \( PH(M) \), and obtained \( y \), without having access to execution details. This step ensures that the commitments are valid under the HALO2 ZK-SNARK protocol used by EZKL.

\section{Validation Robustness against Example Attack Scenarios}
\label{sec:robustnessofzk}

As we stated before, our solution guarantees robust and secure model validation through the use of zero-knowledge proofs.
In particular, such proofs certify that each validator has used a defined set of input data, has tested target models, and has generated specific outputs during its validation activity. 
To demonstrate the efficacy of our solution, in this section, we focus on four possible attack strategies that are derived from the attack model described in Section \ref{sec:attackModel}.
In particular, we focus on scenarios where the attacker controls some validator nodes and tries to compromise their standard behavior.
To achieve this objective, the attacker can try to exploit the following components: {\em (i)} input data; {\em (ii)} secure parameters in the Proof; {\em (iii)} model weights; {\em iv} public output.

Without losing generality, we focus on the case in which the attacker controls a single validator, and we consider four scenarios, namely Scenario $1$, Scenario $2$, Scenario $3$ 
and Scenario $4$. In each scenario, the adversary focuses on one of the above-mentioned components to forge the attack. 
The adversary objective 
is to elude the {\tt Verifier SC} so as to jeopardize the estimated trust score for target models. This causes model exclusion (resp., inclusion) in the subsequent aggregation step.

\begin{table}[]
\scriptsize
\centering
\caption{Scenario 1} \label{tab:validation-protection-scenario-1}
\begin{tabular}{l|l|l|}
\cline{2-3} 
&\textbf{Valid} & \textbf{Not valid} \\ 
\hline
\multicolumn{1}{|l|}{\textbf{Data Hash}}  & cfbacc & 0fbacc \\\hline
\multicolumn{1}{|l|}{\textbf{Proof Bytes}} & 0x10513eb\dots & 0x10513eb\dots \\\hline
\multicolumn{1}{|l|}{\textbf{Output}} & 1 & The constraint system is not satisfied\\\hline
\end{tabular}
\end{table}

\par
In Scenario $1$, a malicious validator tries to deceive the {\tt Verifier SC} by exploiting different input data to generate the public output for a target model. 
The rationale behind this strategy is to penalize 
a target model by using special input data to validate it.
As discussed in Section \ref{security-analysis-of-zkp-operations}, changing the input to the proof will generate a different commitment value (see Equation \ref{eq:commitX}. Therefore, our {\tt Verifier SC} will return an error as visible in Table \ref{tab:validation-protection-scenario-1}. 

In Scenario $2$, a malicious validator tries to tamper with the proof by changing the Elliptic Curve Points~(ECP). However, again, due to the properties of the ZKP Scheme and Hash Function, this action will generate different values with respect to the content of the SRS, thus making the proof invalid. Therefore, once again, 
the {\tt Verifier SC} will return an error message as visible in Table \ref{tab:validation-protection-scenario-2}.

\begin{table}[]
\scriptsize
\centering
\caption{Scenario 2} \label{tab:validation-protection-scenario-2}
\begin{tabular}{l|l|l|}
\cline{2-3}
&\textbf{Valid} & \textbf{Not valid} \\\hline
\multicolumn{1}{|l|}{\textbf{ECP}} & $30,189$; \dots      & $40,189$; \dots \\\hline
\multicolumn{1}{|l|}{\textbf{Output}} & 1 & The constraint system is not satisfied\\\hline
\end{tabular}
\end{table}

\begin{table}[]
\scriptsize
\centering
\caption{Scenario 3} 
\label{tab:validation-protection-scenario-3}
\begin{tabular}{l|l|l|}
\cline{2-3}
&\textbf{Valid} & \textbf{Not valid} \\ \hline
\multicolumn{1}{|l|}{\textbf{Model Hash}} & 96cbd4d3 & 06cbd4d3 \\\hline
\multicolumn{1}{|l|}{\textbf{Proof Bytes}} & 0x10513eb\dots & 0x10513eb\dots \\\hline
\multicolumn{1}{|l|}{\textbf{Output}} & 1 & The constraint system is not satisfied\\\hline
\end{tabular}
\end{table}

In Scenario $3$, a malicious validator attempts to use a different model to generate results. 
This action will cause the violation of the Model Commitment (see Equation \ref{eq:commitM}), which will hence make the proof not valid. The output produced by the {\tt Verifier SC} is visible in Table \ref{tab:validation-protection-scenario-3}. 

Finally, in Scenario $4$, a malicious validator tries to modify the public output produced by the execution of a target model before submitting it to the blockchain. 
However, because this forged output is not generated by the given input data hash and model, such modification would violate the Public Output Commitment (see Equation \ref{eq:commitY}), which will, hence, invalidate the proof.
In such a case, the {\tt Verifier SC} will return the error visible in Table \ref{tab:validation-protection-scenario-4}.

\begin{table}[]
\scriptsize
\centering
\caption{Scenario 4} \label{tab:validation-protection-scenario-4}
\begin{tabular}{l|l|l|}
\cline{2-3}
&\textbf{Valid} & \textbf{Not valid} \\\hline
\multicolumn{1}{|l|}{\textbf{Public Output}} & $0$ & $1E+63$ \\\hline
\multicolumn{1}{|l|}{\textbf{Proof Bytes}} & 0x1308d6f\dots & 0x1308d6f\dots \\\hline
\multicolumn{1}{|l|}{\textbf{Output}} & $1$ & The constraint system is not satisfied\\\hline
\end{tabular}
\end{table}

\section{Execution performance for ZKP operations}
\label{sec:ZKPTime}
To provide insights about the execution time of the cryptographic operations included in our EZKL-based solution, We conducted a final experiment to measure the execution times of even distinct actions from two key actors: Validators and Learners. In particular, we focused on the following operations.

\begin{enumerate}
    \item  Calibration of input data. (Learner)
    \item  Circuit Compiling. (Learner)
    \item  Circuit Setup. (Learner)
    \item  Circuit Deployment on the blockchain.(Learner)
    \item  Witness File Generation. (Validator)
    \item  Proof Generation. (Validator)
    \item  Proof Verification on the blockchain. (Validator)
\end{enumerate}

We ran this experiment by using our previously discussed neural network on a 2021 Apple M1 Pro with 8 CPU cores at 3200 MHz and 16 GB RAM. Our analysis found out that, without considering ZKP operations, each client spent approximately 100 seconds for a global FL round, while each validator spent around 95 seconds. Table \ref{tab:validatorLearner-zk-executions} provides a comprehensive breakdown of the time duration associated with the various operations performed during the execution of ZKP processes by both Learner and Validator entities. In particular, for Learner operations, it shows that the most time-intensive task is data calibration, requiring an average of 223.14 seconds. Conversely, the compilation of the circuit demonstrates negligible duration, taking merely 0.0052 seconds. Setting up the circuit follows as the next significant operation, consuming an average of 72.89 seconds. The circuit is deployed, and the final operation is completed within an average time of 12 seconds. When it comes to validator operations, witness generation is the quickest task, requiring an average of 0.39 seconds. Conversely, the process of proving is the most time-consuming operation, averaging 83.25 seconds, followed by the verification process, which takes an average of 12 seconds.

We are also aware of other zero-knowledge protocols commonly adopted by researchers and developers, such as Orion and Risc0 \cite{xie2022orion,tairi20212}. According to the EZKL official GitHub repository, in which they compare different frameworks and their benchmark results on various models, EZKL uses $63.95\%$ less memory than Orion and $99.14\%$ less than RISC0. Additionally, it is $2.92$ times faster in proving compared to Orion and $77.29$ times faster compared to RISC0.

Our implementation ensured that both privacy and efficiency were upheld during the proof generation and verification processes. Through a series of experiments, we validated the effectiveness of our framework 
while preventing any attempts by dishonest provers to deceive the verifier. In cases where the proof held true, honest provers were able to demonstrate its truth to the verifier. Conversely, attempts by dishonest provers to falsify proofs were effectively thwarted. 
The cryptographic capability of EZKL's underlying algorithms acted as a robust safeguard against tampering and falsifying proofs, ensuring our framework's integrity and security. Furthermore, EZKL's sophisticated cryptographic techniques obscured the underlying data, preventing any leakage of additional information beyond the truth of the statement during the verification process.

\begin{table}[]
\scriptsize
\centering
\caption{Times taken for learner and validator operations} 
\label{tab:validatorLearner-zk-executions}
\begin{tabular}{ |l|l|l| } 
\hline
 & \textbf{Operation} & \textbf{Time (seconds)} \\
\hline
\multirow{3}{4em}{\textbf{Learner}} &Data Calibration & $223.14$s\\ \cline{2-3}
& Compile Circuit & $0.0052$s \\\cline{2-3}
& Setup Circuit & $72.89$s \\\cline{2-3}
& Deployment & $12$s \\\hline \hline
\multirow{3}{4em}{\textbf{Validator}} & Witness Generation & $0.39$s\\ \cline{2-3}
&Prove & $83.25$s \\\cline{2-3}
&Verify & $12$s \\
\hline
\end{tabular}
\end{table}

\section{Conclusion}
\label{sec:conclusion}

CTI Sharing provides businesses with access to CTI data that ordinarily they may not have been able to obtain without collaboration with other organizations.
This information can be exploited to improve their overall security posture by using the knowledge, experience, and capabilities of the participating
entities. This ensures that the detection and previous knowledge of one organization becomes the future prevention of another one. Unfortunately,
most organizations are hesitant to share their private CTI data for several reasons, such as the 
possible loss of credibility, the lack of trust in other peer organizations, possible risks in using external data that may be false or wrong, and strict law regulations. 

To provide a contribution in this setting, we propose a complete framework called SeCTIS (Secure CTI Sharing) that aims to tackle different challenges. Firstly, it performs a collaborative training of ML models between different organizations through a Swarm Learning approach. 
Furthermore, SeCTIS assesses models and data quality through the use of {\em validator} nodes and the Zero-Knowledge Proof, thus developing a robust reputation Model to estimate the trustworthiness of all participants. To evaluate our reputation model experimentally, we introduced Byzantine clients with noisy labels and simulated such an attack scenario. Our results show that integrating a reputation scheme into Swarm Learning effectively mitigates the influence of noisy labels. Moreover, this integration streamlines the identification and exclusion of low-quality models during aggregation, thereby strengthening the collective performance and resilience of collaborative learning.

To the best of our knowledge, our proposal is novel, and SeCTIS is the first framework to provide complete assurance of data confidentiality, Organization privacy, data, and model quality, and the trustworthiness of participants.

As a limitation of our approach, we identify two main points. First, our framework relies on the EZKL library to generate and verify the required cryptographic proofs. This library allows our framework to run on any blockchain compatible with the Ethereum Virtual Machine (EVM). However, in the current most diffused public blockchains, such as Ethereum, the use of such technology leads to an increase in gas costs for the execution of involved Smart Contracts.
On the other hand, opting for a private permissioned blockchain would require additional solutions for basic operations included in our solution. For example, the simple random selection of validators and aggregators among available clients may become problematic due to the lack of third-party services, such as Chainlink, that provide verifiable random numbers. This would require the deployment of additional, possibly costly, solutions.
As a second point, we showed that secure operations included in EZKL require relatively high execution times.
For both the points above, we underlying that, in our application context in which (even big) organizations are involved, such limitations appear negligible. However, they could become impacting if our solution should be extended to other application contexts, such as scenarios in which Internet of Things devices are directly exposed and involved in the learning task.

In the future, we aim to improve our SeCTIS framework by considering also Organization-Specific Threat Intelligence. In particular, we want to help Security Operations Center (SoC) analysts exchange and train ML models to suit a peculiar organization's needs. 
As for the limitations mentioned above, we are also planning to adapt and test our solution with newer layer 2 blockchains, like Arbitrum, Optimism, and Base, which could allow us to enhance the obtainable performance.

\section*{Acknowledgments}

\begin{figure}[ht]
    \centering
    \subfigure
    {
    \includegraphics[width=0.2\textwidth]{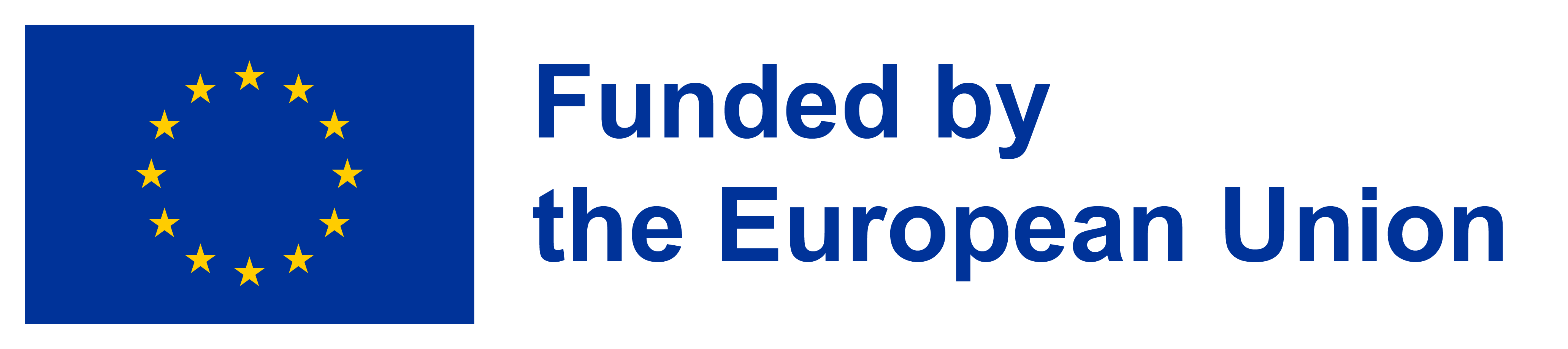}
    } 
    \subfigure{
    \includegraphics[width=0.2\textwidth]{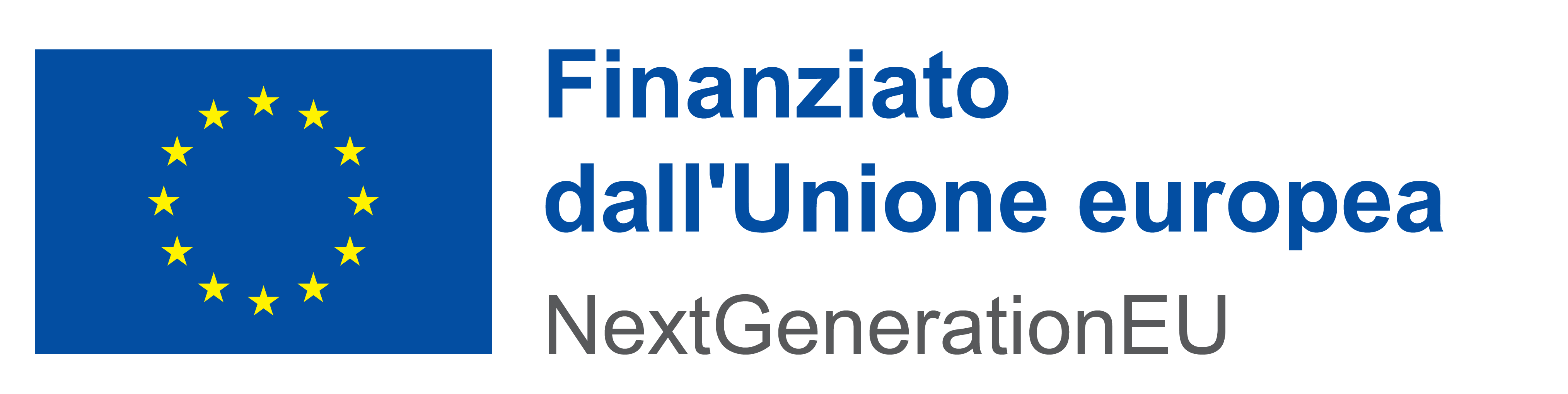}
    } 
\end{figure}

\noindent
This work was supported in part by the following projects: 

{\em (i)} the HORIZON Europe Framework Programme partly supported this work through the project ``OPTIMA - Organization sPecific Threat Intelligence Mining and sharing" (101063107). 

{\em (ii)} The PRIN 2022 Project ``HOMEY: a Human-centric IoE-based Framework for Supporting the Transition Towards Industry 5.0'' (code: 2022NX7WKE, CUP: F53D23004340006) funded by the European Union - Next Generation EU. 

{\em (iii)} The project SERICS (PE00000014) under the NRRP MUR program funded by the EU - NGEU. Views and opinions expressed are however those of the authors only and do not necessarily reflect those of the European Union or the Italian MUR. Neither the European Union nor the Italian MUR can be held responsible for them.
\balance

\end{document}